\documentclass[longauth]{aa}
\usepackage{amsmath}
\usepackage[varg]{txfonts}
\usepackage{hyperref}
\usepackage{cleveref}

\usepackage{natbib}
\bibpunct{(}{)}{;}{a}{}{,} 

\def\lsim{\mathrel{\rlap{\lower4pt\hbox{\hskip1pt$\sim$}}
    \raise1pt\hbox{$<$}}}                
\def\gsim{\mathrel{\rlap{\lower4pt\hbox{\hskip1pt$\sim$}}
    \raise1pt\hbox{$>$}}}                

\def\orcid#1{\href{https://orcid.org/#1}{\includegraphics[scale=1]{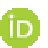}}}

\def\plottwo#1#2{\centering \leavevmode
    \includegraphics[angle=0,width=0.98\columnwidth]{#1} \hfil
    \includegraphics[angle=0,width=0.98\columnwidth]{#2}}

\def\plottwoalmostspecial#1#2{\centering \leavevmode
    \includegraphics[angle=0,width=0.74\columnwidth]{#1} \hfil
    \includegraphics[angle=0,width=1.24\columnwidth]{#2}}

\def\plottwosuperspecial#1#2{\centering \leavevmode
    \includegraphics[angle=0,width=0.98\columnwidth]{#1} \hfil
    \includegraphics[angle=0,width=1.01\columnwidth]{#2}}

\def\plottwonotsospecial#1#2{\centering \leavevmode
    \includegraphics[angle=0,width=0.84\columnwidth]{#1} \hfil
    \includegraphics[angle=0,width=1.13\columnwidth]{#2}}

\def\plotone#1{\centering \leavevmode
    \includegraphics[angle=0,width=1.0\columnwidth]{#1}}

\def\eps@scaling{1.0}%
\newcommand\epsscale[1]{\def\eps@scaling{#1}}%

\hypersetup{
   colorlinks=true,
   linkcolor=blue,
   citecolor=blue,
   urlcolor=cyan, 
}

\usepackage{marginnote}
\usepackage{color}
\begin{document}

\title{The VIMOS Ultra Deep Survey: The reversal of the star-formation rate $-$ density relation at $2 < z < 5$\thanks{This paper is dedicated to Dr.\ Olivier Le F\`evre. Though he was unable to witness its completion, it was his vision and tenacity that made this work, among countless others, possible.}}

\author{B.~C.\ Lemaux\orcid{0000-0002-1428-7036}\inst{\ref{inst1},}\inst{\ref{inst2},}\inst{\ref{inst3}}  \and 
O. Cucciati\orcid{0000-0002-9336-7551}\inst{\ref{inst4}} \and 
O. Le F\`evre\orcid{0000-0001-5891-2596}\inst{\ref{inst1},\ref{*}} \and
G. Zamorani\orcid{0000-0002-2318-301X}\inst{\ref{inst4}} \and 
L.~M. Lubin\orcid{0000-0003-2119-8151}\inst{\ref{inst2}} \and 
N. Hathi\orcid{0000-0001-6145-5090}\inst{\ref{inst5}} \and 
O. Ilbert\orcid{0000-0002-7303-4397}\inst{\ref{inst1}} \and 
D. Pelliccia\orcid{0000-0002-3007-0013}\inst{\ref{inst2},} \inst{\ref{inst6}} \and 
R. Amor\'{i}n\orcid{0000-0001-5758-1000}\inst{\ref{inst7},} \inst{\ref{inst8}} \and 
S. Bardelli\orcid{0000-0002-8900-0298}\inst{\ref{inst4}} \and 
P. Cassata\orcid{0000-0002-6716-4400}\inst{\ref{inst9},} \inst{\ref{inst10}} \and 
R.~R. Gal\orcid{0000-0001-8255-6560}\inst{\ref{inst11}} \and
B. Garilli\orcid{0000-0001-7455-8750}\inst{\ref{inst12}} \and 
L. Guaita\orcid{0000-0002-4902-0075}\inst{\ref{inst13},} \inst{\ref{inst14}}\and 
M. Giavalisco\orcid{0000-0002-7831-8751}\inst{\ref{inst15}} \and 
D. Hung\orcid{0000-0001-7523-140X}\inst{\ref{inst11}} \and
A. Koekemoer\orcid{0000-0002-6610-2048}\inst{\ref{inst5}} \and 
D. Maccagni\orcid{0000-0002-3978-3503}\inst{\ref{inst12}} \and 
L. Pentericci\orcid{0000-0001-8940-6768}\inst{\ref{inst12}} \and 
B. Ribeiro\orcid{0000-0002-4844-0414}\inst{\ref{inst16}} \and 
D. Schaerer\orcid{0000-0001-7144-7182}\inst{\ref{inst17}} \and 
E. Shah\orcid{0000-0001-7811-9042}\inst{\ref{inst2},}\inst{\ref{inst18}} \and
L. Shen\orcid{0000-0001-9495-7759}\inst{\ref{inst2},} \inst{\ref{inst19},} \inst{\ref{inst20}} \and
P. Staab\orcid{0000-0002-8877-4320}\inst{\ref{inst2}} \and
M. Talia\orcid{0000-0003-4352-2063}\inst{\ref{inst4}} \and 
R. Thomas\orcid{0000-0001-8385-3276}\inst{\ref{inst21}}\and
A.~R. Tomczak\orcid{0000-0003-2008-1752}\inst{\ref{inst2}} \and 
L. Tresse\orcid{0000-0001-8776-0958}\inst{\ref{inst1}} \and
E. Vanzella\orcid{0000-0002-5057-135X}\inst{\ref{inst4}} \and 
D. Vergani\orcid{0000-0003-0898-2216}\inst{\ref{inst4}} \and 
E. Zucca\orcid{0000-0002-5845-8132}\inst{\ref{inst4}}}



\institute{Aix-Marseille Univ, CNRS, CNES, Laboratoire d'Astrophysique de Marseille, Marseille, France, \email{brian.lemaux@noirlab.edu}\label{inst1}
\and
Department of Physics and Astronomy, University of California, Davis, One Shields Ave., Davis, CA 95616, USA \label{inst2}
\and
Gemini Observatory, NSF's NOIRLab, 670 N. A'ohoku Place, Hilo, Hawai'i, 96720, USA \label{inst3} 
\and
INAF - Osservatorio di Astrofisica e Scienza dello Spazio di Bologna, via Gobetti 93/3 - 40129 Bologna - Italy\label{inst4}
\and
Space Telescope Science Institute, 3700 San Martin Drive, Baltimore, MD 21218, USA\label{inst5}
\and
UCO/Lick Observatory, Department of Astronomy \& Astrophysics, UCSC, 1156 High Street, Santa Cruz, CA, 95064, USA\label{inst6}
\and
Instituto de Investigac\'{i}on Multidisciplinar en Ciencia y Tecnolog\'{i}a, Universidad de La Serena, Raul Bitr\'{a}n 1305, La Serena, Chile\label{inst7}
\and 
Departamento de F\'{i}sica y Astronom\'{i}a, Universidad de La Serena, Av. Juan Cisternas 1200 Norte, La Serena, Chile\label{inst8}
\and
Dipartimento di Fisica e Astronomia, Universit\`a  di Padova, Vicolo dell'Osservatorio, 3 35122 Padova, Italy\label{inst9}
\and 
INAF Osservatorio Astronomico di Padova, vicolo dell'Osservatorio 5, I-35122 Padova, Italy\label{inst10}
\and
University of Hawai'i, Institute for Astronomy, 2680 Woodlawn Drive, Honolulu, HI 96822, USA\label{inst11}
\and
INAF--IASF, via Bassini 15, I-20133, Milano, Italy\label{inst12}
\and
INAF--Osservatorio Astronomico di Roma, via di Frascati 33, I-00040, Monte Porzio Catone, Italy\label{inst13}
\and
N\'ucleo de Astronom\'ia, Facultad de Ingenier\'ia, Universidad Diego Portales, Av. Ej\'ercito 441, Santiago, Chile\label{inst14}
\and
Astronomy Department, University of Massachusetts, Amherst, MA, 01003 USA\label{inst15}
\and
Leiden Observatory, Leiden University, PO Box 9513, 2300 RA Leiden, The Netherlands\label{inst16}
\and
Geneva Observatory, University of Geneva, ch. des Maillettes 51, CH-1290 Versoix, Switzerland\label{inst17}
\and
LSSTC DSFP Fellow\label{inst18}
\and
CAS Key Laboratory for Research in Galaxies and Cosmology, Department of Astronomy, University of Science and Technology of China, Hefei 230026, China\label{inst19}
\and
School of Astronomy and Space Sciences, University of Science and Technology of China, Hefei, 230026, China\label{inst20}
\and
European Southern Observatory, Av. Alonso de C\'{o}rdova 3107, Vitacura, Santiago, Chile\label{inst21}
\and
Deceased\label{*}
}

\abstract{Utilizing spectroscopic observations taken for the VIMOS Ultra-Deep Survey (VUDS), new observations from Keck/DEIMOS, and publicly available observations of large samples of 
star-forming galaxies, we report here on the relationship between the star-formation rate ($\mathcal{SFR}$) and the local environment ($\delta_{gal}$) of galaxies in the 
early universe ($2<z<5$). Unlike what is observed at lower redshifts ($z\lsim2$), 
we observe a definite, nearly monotonic increase in the average $\mathcal{SFR}$ with increasing galaxy overdensity over more than an order of magnitude in $\delta_{gal}$. The robustness 
of this trend is quantified by accounting for both uncertainties in our measurements and galaxy populations that are either underrepresented or not present in our sample 
(e.g., extremely dusty star-forming and quiescent galaxies), and we find that the trend 
remains significant under all circumstances. This trend appears to be primarily driven by the fractional increase of galaxies in high-density environments that are more massive in 
their stellar content and are forming stars at a higher rate than their less massive counterparts. We find that, even after stellar mass effects are accounted for, there remains a weak 
but significant $\mathcal{SFR}$-$\delta_{gal}$ trend in our sample implying that additional environmentally related processes are helping to drive this trend. We also find clear 
evidence that the average $\mathcal{SFR}$ of galaxies in the densest environments increases with increasing redshift. These results lend themselves to a picture in which massive gas-rich 
galaxies coalesce into proto-cluster environments at $z\gsim3$, interact with other galaxies or with a forming large-scale medium, subsequently using or losing most of their gas in the process, 
and begin to seed the nascent red sequence that is present in clusters at slightly lower redshifts.} 
\titlerunning{The Reversal of the $\mathcal{SFR}$-Density Relation}
\authorrunning{B.~C.\ Lemaux et al.}
\maketitle

\section{Introduction}




It has now been firmly established that dense environments at low to intermediate redshifts ($z\la1.5$) are generally hostile to star-formation
activity. While the mechanisms underyling the establishment of such trends are still debated, observations have clearly shown that galaxies that inhabit low- to intermediate-redshift 
clusters are generally redder (e.g., \citealt{hansen09, peng10, mcoopz10, lem12, kovac14, balogh16, nantais16, nantais17, olga17,lem19,vanderberg20}), are less star-forming 
(e.g., \citealt{gomez03, vonderlinden10,muzz12, AA19, old20}), and are more likely to have undergone recent quenching events  
(e.g., \citealt{tran03,pwu14,lem17,mildmanneredmiguel18, owers19,paccagnella19}) than coeval field samples. As a result, the general relationship between 
the average star-formation rate ($\mathcal{SFR}$) of galaxies and galaxy density ($\delta_{gal}$) up to $z\sim1.5$ appears to be one of an anticorrelation. 
Furthermore, the presence of massive, red galaxies appears to persist in some overdense environments at least to $z$$\sim$2 (e.g., \citealt{kodama07, strazz13, dutifuldiener15}), 
However, there are hints that this relationship between $\mathcal{SFR}$ and $\delta_{gal}$ observed at lower redshifts begins to break down or even reverse by $z\sim1-2$. At these redshifts, several case
studies from individual structures or relatively small field surveys have shown intriguing indications of this reversal, with galaxies that inhabit denser environments 
showing increased star formation activity relative to counterparts in more rarefied environments (e.g., \citealt{olga06, elbaz07, tran10, santos14, santos15}, though see also 
the discussion in \citealt{mcoopz07} regarding the results presented in \citealt{olga06}). In a larger sample of $\sim$16,000 galaxies spanning 2.8 square degrees taken 
from the Deep Extragalactic Evolutionary Probe 2 (DEEP2; \citealt{davis03, new13}), \cite{mcoopz08} found a clear reversal of the relationship between 
$\mathcal{SFR}$ and $\delta_{gal}$ at $z\sim1$ relative to that observed in the local universe. Galaxies that inhabit the most extreme environments (i.e., high-mass groups 
and clusters of galaxies) were, however, largely absent from the sample presented in \cite{mcoopz08}. Recent results from \cite{AA19}, in an analysis of $\sim$10,000 galaxies taken from
the Observations of Redshift Evolution in Large Scale Environment (ORELSE; \citealt{lub09}) survey that spanned field, group, and cluster environments at $z\sim1$, found no
such positive correlation between $\mathcal{SFR}$ and $\delta_{gal}$ in aggregate. At slightly higher redshifts ($z\sim1.2$), \cite{old20} also found a flat or negative
trend between $\mathcal{SFR}$ and $\delta_{gal}$ from $\sim$1500 cluster and field galaxies drawn from the Gemini Observations of 
Galaxies in Rich Early ENvironments (GOGREEN; \citealt{balogh17, balogh21}) survey, with a stronger anticorrelation seen in the lower redshift portion of their sample ($z\sim1$). 
Such seemingly conflicting results speak to the complex relationship between the $\mathcal{SFR}$ of galaxies and 
their environment at intermediate redshift, which appears to vary considerably depending on the range of environments probed and across different structures. 
Clearly, however, this relationship has evolved dramatically by $z\sim1$ relative to the local relation, and
indications of a reversal at intermediate redshift are consistent with the early formation epochs inferred for massive galaxies in intermediate-redshift clusters \citep{hilton09, ret10, rai11, lem12, fumagalli16}, 
as well as with the mere presence of massive and evolved galaxies at $z\sim5$ (e.g., \citealt{maw16, lem18}). These lines of evidence strongly indicate that the peak of 
star-formation activity for massive cluster galaxies broadly lies at early epochs ($z>2$).
 
This observational picture is corroborated by results from simulations in which it is estimated that forming clusters (i.e., proto-clusters)
form the vast majority of their stellar content at early times, $\sim$50\% during the 1.5 Gyr period from $2<z<4$ \citep{coolchiang17}. During
these epochs, it is predicted that proto-cluster environments become an important contribution to the overall comoving cosmic $\mathcal{SFR}$ density ($\mathcal{SFRD}$),
which is seen to peak at these redshifts \citep{lunabarnfriendz14}. While proto-clusters fill only $\sim$3\% of the comoving volume of the universe at
these epochs, they are estimated to contribute 20-30\% to the overall $\mathcal{SFRD}$ \citep{coolchiang17, muldrew18}, implying a rate of stellar mass assembly far outpacing
that of the field. Concurrent to the peak in the cosmic $\mathcal{SFRD}$, the rate of galaxies undergoing major mergers also appears to increase to these
redshifts (e.g., \citealt{carlos13, tizzytasca14}), suggesting that the seeds of the most massive cluster galaxies observed today are sewn both through
\emph{in situ} star formation and \emph{ex situ} through galaxy merging in the extended proto-cluster environment. 

While this picture is extremely exciting, definitive observational evidence at these redshifts has been lagging. Many searches for proto-clusters at such redshifts
have been attempted. However, such searches are extremely challenging due to the extreme faintness of the galaxy populations. While photometric redshifts
can be used to identify proto-cluster candidates in blank fields (e.g., \citealt{dutifuldiener13, coolchiang14}) or around radio
quasars (e.g., \citealt{hatch14}), without spectroscopic confirmation the authenticity of these candidates is suspect. Spectroscopic searches for proto-clusters
are typically targeted at Lyman-alpha emitters (LAEs; \citealt{tersetoshikawa14,tersetoshikawa16, tersetoshikawa18, tersetoshikawa20, dey16}) or Submillimeter galaxies
(SMGs; \citealt{connivingcasey15, vernesa17, greenslade18, lewis18, oteo18, cheng20, long20}), galaxies that can be either subdominant or extremely biased relative to the overall galaxy
population (e.g., \citealt{paolo15, miller15}) and may trace overdensities differently than more typical populations (e.g., \citealt{shi19, loopylucia20}). 
The inhomogeneity inherent in this selection, combined with a lack of large, comparable field samples,
makes any study of environmentally driven evolution difficult to interpret. 

Still, there exist exciting case studies of high-redshift proto-cluster 
systems detected through these search techniques, where star-formation activity among member galaxies is seen to far outpace that of the coeval field 
(e.g., \citealt{greenslade18,miller18, rhythm18, noirot18,cheng19, cheng20,ryley20}). Several case studies probing more typical galaxies in proto-cluster 
environments at $z>2$ also show similar behavior (e.g., \citealt{taowang16, rhythm18, shi20}). Additionally, analysis of the ultraviolet luminosity function of 
photometrically selected proto-cluster candidate galaxies suggests that such behavior might be a general property of proto-cluster galaxies
at high redshift \citep{ito20}. However, in order to confirm that this trend is indeed general in the high-redshift universe, rather than limited to a small 
number of proto-clusters or proto-clusters in a rare phase of their evolution, a spectroscopic census of a representative galaxy population in a large 
ensemble of forming overdensities in the early universe, as well as a comparably selected field sample, is required.

In this paper we present an investigation of the relationship between the $\mathcal{SFR}$ and galaxy density of a large sample of spectroscopically confirmed 
galaxies in the early universe ($2\le z \le 5$). This sample is primarily drawn from observations taken with VIsible Multi-Object Spectrograph (VIMOS, \citealt{dong03})
as part of the VIMOS Ultra-Deep Survey (VUDS, \citealt{dong15}) across three well-studied extragalactic fields. Additional spectroscopic data were drawn from the VIMOS VLT 
Deep Survey (VVDS; \citealt{dong13}), the zCOSMOS survey (\citealt{lilly07,lilly09}, Lilly et al.\ \emph{in prep}), a variety of other smaller public surveys, and 
new observations from the Keck DEep Imaging Multi-Object Spectrograph (DEIMOS, \citealt{fab03}) to create a sample of 6730 spectroscopically confirmed star-forming galaxies
extending over a footprint of $\sim$3 deg$^2$. These spectral observations were complemented by deep multiband imaging observations, which were used both for the estimate 
of high-quality photometric redshifts and an estimate of the instantaneous $\mathcal{SFR}$s of galaxies. A weighted combination of spectroscopic and photometric information
was used to create high-fidelity galaxy density maps with which to investigate the effect of environment on star formation activity at these redshifts.

The structure of the paper is as follows. In \S\ref{obsnred} we discuss the imaging and spectroscopic observations used for this study, the methods for estimating
photometric redshifts, $\mathcal{SFR}$s, and environment, as well as the representativeness of our spectral sample. In \S\ref{SFRdensity} we investigate
the general relationship between $\mathcal{SFR}-\delta_{gal}$ and stellar mass-$\delta_{gal}$ for our final sample, including the behavior of $\mathcal{SFR}-\delta_{gal}$
as a function of stellar mass and of redshift, 
and explore scenarios that may give rise to the observed trends. In \S\ref{conclusions} we summarize our results and present our main
conclusions. Throughout this paper all magnitudes, including those in the infrared, are presented in the AB system \citep{okengunn83, fukugita96} and
distances are given in proper rather than comoving units. We adopt a concordance $\Lambda$CDM cosmology with $H_{0}$ = 70 km s$^{-1}$ Mpc$^{-1}$, 
$\Omega_{\Lambda}$ = 0.73, and $\Omega_{M}$ = 0.27. While abbreviated for convenience, throughout the paper stellar masses are presented in units of $h_{70}^{-2} 
\mathcal{M}_{\ast}$, star formation rates in units of $h_{70}^{-2} \mathcal{M}_{\ast}$ yr$^{-1}$, proper distances in units of $h_{70}^{-1}$ kpc or Mpc, 
where $h_{70}\equiv H_{0}/70$ km$^{-1}$ s Mpc.

\section{Observations and spectral energy distribution fitting}
\label{obsnred}

The primary observations used for this study are drawn from VUDS \citep{dong15}, a massive 640-hour ($\sim$80 night) spectroscopic 
campaign reaching extreme depths ($i^{\prime}\la25$) over three well-studied extragalactic fields: the Cosmic Evolution Survey (COSMOS, \citealt{scoville07a}) field,
the Extended Chandra Deep Field South (ECDFS, \citealt{lehmer05}), and the first field of the Canada-France-Hawai'i Telescope Legacy Survey		
(CFHTLS-D1\footnote{\url{http://www.cfht.hawaii.edu/Science/CFHTLS/}}) also known as the VVDS-02h field. These 
observations are supplemented by a variety of publicly available imaging and spectroscopic data sets taken in the same three fields as well as our own follow-up 
observations. We describe the basic properties of these data below.  

\subsection{Imaging data and photometry}
The imaging data relevant to this study have been discussed in detail in other papers and, as such, are only briefly described here. These data broadly include 
extremely deep ($m_{AB}\sim24-27$, 5$\sigma$ completeness limit) $\ge$10$-$band imaging spanning from the observed-frame ultraviolet (UV) to the near-infrared (NIR) in 
addition to imaging data that span a variety of other wavelengths that are generally not used in this study. For the CFHTLS-D1 field we use the UV/optical/NIR
imaging data described in \citet{lem14a,lem14b}. These data include deep ground-based $u^{\ast}g^{\prime}r^{\prime}i^{\prime}z^{\prime}$ imaging with the 
Canada-France-Hawai'i Telescope (CFHT)/MegaCam \citep{boulade03} as part of the CFHTLS, 
as well as deep $JHK_{s}$ imaging in the NIR from WIRCam \citep{puget04} taken for the WIRCam Deep Survey (WIRDS; \citealt{bielby12}) and [3.6][4.5] $\mu$m 
imaging from the InfraRed Array Camera (IRAC; \citealt{fazio04}) on board the \emph{Spitzer} Space Telescope taken as part of the Spitzer Extragalactic Representative 
Volume Survey (SERVS; \citealt{mauduit12}). 

For the ECDFS field, we draw on the imaging data described in \citet{car10} and references therein. These data include deep 
$UBVRIz$ imaging taken from the Garching-Bonn Deep Survey (GaBoDS; \citealt{hil06}) and the Multiwavelength Survey by Yale-Chile (MUSYC; \citealt{fukinericgawiser06}), 
as well as deep imaging in the NIR ($JHK$[3.6][4.5][5.8][8.0]) from MUSYC, \citet{moy03}, and the Spitzer IRAC/MUSYC Public Legacy Survey in the Extended CDF-South 
(SIMPLE; \citealt{damen11}). Though this field is covered by \emph{Hubble Space Telescope} (\emph{HST}) observations with the Cosmic Assembly Near-infrared Deep Extragalactic 
Legacy Survey (CANDELS; \citealt{grogin11,anton11}), only a subsection of the area that is spanned by the spectroscopic data used in this study is covered by these observations. 
Moreover, we have found that none of the physical parameters presented in this study (see \S\ref{SED}) are, on bulk, appreciably affected by fitting to the deeper CANDELS data as opposed to the 
primarily ground-based imaging of \citet{car10}\footnote{More specifically we find a median difference of only 0.09 and 0.02 dex for stellar mass and $\mathcal{SFR}$, respectively, for galaxies		
that are detected in both catalogs, with fitting run on MUSYC photometry generally returning larger values.}. 

For the COSMOS field we draw on two different data sets, which are described in detail in \citet{lem18}. For the vast majority 
of the objects studied in this paper we use the ``COSMOS2015" imaging data compiled by \citet{laigle16}, which include imaging in the near-ultraviolet (NUV) from the GALaxy 
Evolution eXplorer (\emph{GALEX}; \citealt{martin05}), ground-based $u^{\ast}BVri^{+}z^{++}$ UV/optical imaging from CFHT/MegaCam and Subaru/Suprime-Cam \citep{miyazaki02}, 
ground-based $YJHK$ imaging from Subaru/Hyper-Suprime-Cam (HSC, \citealt{miyazaki12}), the UltraVISTA survey \citep{mccracken12}, and CFHT/WIRCam, as well as extremely deep 
[3.6][4.5][5.8][8.0] imaging from \emph{Spitzer}/IRAC. All galaxies in our COSMOS spectroscopic sample (see \S\ref{spec}) were matched to the COSMOS2015 catalog using nearest-neighbor matching
with a maximum radius of 0.75$\arcsec$ (median offset 0.07$\arcsec$). For the small subset of galaxies with secure spectroscopic redshifts (hereafter $z_{spec}$, see \S\ref{spec} for 
the operational definition of this 
phrase) in the COSMOS field for which we were not able to find a photometric counterpart in the COSMOS2015 catalog, we used v2.0 of the photometric catalog presented in 
\citet{capak07}\footnote{Different detection bands and photometric masking methods resulted in some objects being present in the \citet{capak07} catalog that are not in the
COSMOS2015 catalog. As in \citet{lem18}, we exclude the \emph{Spitzer}/IRAC cryogenic bands ([5.8]/[8.0]) presented in the \citet{capak07} catalog from all analysis.}, 
which uses a subset of the imaging in the COSMOS2015. For those objects that had counterparts in both catalogs, we noticed no systematic offset between parameters derived using the 
two different sets of photometry for any of the parameters presented in this study. Further details on these observations, their reduction, source detection, and 
magnitude measurements including point spread function (PSF) homogenization and the various methods used to apply aperture corrections can be found in \citet{lem14a,lem14b}, 
\citet{car10}, \citet{capak07}, \citet{laigle16} and references therein. 




\subsection{Spectroscopic data}
\label{spec}
The spectroscopic data employed in this study were drawn from a variety of different surveys. For all redshift surveys we selected only
those galaxies with secure $z_{spec}$ measurements, the definition of which differs slightly based on the survey being used, and limited the redshift range to $2<z_{spec}<5$. The majority of our
$z_{spec}$ values (53.7\%) were taken from VUDS. This survey primarily utilizes photometric redshifts (hereafter $z_{phot}$) drawn from the photometric catalogs
described above\footnote{Because the COSMOS2015 catalog was not available at the inception of the VUDS survey, the \citet{capak07} photometry was used to measure $z_{phot}$ values.} to 
select targets for spectroscopic follow up using VIMOS \citep{dong03} mounted on the Nasmyth platform of the 8.2-m Very Large Telescope (VLT) Unit
Telescope 2 (UT2/Melipal) at Cerro Paranal. The spectroscopic targets are largely limited to $i^{\prime}/i^{+}/I<25$ and those objects whose primary or secondary $z_{phot}$ 
solution satisfies $z_{phot}+1\sigma>2.4$. Spectroscopic observations consisted of approximately 50400s of integration across the wavelength range $3650 \leq \lambda_{obs} \leq 9350$\AA\ 
at a spectral resolution of $R=230$. Further discussion of the survey design, observations, reduction, redshift determination, and the properties of the full 
VUDS sample can be found in \citet{dong15}. Details of the first VUDS data release\footnote{\url{http://cesam.lam.fr/vuds/DR1/}} can be found in \citet{tizzytasca17}.  

The flagging code for VUDS is discussed in \citet{dong15}. We take two different approaches regarding galaxies with different flags in this paper. For the selection of our sample as 
well as spectral energy distribution (SED) fitting, we consider those galaxies with flags = X2, X3, X4, \& X9 as being ``secure," where X=0, 2, or 3\footnote{X=0 is reserved for target galaxies,		
X=2 for non-targeted objects that fell serendipitously on a slit at a spatial location separable from the target, and X=3 for non-targeted objects that were serendipitously subtended 
slit at the same spatial location as the target. Galaxies identified spectroscopically as broadline active galactic nuclei, X=1 in the VUDS 
flagging system, are excluded from our analysis, due to the difficulty of measuring their physical parameters through spectral energy distribution fitting with the standard approach taken in 
this study. For more details on the fiducial probability of a correct redshift for a given flag, see \citet{dong15}}. According to \citet{dong15}, galaxies with these flags have a redshift 
that is reliable at the 75-99.3\% level. In \S\ref{voronoi} we quantify the reliability of the different flags for VUDS and in the remaining part of our analysis that employs $z_{spec}$ measurements, 
the creation of our metric of environment, the $z_{spec}$ value for galaxies of differing flags are treated probabilistically using these reliability estimates. 

The next largest contribution to our sample falls in the COSMOS field and is drawn from the Bright and Deep phases of the zCOSMOS survey 
(\citealt{lilly07,lilly09}, Lilly et al.\ \emph{in prep}, \citealt{dutifuldiener13, dutifuldiener15}). The galaxies drawn from these surveys contribute 26.1\% of our total sample. In 
the COSMOS field, a small number of additional redshifts were taken from \citet{connivingcasey15}, \citet{coolchiang15}, and \citet{dutifuldiener15} at $z\sim2.5$. In the ECDFS field,			
a list of publicly available redshifts was compiled by one of the authors (NPH) taken from a variety of different surveys (e.g., \citealt{eros08, eros09, nimish09, straughn09, balestra10, cooper12, kurk13, trump13, morris15}) and 
contribute 13.2\% to our total sample. In the CFHTLS-D1 field, galaxies were drawn from the Deep and UltraDeep phases of the VVDS \citep{dong05, dong13} and 
contributed 6.0\% to our total sample. For each survey listed here we attempted as best as we were able to homogenize the flagging system relative to the VUDS flagging system and select only
those galaxies that had flags equivalent to those considered secure for VUDS\footnote{Broadly, the statistics of the flagging system in the zCOSMOS and VVDS surveys are on the same basis 
as those in VUDS making this task trivial for the bulk of these galaxies.}. All galaxies observed in two or more surveys were assigned a single redshift based on the $z_{spec}$ with
the highest confidence flag, with VUDS redshifts being used in the case of a tie. A comparison of duplicate observations where at least one of those observations yielded a spectrum with a
flag of X3 or X4 resulted in similar reliability statistics to those stated in \S3.3 of \citet{dong15} for flags X1, X2/X9, and X3/X4. 

\begin{figure*}
\plottwonotsospecial{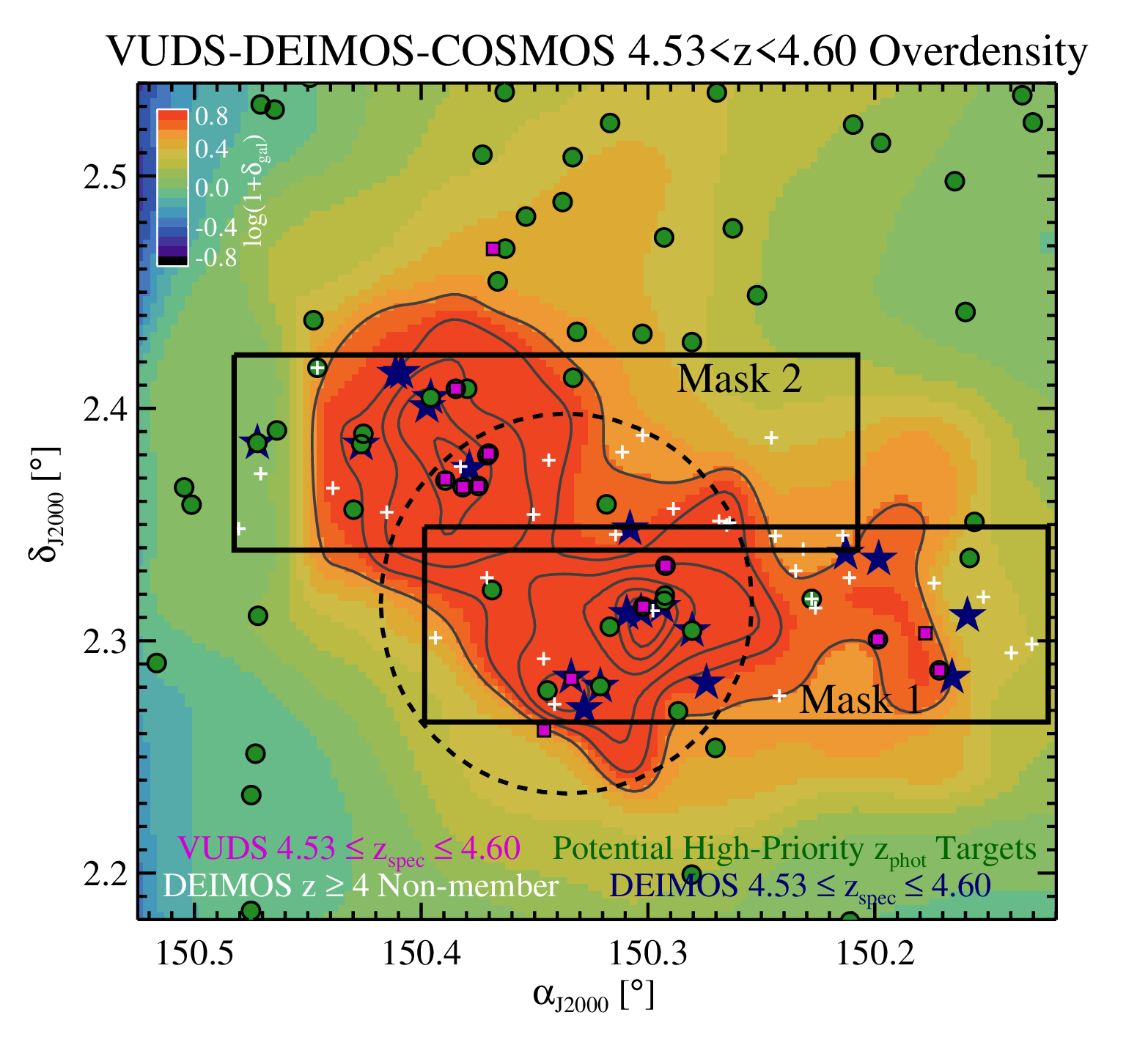}{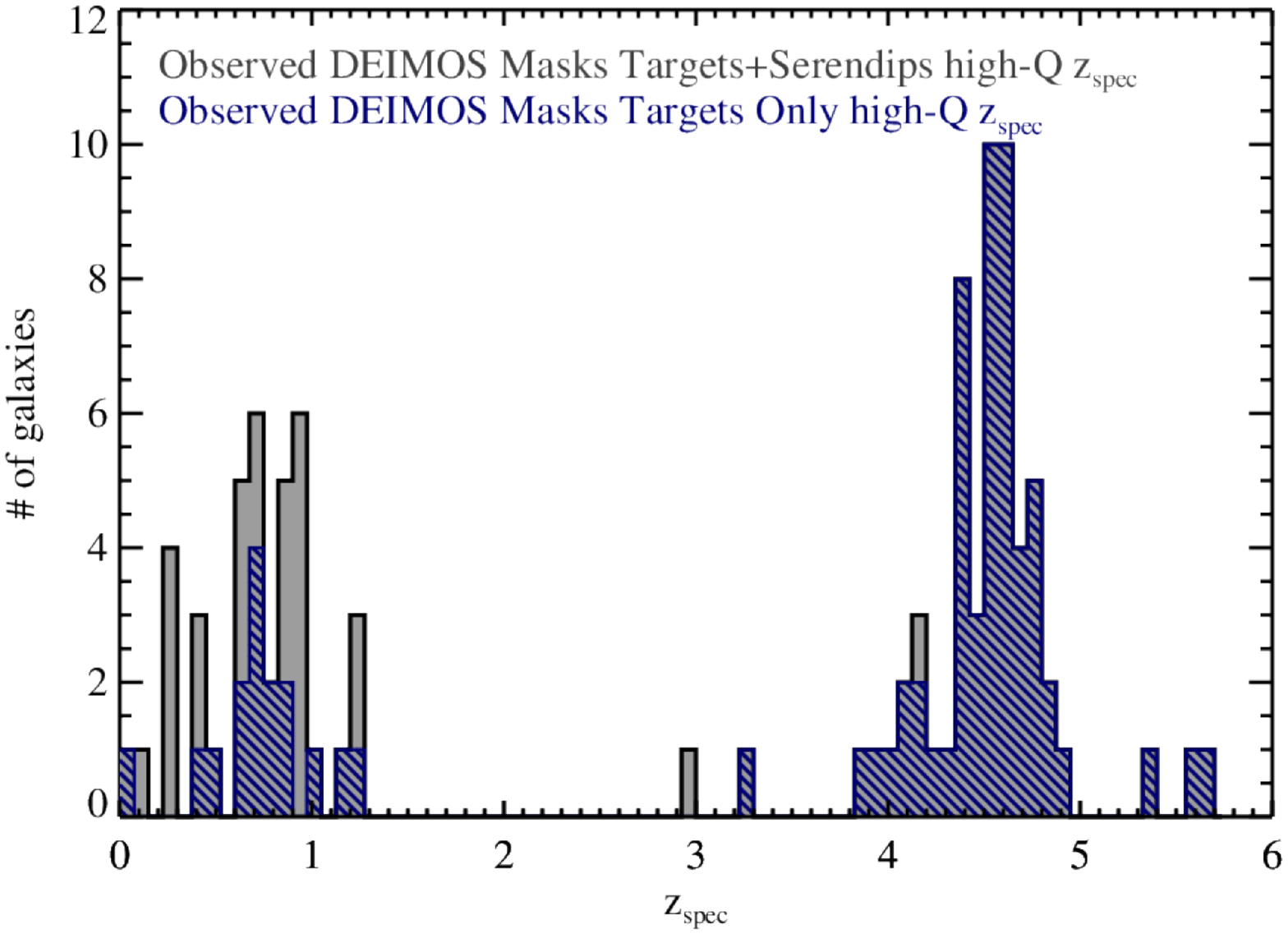}
\caption{View of the C3VO observations of the PCl J1001+0220 proto-cluster at $z\sim4.57$. \emph{Left:} Layout of the two masks observed with DEIMOS as part of the C3VO survey in the vicinity of the 
PCl J1001+0220 proto-cluster against the backdrop of a two-dimensional overdensity 
map generated using the methodology described in \S\ref{voronoi}, though here the overdensity is calculated over the entire redshift extent of PCl J1001+0220 indicated at the top of the
plot. The two completed DEIMOS slitmasks are outlined in black. The dashed circle denotes $R_{proj}=2$ Mpc centered on the 
barycenter of the detection in the overdensity map. The highest priority DEIMOS targets ($i^{+}<25.3$ and $4.44 \le z_{phot} \le 4.70$) in the vicinity of PCl J1001+0220 are shown as 
filled dark green circles. Galaxies with high-quality $z_{spec}$ measurements from VUDS and from our new DEIMOS observations within the redshift range of PCl J1001+0220 are 
shown as purple squares and blue stars, respectively. Galaxies with high-quality $z_{spec}$ measurements from the DEIMOS observations at $z>4$ but not within the redshift range of 
PCl J1001+0220 are shown as white crosses. Galaxy overdensity contours indicate levels of $\sim$3, 4, 5, 6, 7, 8$\sigma_{NMAD}$ in excess of the background. \emph{Right:} 
Redshift histogram of all objects with secure $z_{spec}$ measurements from Keck/DEIMOS observations. The blue hatched histogram shows the $z_{spec}$ distribution of all targeted 
galaxies with a secure $z_{spec}$, while the gray solid histogram also includes those galaxies 
that serendipitously fell on in the area subtended by the slits in the two masks. These DEIMOS observations add considerably to the high-redshift portion of the sample presented
in this study, especially at the high-density end.}
\label{fig:dongo}
\end{figure*}
 
Finally, we incorporate new follow-up DEIMOS \citep{fab03} observations of the PCl J1001+0220 $z\sim4.57$ proto-cluster located in
the COSMOS field \citep{lem18}. These observations are part of the Charting Cluster Construction with VUDS and ORELSE (C3VO) survey, an ongoing campaign that 
involves observations with both DEIMOS and the Multi-Object Spectrometer For Infra-Red Exploration (MOSFIRE; \citealt{mclean12}) on the 
Keck \ion{}{I}/\ion{}{II} telescopes. The Keck component of the C3VO survey is designed to provide a nearly complete mapping of the five most significant overdensities detected in VUDS, including those 
reported in \cite{lem14b}, \cite{olga14}, \cite{lem18}, and \cite{olga18}, by targeting star-forming galaxies of all types to $i_{AB}<25.3$ (or $\sim L^{\ast}_{FUV}$ at 
$z\sim4.5$ and $<L^{\ast}_{FUV}$ at $z\sim2.5$) and Lyman-$\alpha$ (hereafter Ly$\alpha$) emitting galaxies to fainter magnitudes. 

The Keck-C3VO sample presented here comprises galaxies from the only two DEIMOS slitmasks that have been fully observed and analyzed at the time of writing. Including these 
observations allowed us to bolster the high-redshift end of our sample appreciably, as $\sim$10\% of our high$-z$ ($z>4$) galaxy sample are Keck/DEIMOS galaxies. These 
observations additionally allowed us to 
populate the sample of galaxies in high-density environments at these redshifts more fully, as 20 new members of PCl J1001+0220 were discovered by these observations, which
results in a combined sample that is more than triple the previously known member sample and provides an appreciable fraction ($\sim$20\%) of the high$-z$, high-density 
sample ($\log(1+\delta_{gal})>0.22$, see \S\ref{voronoi}) presented in this work\footnote{In \S\ref{SMnzeffects} we state, however, that the trends seen in our analysis 
hold if the entire population of the PCl J1001+0220 proto-cluster, including the DEIMOS sample, is excluded and are, thus, not dominated by a single structure at high redshift.}  
The two DEIMOS masks included here were observed on December 22, 2016, and December 26, 2017, for $\sim$4 hours each under		
nearly photometric conditions and 0.7$-$0.8\arcsec seeing. For each mask we used the		
600 l mm$^{-1}$ grating in conjunction with the GG455 order blocking filter and a central wavelength of $\lambda_{c}=7200$\AA. All slitmasks were milled with 1$\arcsec$-wide 
slits. This setup resulted in a plate scale of 0.66\AA\ pix$^{-1}$, an $R\sim2500$ ($R=\lambda$/$\theta_{FWHM}$, where $\theta_{FWHM}$ is the full-width half-maximum 
resolution), and a wavelength coverage of $4600\rm{\AA}\la\lambda\la9800$\AA. The two masks were centered at [$\alpha_{J2000}$, $\delta_{J2000}$] = [10:01:02.25, 2:17:20.0] and 
[10:01:22.92, 2:21:41.6] with a position angle of 90$^{\circ}$, thus subtending a considerable fraction of the proto-cluster area as defined in \cite{lem18} 
(see Figure \ref{fig:dongo} for the layout of these observations). 

Targets were selected using a combination of magnitude and $z_{phot}$ cuts. Primary targets were selected using the criteria 
$4.53 - 1\sigma_{\Delta z/(1+z)}(1+z_{PCl}) \le z_{phot} \le 4.60 + 1\sigma_{\Delta z/(1+z)}(1+z_{PCl})$ and $i_{AB}<25.3$, where $z_{PCl}=4.57$, the systemic 
redshift of PCl J1001+0220, and $\sigma_{\Delta z/(1+z)}=0.016$, which is a rough estimate of the photometric redshift precision in COSMOS at these redshifts. We note that this 
value of $\sigma_{\Delta z/(1+z)}$ was based on analysis presented in \citet{vernesa17} using a different sample selection and methodology than that used for the full sample in this paper. 
It is considerably smaller than the equivalent value estimated using our sample (see \S\ref{SED}), though, in practice, this discrepancy did not meaningfully affect our target selection process 
as primary targets were always subdominant on the DEIMOS masks due to their relatively low surface density resulting from the choice of $\sigma_{\Delta z/(1+z)}$.
As such, lower priority targets were accommodated on our masks with relative ease, which included true member galaxies that were relegated to lower priority targets 
due to the restrictive $z_{phot}$ cut used to select primary targets. 

Objects with detections in X-ray and/or radio imaging at any $z_{phot}$ value that lacked a secure $z_{spec}$ were highly prioritized. Lower priority targets 
included objects in a similar magnitude range but within a larger $z_{phot}$ range centered on the systemic redshift of PCl J1001+0220, objects in the same $z_{phot}$ range
but at fainter magnitudes ($25.3 < i_{AB} \le 26.5$), brighter objects ($i_{AB}<25.3$) likely in the background of the proto-cluster, and galaxies previously targeted
by VUDS. These magnitude limits were 
imposed such that continuum observations at modest signal-to-noise ratios (S/N) of $\sim$1 per native resolution element could be achieved for our primary target sample such 
that redshifts were obtainable in the presence or absence of Ly$\alpha$ in emission. 

Redshifts were evaluated with a modified version of the \texttt{zspec} code 
\citep{new13} which incorporated empirical high-redshift galaxy templates from the VUDS and VVDS surveys as well as high-resolution empirical Ly$\alpha$ templates from 
\cite{lem09}. Each object inspected was assigned a quality code, $Q$, broadly following the convention of \cite{new13}. The one modification made to this convention was 
to assign a secure redshift ($Q=3$) to those galaxies that exhibited a single redward skewed emission line indicative of Ly$\alpha$, In total 256 objects were 
targeted across the two masks, yielding 73 high-quality redshifts, of which 50 were in the redshift range $4 < z < 5$. In addition, 25 
high-quality redshifts were obtained for objects that serendipitously fell on the otherwise empty regions of our slits. These serendipitous detections were identified 
and extracted in the manner described in \cite{lem09}. Figure \ref{fig:dongo} shows the spatial location of the two DEIMOS masks incorporated in this study, the locations  
of newly confirmed $z>4$ galaxies, and the distribution of high-quality $z_{spec}$ measurements on both masks. 

Combining all samples together, removing duplicates, and imposing the NIR magnitude cuts that are used for each field to define the final sample both for photometric 
and spectroscopic objects (see \S\ref{voronoi}), resulted in a total of 6730 galaxies with high-quality $z_{spec}$ values. We note here, and will show later 
(see \S\ref{represent}), that this sample is largely devoid of extremely dusty star-forming and quiescent galaxies, an issue that we will attempt to account for in later 
sections. This sample of 6730 galaxies will be referred to hereafter as the final VUDS+ spectral sample or simply the VUDS+ sample. We limit our main analysis 
to this purely spectroscopically selected sample of galaxies, rather than attempting to additionally include objects with photometric redshifts, as we cannot accurately or precisely 
assign environment measures to the latter population (see \S\ref{voronoi}) given the large photometric redshift uncertainties at these redshifts. Additionally, uncertainties in 
estimates of galaxy parameters (see \S\ref{SED}) for galaxies without a spectroscopic redshift do not incorporate the uncertainties in the photometric redshift, which is problematic for 
certain aspects of our analysis aimed at testing the robustness of the observed trends.


\subsection{Synthetic model fitting}
\label{SED}
 
The photometry measured on the imaging observations described in the previous section was used as input for two different forms of SED fitting. In the first form of the SED fitting,
 aperture-uncorrected magnitudes were input to the code {\textsc Le Phare}\footnote{\url{http://cfht.hawaii.edu/~arnouts/LEPHARE/lephare.html}} \citep{stephane99, 
dreadolivier06, dreadolivier09} in order to determine $z_{phot}$ values and associated probability density function (PDFs) for all sources detected in each field. The 
methodology used for computing $z_{phot}$ follows that described in \citet{dreadolivier09, dreadolivier13}, with the final $z_{phot}$ for each object being derived from 
the median of the resultant PDFs, with their associated errors set by the 16th and 84th percentile of each PDF.
The precision and accuracy of the $z_{phot}$ estimates in each field over the redshift range $2\le z \le 5$, at least as they are able to be measured by galaxies with secure $z_{spec}$, 
are discussed in detail in \citet{lem14a, lem18, lem19, vernesa17, laigle16}. These numbers range from $\sigma_{\Delta z/(1+z)}=0.032-0.035$ and $\eta=20.2-26.3$\%, where $\eta$ is the 
percentage of catastrophic outliers defined as $|z_{spec}-z_{phot}|/(1+z_{spec})>0.15$, with no appreciable bias for a sample limited to $2 < z_{spec} < 5$ and $K_{s}<24.1$/[3.6]$<$23.1, 
[3.6]$<$24.8, and [3.6]$<$25.3 in the CFHTLS-D1, ECDFS, and COSMOS fields, respectively (see \S\ref{voronoi} for the meaning behind these photometric cuts). We note that these
estimates are different than some previous estimates taken from the literature because of the differing spectroscopic sample, photometric cuts, and redshift range considered. 
The above numbers are provide only a broad view of the quality of the photometric redshifts in each field, and it is rather through
use of the reconstructed PDF for \emph{each individual} $z_{phot}$ object that the precision and accuracy of the $z_{phot}$ measurements manifest themselves 
statistically in our analysis (see \S\ref{voronoi}). 

The second form of the SED fitting used the aperture corrected ``total" magnitudes as input to {\textsc Le Phare} in order to estimate physical parameters of the sample. For the 
purposes of this paper, we used only the fitting for those galaxies with a secure $z_{spec}$. For these galaxies, the redshift was fixed to the spectral redshift prior to fitting. 
Input photometry was fit to a variety of different \citet{BC03} synthetic models generated from exponentially declining and delayed star-formation histories (SFHs) using a 
\citet{chab03} initial mass function (IMF) and varying dust contents and stellar-phase metallicities. More details on this fitting, including the parameter set used
and a brief comparison to an alternative approach, can be found in \S\ref{LePharedeetz}. For each galaxy, the median of the marginalized PDF for each physical parameter 
was adopted as the value of that parameter, with the formal uncertainty adopted from the values of each parameter at the 16th and 
84th percentile of the PDF. 

We note that while SED-fit $\mathcal{SFR}$s have considerable associated uncertainties, they are found to correlate well, on average, with independent 
measures of $\mathcal{SFR}$s at intermediate and high redshift, at least for galaxies with $\mathcal{SFR}$s in the range $5\la\mathcal{SFR}\la200$ $\mathcal{M_{\odot}}$ yr$^{-1}$ 
(e.g., \citealt{wuyts11, mostek12, stephane13, rodighiero14, marghe15, shivaei16, schaerer20}, though see \citealt{lower20} for an alternate view). These independent measures
include extinction-corrected rest-frame optical recombination line strengths as well as far-infrared and radio measurements among others, all of which are considerably less subject to the 
extinction and SFH concerns that can bias for UV-to-NIR SED-fit $\mathcal{SFR}$s. The $\mathcal{SFR}$ range quoted above encompasses 
the vast majority of our sample. 

In addition, all of our analysis is robust to the statistical incorporation of the formal random uncertainties in the derived $\mathcal{SFR}$s 
(see \S\ref{generalSFRdens}). This statement is only conclusive evidence of the robustness of our results if the error budget of our SED-fit $\mathcal{SFR}$ estimates is 
not dominated by systematics and the statistical errors from the SED fitting process are properly estimated from the SED fitting process. However, it is only if
the misestimation of the statistical errors or if those systematics are a strong function of the true $\mathcal{SFR}$ or of environment that they would meaningfully
modify any of the results presented in this study. Regarding the statistical errors, the marginalized SFR PDFs are consistent with independent error estimates, such as
those employing a Monte Carlo approach (as in, e.g., \citealt{wrustle14, victoria21}), which lends credence to the efficacy of the method on an absolute scale irrespective of
variables like dust content and environment. Regarding the systematics, it is unlikely 
the level of extinction has a strong effect for any galaxies other than those with the largest dust content 
(e.g., \citealt{rodighiero14}), a population that is mostly absent from our VUDS+ sample and one that we attempt to account for separately in \S\ref{generalSFRdens}. 
There is no \emph{a priori} reason to believe that the accuracy of SED-fit $\mathcal{SFR}$s is a function of environment.

Despite attempting to homogenize the data as much as possible, a few differences were noticed in the distributions of the SED-fit parameters from field to field likely due to issues of 
attempting to fit both different sets of filters as well as different treatment of the photometry for each of the various surveys (see \\S\ref{SEDsamplevariance}). For the SED-fit parameters 
employed in this study, $\mathcal{M}_{\ast}$ and $\mathcal{SFR}$, our full spectroscopic sample exhibited a median offset of +0.03/+0.07 dex and -0.13/+0.14 dex for the two parameters,
respectively, for the ECDFS and CFHTLS-D1 fields, respectively, relative to the COSMOS field. The COSMOS field is chosen as a reference field because it is the largest of the three VUDS fields,
and choosing it serves to mitigate sample variance and allows for the tests that are performed in \S\ref{SEDsamplevariance}. While it is possible this offset is physical and related to sample variance 
of the galaxy populations or differing selection functions across the different fields, this is not likely as a similar field-to-field offsets were seen in smaller redshift bins 
spanning the range of the full sample. The possibility that sample variance is primarily responsible for inducing these offsets is discussed further in \S\ref{SEDsamplevariance}. 

We correct for this effect by subtracting the bulk offset 
observed in the ECDFS and CFHTLS-D1 fields from the individual $\mathcal{M}_{\ast}$ and $\mathcal{SFR}$ values of all galaxies in those fields. As the offsets were measured by comparing to 
the average values in the COSMOS field, average $\mathcal{M}_{\ast}$ and $\mathcal{SFR}$ values for the galaxies in the COSMOS field were unchanged by this process. These corrected
values in the CFHTLS-D1 and ECDFS fields are used in all subsequent analysis. We note that none of the results in this paper are changed appreciably if we do not make this correction. 
Additionally, the main results also persist at a slightly reduced significance 
if we instead limit our sample to the COSMOS field that is uncorrected for 
this effect.

\begin{figure*}
\epsscale{1}
\plottwoalmostspecial{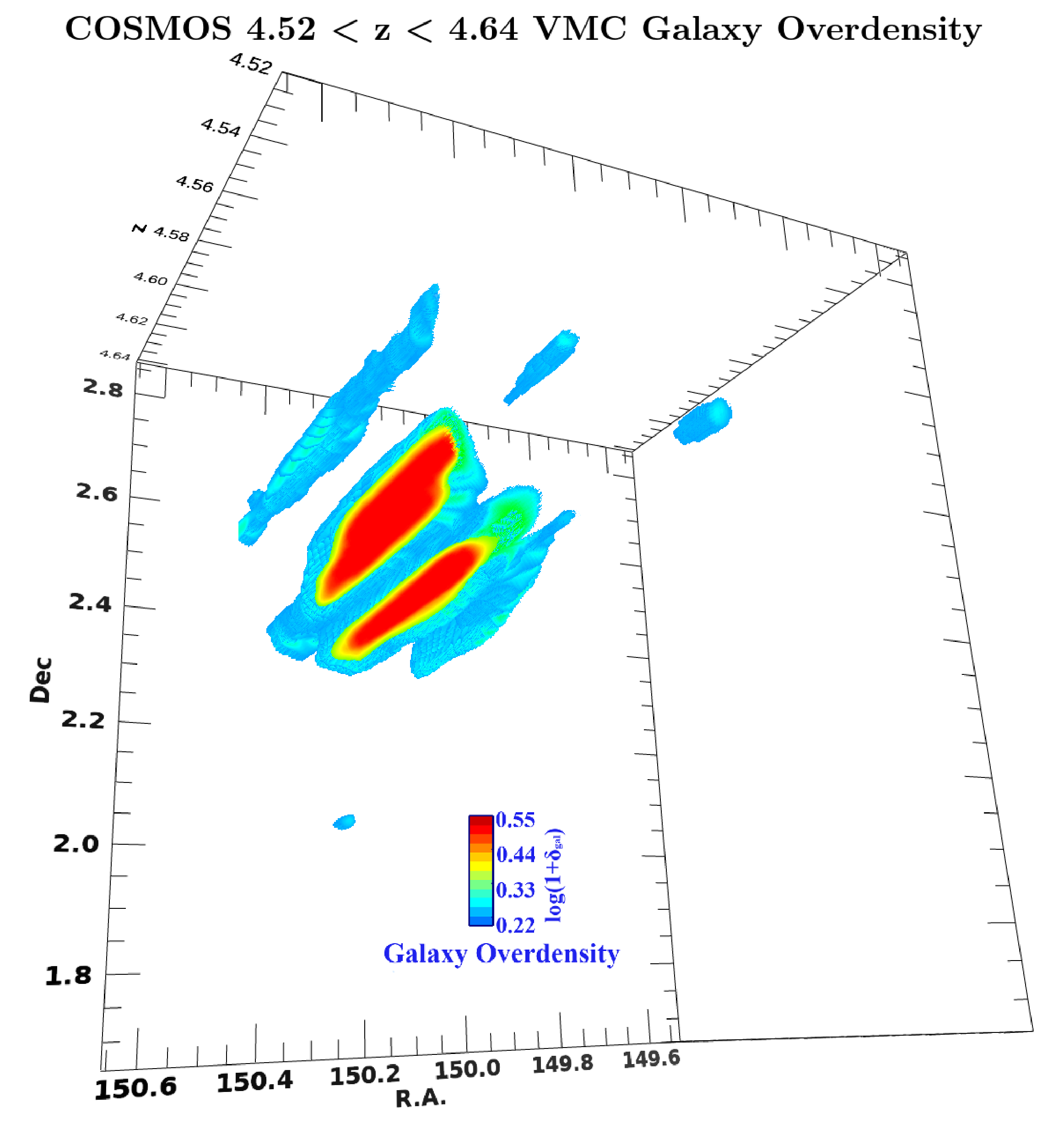}{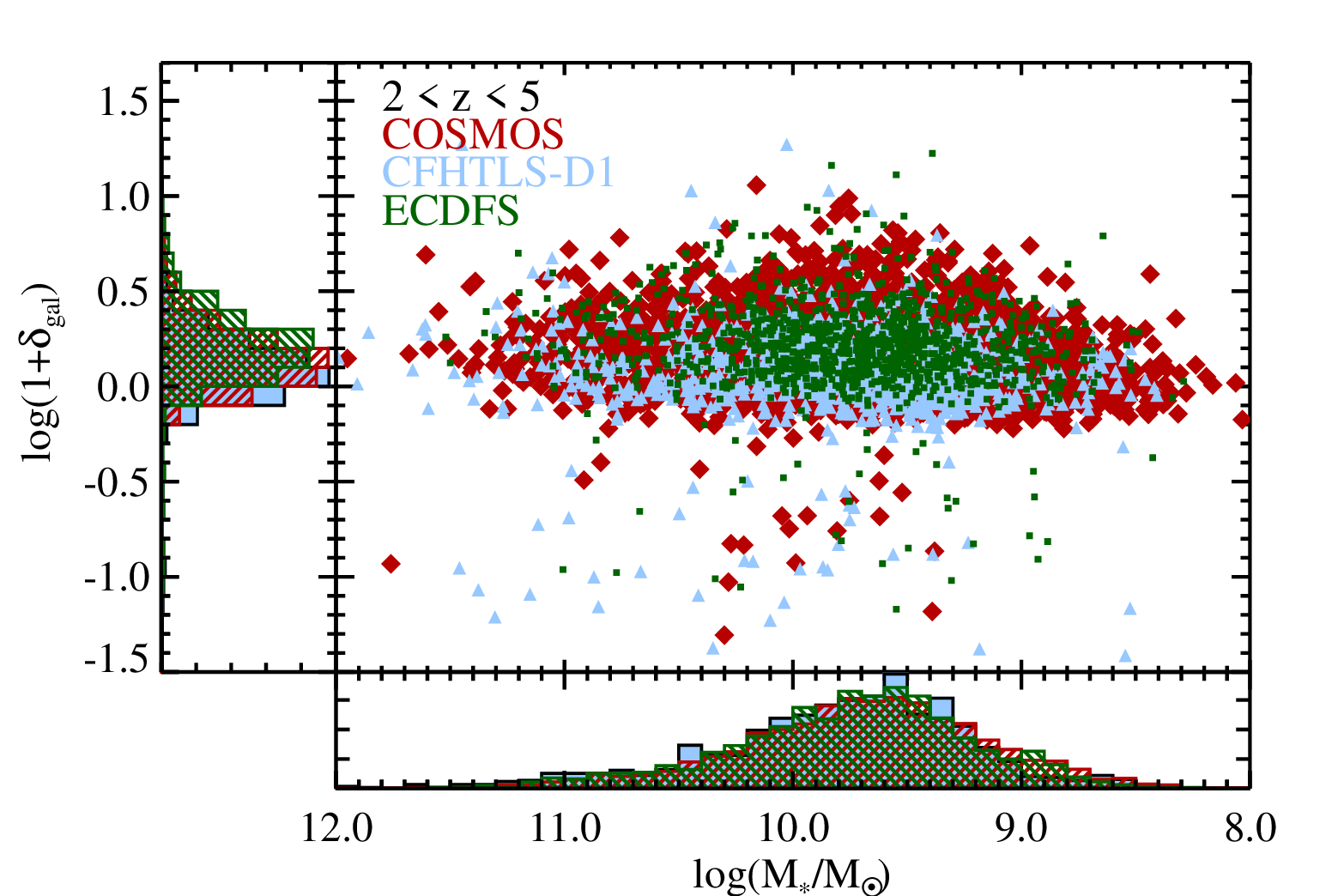}
\caption{Galaxy overdensity reconstruction using the VMC mapping method. \emph{Left:} Example of our VMC overdensity mapping (see \S\ref{voronoi}) as applied to the COSMOS field in the redshift range
$4.52 \le z \le 4.64$, nearly an identical range as is shown in Figure \ref{fig:dongo}. The main structure seen in the mapping is again the PCl J1001+0220 proto-cluster
at $z\sim4.57$. Here, in addition to showing the transverse dimensions, we also show the extent of the structure in the third dimension, redshift, which extends in this cube
between $4.52\le z \le 4.64$. The redshift dimension is added by rendering the narrow redshift slices used in our analysis to define galaxy overdensity rather than employing
a single slice over the redshift extent of the proto-cluster as was done in Figure \ref{fig:dongo}.  
The scale bar in the lower right shows the galaxy overdensity values, with the lower and upper end of the dynamic range corresponding the ``Intermediate" and ``Peak" regions 
as defined in \S\ref{generalSFRdens}. \emph{Right:} Distribution of the full VUDS+ spectral sample in $\mathcal{M}_{\ast}$ and
$\log(1+\delta_{gal})$ for the three separate fields targeted by VUDS (ECDFS: small green squares, CFHTLS-D1: light blue triangles, COSMOS: large red diamonds). The redshift range
imposed on the plotted galaxies is shown in the top left. Area-normalized histograms for the three fields for both parameters are shown on the side of the plot. The three fields show
generally similar distributions in both parameters, with a slight excess of galaxies at higher $\log(1+\delta_{gal})$ observed in COSMOS and ECDFS.}
\label{fig:voronoistuff}
\end{figure*}

\subsubsection{Details of the \textsc{Le Phare} spectral energy distribution fitting}
\label{LePharedeetz}

In total, we employed seven different BC03 models for the \textsc{Le Phare} SED fitting, 
five with SFHs characterized by an exponentially decaying tau model of the form $\psi(t)\propto \tau^{-1}e^{-t/\tau}$ and two delayed exponentially
decaying tau models of the
form $\psi(t)\propto \tau^{-2}te^{-t/\tau}$. All templates are considered at 43 possible ages between 50 Myr and 13.5 Gyr with the
constraint that the age of the model fit to a given galaxy cannot be older than the age of the universe
$t_{H}$ at the redshift of that galaxy. The delayed tau models are included as it has been suggested that high--redshift galaxies have SFHs that
may deviate considerably from the simple exponentially decaying tau models (e.g., \citealt{maraston10,schaerer13}). Values of $\tau$ range from 0.1 to
30 Gyr in roughly logarithmically equal time steps and delay times are set to 1 and 3 Gyr. Each BC03 model employs a \cite{chab03} initial
mass function and two values of stellar-phase metallicity, 0.4$Z_{\odot}$ and $Z_{\odot}$. Stellar extinction is allowed to vary between
$E_{s}(B-V)=0$ and 0.5 in steps of 0.05, with the prescription also allowed to vary between the \cite{calzetti00} starburst law and a Small
Magellanic Cloud--like law \citep{prevot84,stephane13}. Several prominent nebular emission lines are also added to the templates following the
methodology of \citet{dreadolivier09}. The median of the marginalized PDF for each parameter for each galaxy was adopted as the parameter value, with the uncertainties
set by the 16th/84th percentile of the marginalized PDF. We note that, for this fitting, we only used the broadbands available in each field to perform the fitting
in order to homogenize the photometric constraints as much as possible across the three fields, while for the photometric redshift estimation all available bands
(i.e., medium and broadbands) were used.

We compared the values of $\mathcal{M}_{\ast}$ and $\mathcal{SFR}$ estimated by our \textsc{Le Phare} fitting to those estimated using the Code Investigating
GALaxy Emission (\textsc{CIGALE}; \citealt{denis05, noll09, mederic19})\footnote{\url{https://cigale.lam.fr/}} imposing an identical parameter space to that employed in the
\textsc{Le Phare} fitting to the best of our ability given the large differences in the codes. We also tested the effects of
including and excluding various bands for subsets of our sample in the COSMOS field at various redshifts (Staab et al.\ \emph{in prep}). Despite the dramatically
different approach taken by \textsc{CIGALE}, we found no appreciable bias between the $\mathcal{M}_{\ast}$ and $\mathcal{SFR}$ values estimated by \textsc{CIGALE} using
all bands available for COSMOS and those estimated by \textsc{Le Phare} using only the broadbands, with an $\sim$0.2 dex scatter for each parameter.

\subsubsection{The dependence of physical parameter offsets on sample variance}
\label{SEDsamplevariance}

As discussed in \S\ref{SED}, the median values of $\mathcal{M}_{\ast}$ and $\mathcal{SFR}$ estimated for galaxies in the three fields studied in this paper varied. Specifically,
for our full spectroscopic sample, there was an offset of +0.03/+0.07 dex and -0.13/+0.14 dex for $\mathcal{M}_{\ast}$ and $\mathcal{SFR}$, respectively, for
galaxies in the ECDFS and CFHTLS-D1 field, respectively, relative to those in the COSMOS field. In this section, we investigate whether these differences may be due to
sample variance effects.

The coverage of our spectral data is the largest in the COSMOS field out of the three fields studied here, with the coverage being roughly two and five times the size of the 
CFHTLS-D1 and ECDFS fields, respectively. Since all photometric data in an individual field is treated in a homogenized manner, we can test the ability of sample variance alone
to reproduce the offsets seen in the $\mathcal{M}_{\ast}$ and $\mathcal{SFR}$ values for the average galaxy in the CFHTLS-D1 and ECDFS fields by sub-sampling the COSMOS data in
regions that are of equivalent sizes to these two fields. To this end, we broke the spectral sample in the COSMOS field into separate contiguous areas containing roughly equal numbers
of galaxies and calculated the dispersion of the median $\mathcal{M}_{\ast}$ and $\mathcal{SFR}$ values in the different regions. This exercise was performed for two, three, four, five,
and six separate regions. The maximum dispersion in this method was 0.04 dex in both parameters, which is less than most of the field-to-field offsets measured in \S\ref{SED}.

It is also possible that the heterogeneous selection of galaxies within our final VUDS+ sample could contribute to the observed offsets. The VUDS survey was selected
nearly exclusively through a homogeneous set of $z_{phot}$ or color criteria across all fields. If selection effects were the primary reason for the observed offset, we should see
offsets that are considerably smaller if only the VUDS galaxies are considered for each fields. However, the median offsets from field to field in both parameters are nearly identical
to those of the full VUDS+ sample. We conclude from these exercises that the observed offsets are unlikely due to sample variance or selection effects, and are rather primarily an
artifact of the heterogeneous imaging observations and treatment of the photometric data across the various fields. None of the results presented in this paper change meaningfully
whether or not the observed offsets in $\mathcal{M}_{\ast}$ and $\mathcal{SFR}$ are applied to the galaxies in the non-COSMOS fields.

\subsection{Local environment}
\label{voronoi}

To define environment in this study we rely on a modified version of the Voronoi tessellation measure of the density field that has been employed by a variety of
other studies that rely nearly exclusively on photometric redshifts (e.g., \citealt{scoville13, darvish15, vernesa17}). The method used here, known as Voronoi Monte 
Carlo (VMC) mapping, which uses a weighted combination of spectroscopic and photometric redshift information, follows the method described in 
\citet{lem18} almost identically. As such, it is only briefly described here. For each field, beginning at $z=2$ and reaching up to $z\sim5$ in steps of 3.75 Mpc 
steps along the line of sight, a suite of 25 Monte Carlo realizations of $z_{spec}$ and $z_{phot}$ catalogs were generated for each step. As in other studies employing
the VMC mapping at high redshift \citep{olga18, lem18}, the $z_{spec}$ values are not considered as absolute truth, but are rather treated statistically 
following the method outlined in Appendix \ref{appendix:A}. At each iteration of each step, a set of objects with and without $z_{spec}$ values were defined 
following the method described in Appendix \ref{appendix:A}. 
The spectral and photometric sample realized for a given iteration and slice was then cut at a magnitude 
limit that was determined uniquely for each field, [3.6]$<$25.3 for COSMOS, [3.6]$<24.8$ for ECDFS, and [3.6]$<23.1$ and $K_{s}<24.1$ for CFHTLS-D1. Detection in two NIR bands
was required for the CFHTLS-D1 field due to the relative shallowness of the IRAC imaging in this field and the slightly different approach taken for measuring IRAC
photometry (see \citealt{dong15}). This requirement is not imposed in the other two fields, as the ground-based NIR 
data were shallower than the IRAC imaging in those fields. As a consequence, for the COSMOS and ECDFS field, the [3.6] cut effectively guaranteed useful information in the other 
NIR bands, which ensured that the 
rest-frame SED of all sources was meaningfully probed over the Balmer/$D_{n}$(4000) break. These photometric limits correspond to the 3$\sigma$ limiting depth of the IRAC/WIRCam images in the 
three fields\footnote{We note that overdensity values in a given field are not a strong function of the precise magnitude cut imposed. For example, adopting a [3.6]$<$24.8 cut in the COSMOS
field, the same cut as was used for ECDFS, does not result in a larger scatter or bias relative to the fiducial map as simply rerunning the VMC mapping at the same magnitude limit.}.
Since these bands probe light redward of the Balmer/4000\AA\ break for the redshifts considered here, the tracer populations used to estimate the local density
can be considered to be broadly stellar mass limited\footnote{As an example, in the COSMOS photometric and spectroscopic sample employed in our analysis, the rest-frame dust-uncorrected 
absolute i-band magnitude, a band that is roughly equivalent to observed frame [3.6] at $z\sim4$, has a Spearman rank correlation coefficient, $\rho$, of -0.88 and -0.91, respectively, 
with a $>>$3$\sigma$ rejection of the null hypothesis that the two parameters are uncorrelated.}. Adopting the methodology used by several other works (\citealt{quadri12, AA14, lem18}, 
see Appendix B of \citealt{lem18} for a description of the method), we estimate that the resultant sample after imposing these cuts is 80\% complete to 
$\log(\mathcal{M}_{\ast}/\mathcal{M}_{\odot})$$\sim$9.2-9.5 across the three fields at $z\sim3$, with the exact limit depending on the NIR imaging data available in a particular field.
These estimates are quoted primarily to provide a rough characterization of the tracer population used in the reconstruction of the density field in order to ease comparison
with other works. 

For each realization of each redshift step, two-dimensional Voronoi tessellation is then performed on all objects whose assigned redshift fell within $\pm$3.75 Mpc of the central
redshift of each bin, in other words, a bin width of $\Delta \chi=7.5$ Mpc or $\Delta z\sim0.015-0.08$ from $z\sim2-5$. The 3.75
Mpc steps between slices along with the slice thickness ensure overlap between successive slices such that we do not miss overdensities by randomly choosing unlucky redshift bounds.
For each realization of each slice, a grid of 75$\times$75 kpc pixels was created to sample the underlying local density distribution. The local density
at each grid value for each realization and slice was set equal to the inverse of the Voronoi cell area (multiplied by $D_{A}^2$, where $D_{A}$ is the angular diameter distance at the
redshift of interest) of the cell that enclosed the
central point of each pixel. Final local densities, $\Sigma_{VMC}$, for each grid point in each redshift slice are then computed by median combining the values
of 25 realizations of the Voronoi maps. The choice of 25 realizations was motivated by balancing computational time with a sufficient number of realizations to
effectively sample the probability distributions of the $z_{phot}$ and $z_{spec}$ objects. In previous tests used for the $z\sim1$ version of this mapping (see, 
e.g., \citealt{lem17, AA17, denise20}) where the computational time to generate maps is considerably less, maps that were created using 25 iterations did not show 
significantly higher variance in their density values relative to different realizations of maps using two or four times more iterations. 

The local overdensity value for each grid point is then computed as
$\log(1+\delta_{gal}) \equiv \log(1+ (\Sigma_{VMC}-\tilde{\Sigma}_{VMC})/\tilde{\Sigma}_{VMC})$, where $\tilde{\Sigma}_{VMC}$ is the median $\Sigma_{VMC}$
for all grid points over which the map was defined (i.e., excluding an $\sim1\arcmin$ wide border region to mitigate edge effects)\footnote{This is
what we refer to as the masking procedure later in the section. Briefly, the masking procedure used for the maps disregards all pixels in the maps that 
have values below a certain value in each field that are calibrated for that particular field to exclude the regions where the spectral and photometric 
coverage begins to fall off. This masking is used only when calculating $\tilde{\Sigma}_{VMC}$ for each VMC slice.}. Uncertainties on the
$\log(1+\delta_{gal})$ values of each pixel are taken from the 16th and 84th percentile of the distribution of $\log(1+\delta_{gal})$ values populated
by the 25 realizations (i.e., the fourth lowest and highest $\log(1+\delta_{gal})$ value for all realizations). Because these maps are, generally, so 
large in the transverse dimensions relative to the size of any contained structures, 
the maps are much less sensitive to the types of issues discussed in previous works that implement the VMC method (e.g., \citealt{olga18}; \citealt{denise20}) that
serve to confuse or bias overdensity measurements in certain redshift slices due to the presence of large structures (i.e., if a given structure is near the size 
of the field, the median $\Sigma_{VMC}$ value for that slice will be artificially increased causing an overall decrease in the density contrast, $\delta_{gal}$,
for that structure). The use of a NIR-selected sample rather than
an optically selected sample also serves to mitigate these issues, as, due to the higher source density at a fixed magnitude limit, a larger fraction of $z_{phot}$ 
objects enter into each VMC map, which serves to smooth
out the density field. As such, it was not necessary to take the approach used in \citet{olga18} and \citet{denise20} to fit the average density value and root mean 
squared (RMS) fluctuations as a function of redshift slice. Rather, here we simply use the median density value measured on the masked maps to calculate 
$\log(1+\delta_{gal})$. Using the approach adopted in \citet{olga18} and \citet{denise20}, while 
only subtly different than the one used here, is crucial when attempting to search for coherent overdensities, such as (proto-)groups and clusters, to 
define the global environment of galaxies. This approach will be used to define this environmental metric in future work presenting the full set of 
overdensity candidates in the VUDS+ sample. We note that adopting this approach has no meaningful consequences for the results presented in this paper.

The main modification to the VMC method made in this work relative to other works that employ this technique is the formalism of the statistical treatment of the 
spectral redshifts in the mapping. This modification is described in Appendix \ref{appendix:A}. In Figure \ref{fig:voronoistuff} we show an example of our VMC
reconstruction of a proto-cluster at $z\sim4.57$ in the COSMOS field as well as the stellar mass versus. overdensity distribution of galaxies with secure $z_{spec}$ values in
each of the three fields. We note that all galaxies with an assigned overdensity value of $\log(1+\delta_{gal})<-0.3$ were removed from the final sample as 
such values are likely attributable to edge effects or other artifacts. Such galaxies constituted a negligible percentage of our original VUDS+ sample ($<1$\%). 
Additionally, our results are not appreciably changed if we instead include these galaxies in the analysis nor if we slightly change the value at which we cut 
(i.e., $\pm$0.1 dex in $\log(1+\delta_{gal})$). As in the case of the SED fitting parameters, we noticed a slight variation in the median 
$\log(1+\delta_{gal})$ values across the spectral sample in the three fields, 
with $\log(1+\tilde{\delta}_{gal})=$0.14, 0.11, \& 0.20 in the COSMOS, CFHTLS-D1, \& ECDFS fields, respectively\footnote{We note that $\log(1+\tilde{\delta}_{gal})$ is not
zero since the value is always measured at the location of a galaxy, which is a denser region than typical in the universe.}. However, these offsets were small and varied 
over the redshift range of the final spectral sample, which implied the genesis of the difference was perhaps astrophysical or due to a subtlety in the VMC
map-making process, for example, different tracer populations, slightly different masking, the size of the field relative to the size of large-scale structure. Therefore, we decided 
not to apply corrections for this effect. As for the correction we made for the physical parameters in the previous section, we note here that the main results of 
this study are broadly invariant to whether or not we make a bulk correction for the differing $\log(1+\delta_{gal})$ values.

\subsection{Representativeness of the spectral sample}
\label{represent}

\begin{figure}
\epsscale{1}
\plotone{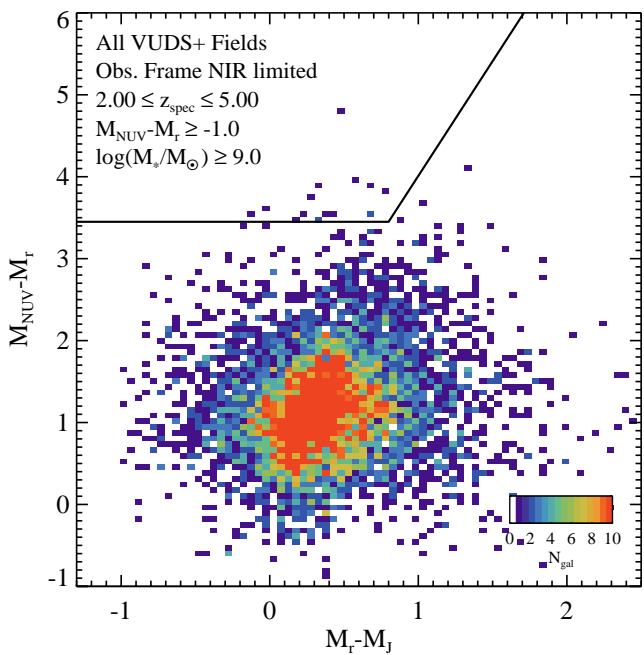}
\caption{Rest-frame $M_{NUV}-M_r$ vs. $M_{r}-M_{J}$ color-color diagram of the galaxies in the final spectral sample that satisfy the stellar mass, redshift, and
$M_{NUV}-M_r$ limits given in the top left as well as the observed-frame NIR photometric cuts mentioned in \S\ref{voronoi}. A color bar on the bottom right indicates 
the color that corresponds to the number of galaxies in each two-dimensional bin. The black lines delineate color-color regions that are inhabited by quiescent (top left) 
and star-forming (bottom) galaxies adopted from regions defined for a sample in a similar redshift in \citet{lem14b}. The sample shown here is primarily composed of 
star-forming galaxies containing low to moderate amounts of dust.} 
\label{fig:specNUVrJ}
\end{figure}

\begin{figure*}
\epsscale{1}
\plottwo{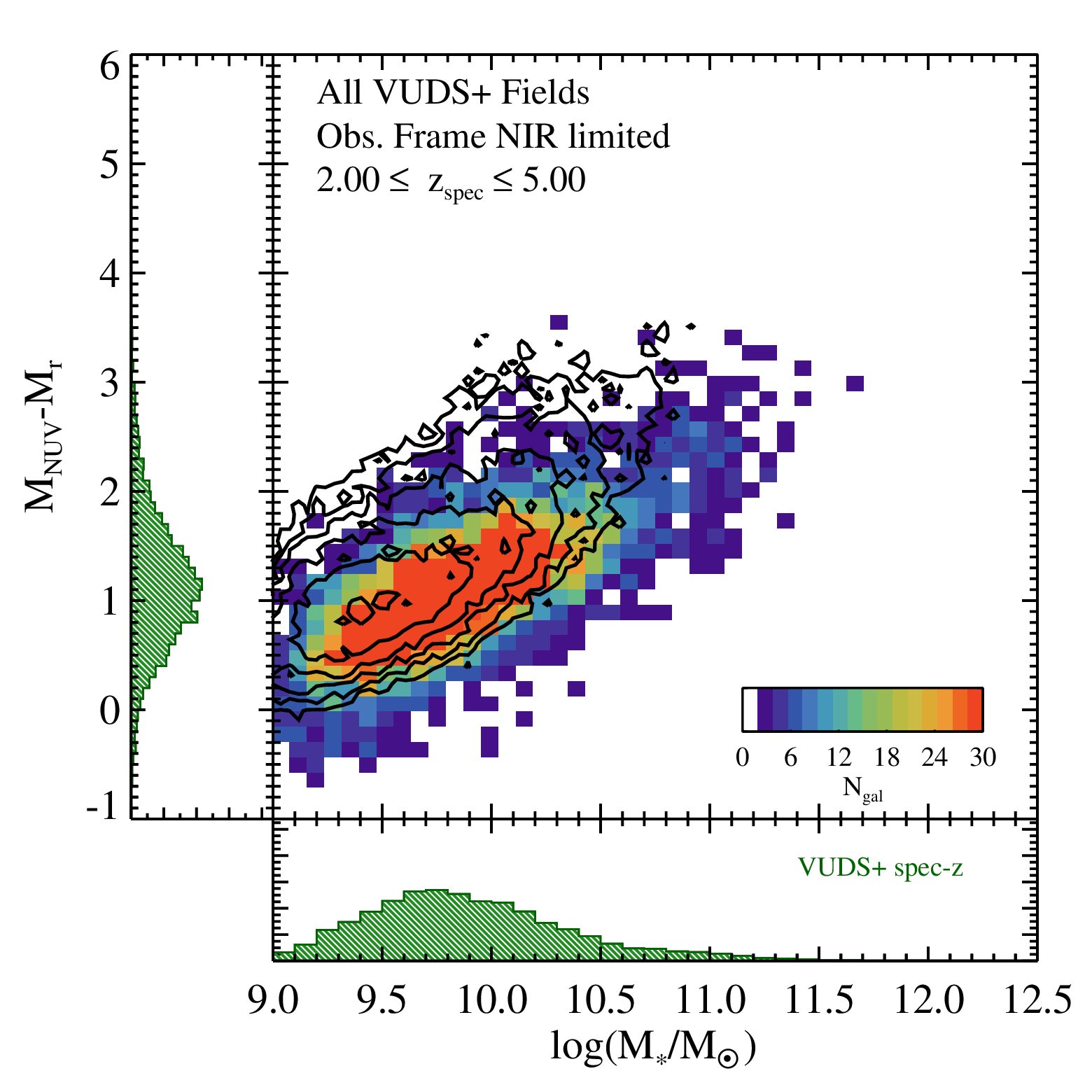}{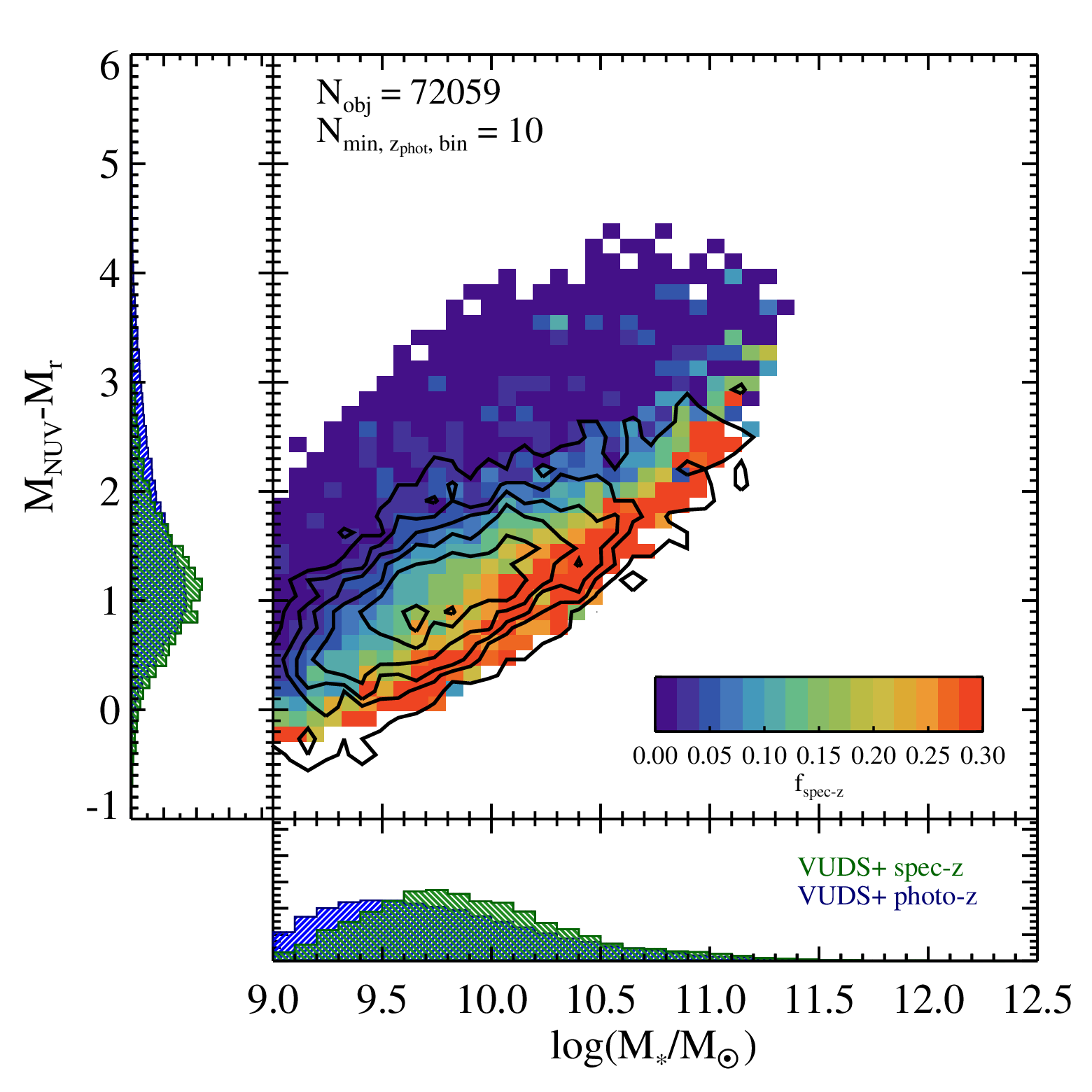}
\caption{Comparison of the VUDS+ spectroscopic sample and the underlying parent sample. \emph{Left:} Rest-frame color-stellar mass of of objects across all three VUDS fields in the range $2 \le z_{spec}\le 5$
that have a secure spectroscopic redshift and meet the observed-frame NIR magnitude cuts discussed in \S\ref{voronoi}. 
The color bar indicates the number of galaxies within
each narrow color and stellar mass bin. Area-normalized histograms of color and stellar mass for the VUDS+ $z_{spec}$ sample are shown on the sides of the plot. The contours
indicate the density of the parent photometric sample in this space and are spaced in equal logarithmic steps of 0.2 dex starting at $N_{phot}$=15. The offset between the peak
of the photometric and spectroscopic samples is $\sim$0.3 dex in stellar mass, with the photometric sample peaking at lower stellar masses. \emph{Right:} 
Same as the left panel, except the color bar now indicates the fraction of $z_{phot}$ objects in each color and stellar mass bin that have secure spectral redshifts. The total 
number of photometric objects and the number of photometric objects required to fall in a two-dimensional bin in order for that bin to be plotted are indicated in the top left. 
Area normalized histograms of the full $z_{phot}$ and VUDS+ $z_{spec}$ sample are shown on the sides of the plot. The offset of $\sim0.3$ dex in stellar mass between
the peak of the photometric and spectroscopic samples is also seen here, with the former peaking at lower stellar masses. The contours indicate the density of the spectral sample
(i.e., the same as is shown in the left panel) and are spaced in equal logarithmic steps of 0.2 dex starting at $N_{spec}$=5. } 
\label{fig:representativeness}
\end{figure*}

In Figure \ref{fig:specNUVrJ} we show the rest-frame $M_{NUV}-M_r$ versus $M_{r}-M_{J}$ color-color diagram of all galaxies in the range 
$2 \le z_{spec} \le 5$ subject to the observed-frame NIR magnitude cut discussed previously (see \S\ref{voronoi} for details on this cut for each field). Additionally, for this exercise 
we impose rest-frame color and stellar mass cuts to the sample of $M_{NUV}-M_r\ge -1$ and $\log(\mathcal{M}_{\ast}/\mathcal{M}_{\odot})\ge 9$. These additional color and stellar 
mass cuts retain the vast majority ($>90$\%) of the galaxies in the final spectral sample and are imposed because our imaging data are severely incomplete outside of 
these limits. Overplotted on 
this figure is the delineation line between regions that 
generally contain quiescent and star-forming galaxies. These regions are adopted from the regions estimated in \citet{lem14b} for galaxies in the redshift range $2 \le z < 4$. 
The rest-frame colors of the galaxies in the full spectral sample are generally consistent with those of star-forming galaxies with minimal levels of extinction 
($A_{\rm{v}}<1.5$).		

That the majority of the galaxies in the final spectral sample appear at these rest-frame colors is perhaps problematic for leveraging this study into a general study 
of galaxy evolution. Though quiescent galaxies are by far a subdominant population over the redshift and stellar mass ranges plotted (e.g., \citealp{dreadolivier13, muz13,AA14}), 
such galaxies are observed to exist over this redshift range (e.g., \citealt{gobat11, paolo13, newman14, taowang16,taowang18}). Further, dusty star-forming galaxies, 
galaxies that are essentially completely absent from the spectral sample, also contribute to the overall galaxy population at these redshifts, particularly 
at the highest stellar masses (e.g., \citealp{connivingcasey14, lem14b, connivingcasey15, vernesa17, vernesa17b}). The contributions of each of these two populations will be considered 
separately when we investigate the $\mathcal{SFR}-\log(1+\delta_{gal})$ relation in earnest later in the paper. For now we consider only the question of whether or not the 
final VUDS+ spectral sample can be considered representative of the full population of non-dusty star-forming galaxies at $2 \le z \le 5$, a population that comprises the 
vast majority of the galaxy population in the early universe. 

In Figure \ref{fig:representativeness} we show a $M_{NUV}-M_r$ vs. $\mathcal{M}_{\ast}$ color-stellar mass diagram. In the left panel of this figure, we show the density 
of the same spectral sample as was shown in Figure \ref{fig:specNUVrJ} in color-stellar mass space. Overplotted contours indicate the density of the underlying
sample of objects in the same $z_{phot}$ range that are significantly detected at or above the 3$\sigma$ limit of the NIR image appropriate for that field (i.e., the photometric 
parent sample) in the same space. The right panel of this figure directly compares the spectral sample with the photometric parent sample.
In all cases we limit the 
sample to $\log(\mathcal{M}_{\ast}/\mathcal{M}_{\odot})\ge 9$ and excise any galaxies with secure spectral redshifts that are outside the range considered here ($2 \le z \le 5$).
Again, this stellar mass limit is imposed as our imaging data are severely incomplete below this limit, which precludes our ability to do the tests that follow. 
We chose to agnostically include galaxies below this stellar mass limit in our main analysis as, due to the lack of imaging depth, we cannot definitively prove or disprove
whether the spectral sample below this limit is representative. However, the vast majority of galaxies in our $z_{spec}$ sample are above this stellar mass limit, and our 
results do not change meaningfully if we instead exclude $z_{spec}$ galaxies below this limit in all subsequent analysis. 

While the left panel of Figure \ref{fig:representativeness} shows the number of galaxies in a given two-dimensional color/stellar mass bin, in the right panel the color bar 
indicates the fraction of $z_{phot}$ objects 
in each bin that have a secure $z_{spec}$. For objects in the $M_{NUV}-M_r$ 
and $\mathcal{M}_{\ast}$ range used to select the final spectral sample it is apparent that $\ga$5\% of objects in each color/stellar mass bin over which the majority 
of the photometric objects lie in this phase space (see the contours in the left panel of Figure \ref{fig:representativeness}) have a secure $z_{spec}$ except 
for the reddest objects, colors that are generally reserved for quiescent and dusty star-forming galaxies. The area normalized histograms 
for the $z_{phot}$ and the final spectral sample generally appear to show a high degree of concordance. However, a deviation in the one-dimensional histograms is observed at the 
lowest stellar masses plotted (i.e., $\log(\mathcal{M}_{\ast}/\mathcal{M}_{\odot})\la9.5$), with an $\sim$0.3 dex offset observed between the peak of the stellar 
mass distribution of the photometric and spectroscopic samples. This deviation can also be seen in the two-dimensional density of the parent photometric and 
spectral sample in the left panel of Figure \ref{fig:representativeness}. Above $\log(\mathcal{M}_{\ast}/\mathcal{M}_{\odot})\ga9.5$), the two distributions are not meaningfully  
different, with a one-dimensional Kolmogorov-Smirnov (KS) test between the two samples at these masses returning no significant ($>$3$\sigma$) evidence to reject the 
null hypothesis that the two distributions are drawn from the same sample.  

There also appears to be a clear selection effect present in the two-dimensional plot in which galaxies of redder colors have $z_{spec}$ fractions lower than those of their bluer 
counterparts at a given stellar mass. While it is apparent from the contours overlaid on the right panel of Figure \ref{fig:representativeness} that the bulk of the galaxies 
within the spectral sample reside at intermediate rest-frame colors at a given $\mathcal{M}_{\ast}$, it is still possible that the over-representation of bluer galaxies at a 
given $\mathcal{M}_{\ast}$ could subtlety bias our investigation of the relationship between $\mathcal{SFR}$ and $\delta_{gal}$. As we will show in \S\ref{altmethod},
this is likely not the case, as homogenizing the $z_{spec}$ fraction for all color-$\mathcal{M}_{\ast}$ that are reasonably well sampled (see \S\ref{altmethod} for the 
precise definition of this term), yields results that are broadly unchanged from those using the full spectral sample. Additionally, we will show in \S\ref{SFRdensity} 
that our results are largely invariant when including the contribution of quiescent and dusty star-forming galaxies under a variety of different assumptions. Thus, 
we conclude that the lack of representativeness 
of our spectral sample does not meaningfully bias any of the results presented in this paper. 
 
\section{The evolution of the $\mathcal{SFR}$-density relation}
\label{SFRdensity}

In the previous sections we have defined a sample of 6730 unique galaxies in the redshift range $2 < z_{spec} < 5$ with estimates of $\mathcal{M}_{\ast}$, 
$\mathcal{SFR}$, and local environment, $\log(1+\delta_{gal})$. For the remainder of the paper, we investigate the relationship of these parameters 
and the consequences of those relationships for environmentally driven galaxy evolution in the early universe. The galaxies in our final spectral sample, 
while not strictly representative of low to moderately dusty star-forming population at these redshifts, do not appear different from such galaxies in any 
way that is meaningful for the results that follow (see \S\ref{altmethod}). Additionally, while quiescent and dusty star-forming galaxies are almost 
completely absent from this sample, we  will show later that our main conclusions are robust to their absence. With those caveats in place, we begin by 
investigating the general relationship between $\mathcal{SFR}$, $\mathcal{M}_{\ast}$, and $\log(1+\delta_{gal})$ in our final spectral sample. 

\subsection{The general $\mathcal{SFR}$$-$density and $\mathcal{M}_{\ast}$$-$density relations at high redshift}
\label{generalSFRdens}

\begin{figure*}
\epsscale{1}
\plottwo{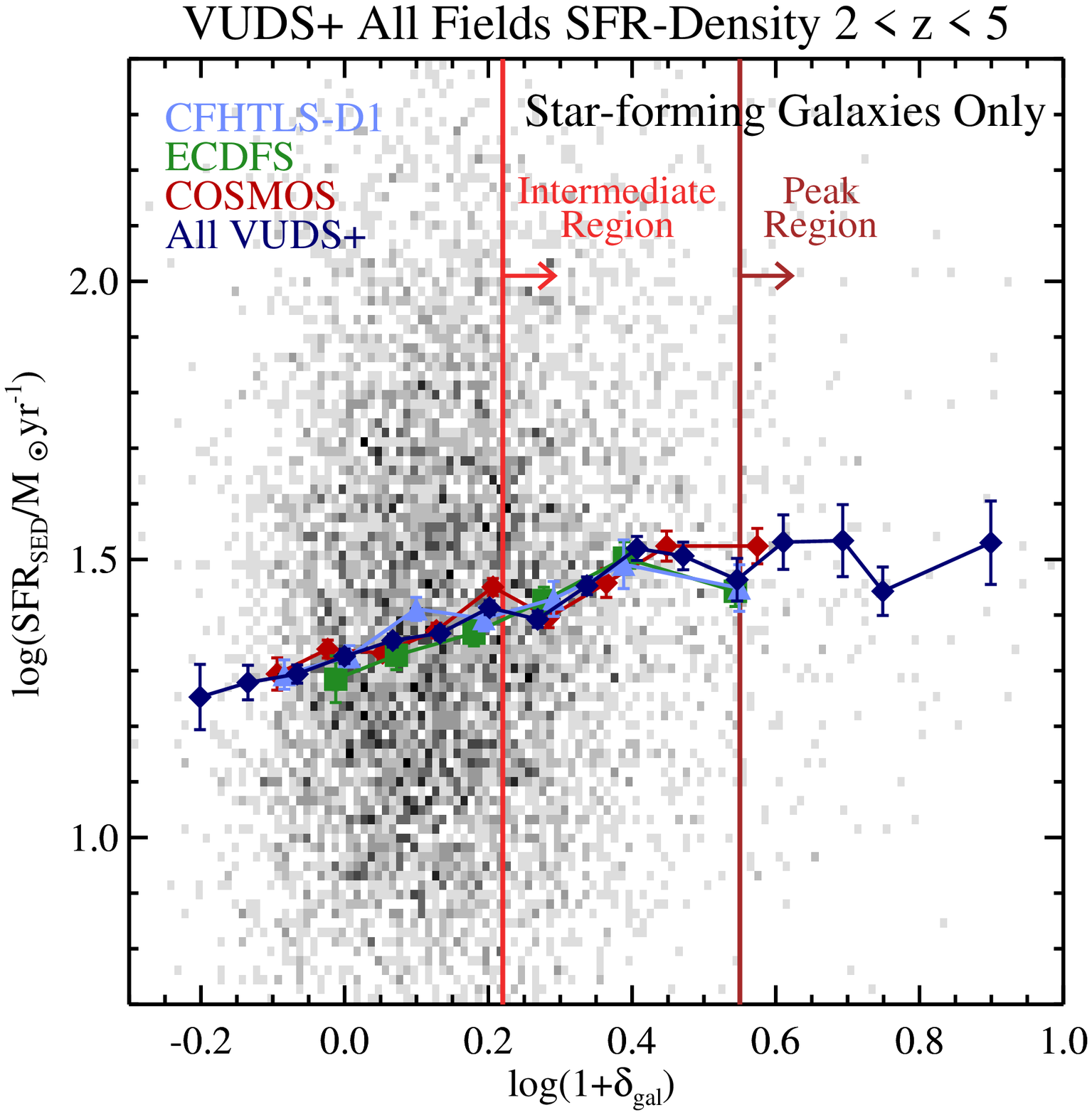}{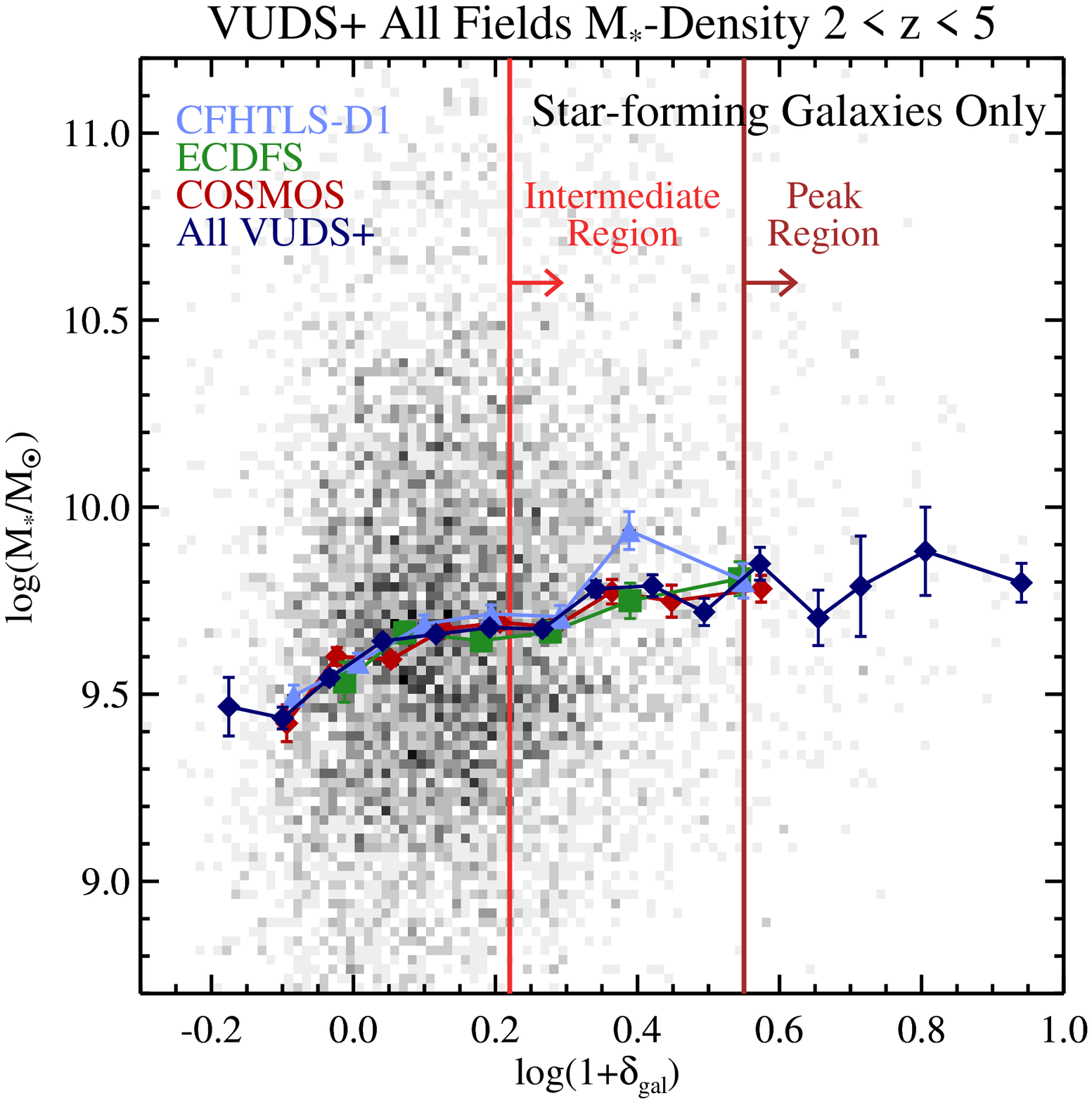}
\caption{Galaxy properties as a function of environment in the VUDS+ sample. \emph{Left:} Relationship between $\mathcal{SFR}$ and $\log(1+\delta_{gal})$ for the 6730 galaxies in the VUDS+ 
spectroscopic sample. The backdrop shows the number density
at a given $\mathcal{SFR}$ and $\log(1+\delta_{gal})$ value for all galaxies in the sample and the light blue, green, red, and dark blue datapoints and lines show median values in
each bin in $\log(1+\delta_{gal})$ for galaxies in the CFHTLS-D1, ECDFS, COSMOS, and the full combined VUDS+ sample, respectively. Uncertainties reflect errors on the median values.
The vertical dashed lines indicate the ``Intermediate" and ``Peak" regions of the $\log(1+\delta_{gal})$ distributions (see \S\ref{generalSFRdens}), with all values lower
than $\log(1+\delta_{gal})=0.22$ being considered ``Field" galaxies. The ``All VUDS+'' sample can be plotted over a larger dyanamic range in $\log(1+\delta_{gal})$ 
relative to the individual field samples due to its larger sample size. 
A weak but highly significant \emph{positive} correlation is observed (see \S\ref{generalSFRdens}).
\emph{Right:} As in the \emph{left} panel, but now plotted for stellar mass ($\mathcal{M}_{\ast}$) and $\log(1+\delta_{gal})$. Again, a weak but highly significant positive
correlation is observed between $\mathcal{M}_{\ast}$ and $\log(1+\delta_{gal})$}.
\label{fig:mainSFRnSMdens}
\end{figure*}

In the left panel of Figure \ref{fig:mainSFRnSMdens} we plot the relationship between $\mathcal{SFR}$ and $\log(1+\delta_{gal})$ for the 6730 galaxies in our final 
VUDS+ sample over the full redshift range of the sample, $2 < z < 5$. The colored $\mathcal{SFR}$ and $\log(1+\delta_{gal})$ points are plotted against the backdrop of
the individual values for the full sample indicate the median $\mathcal{SFR}$ in bins of $\log(1+\delta_{gal})$. The errors associated with each median value are 
calculated by the normalized median of the 
absolute deviations $\sigma_{NMAD}/\sqrt{n-1}$ of the individual $\mathcal{SFR}$ values of each bin, where $n$ is the number of galaxies in each bin (see \citealt{lem18} and 
references therein for details on this error estimate). In addition, three different regions 
are demarcated in the plot, a Field region, defined as $\log(1+\delta_{gal})<0.22$, an Intermediate region, defined as $0.22 \le \log(1+\delta_{gal}) < 0.55$, 
and a Peak region, defined as $\log(1+\delta_{gal}) \ge 0.55$. The latter two of these regions are set to correspond to the outskirts and peak regions of the 
Hyperion proto-supercluster defined in \cite{olga18}, and can more generally be thought of as the outskirts and the core region of the protocluster 
environment\footnote{The term core can be a little deceptive here as such regions can still be quite large, e.g., $R_{proj}=0.5-1$ Mpc \citep{olga18},
values that rival $R_{vir}$ values of massive clusters at $z\sim1$ (e.g., \citealp{lem12}).}. 

In stark contrast to what is observed at lower redshifts ($z<2$), there exists a clear, nearly monotonic increase in the average $\mathcal{SFR}$ with 
increasing $\log(1+\delta_{gal})$. This increase is observed to be $\sim$0.3 dex in $\mathcal{SFR}$ over the full dynamic range of $\log(1+\delta_{gal})$ 
probed by the sample ($-0.3 \le \log(1+\delta_{gal})\le 1$). This relation is also plotted for the subsamples that are 
drawn from the three individual fields and mirrors the behavior of the full VUDS+ sample without 
exception. 
The consistency of the relation across all three fields is important to lend credence to these results. To further formalize this result, we calculated the 
Spearman rank correlation coefficient, $\rho$, between $\mathcal{SFR}$ and $\log(1+\delta_{gal})$ for the full VUDS+ sample, finding $\rho_{\mathcal{SFR}-\delta_{gal}}=0.13$, which
indicates a weak but highly significant positive correlation, with a rejection of the null hypothesis of lack of correlation between the two variables at the $\sim10\sigma$ level. We note that
the Spearman coefficient was calculated on the individual data points in the full VUDS+ sample rather than the values in the binned data points shown in Figure \ref{fig:mainSFRnSMdens}.
This coefficient was also calculated for the COSMOS-only sample resulting in $\rho_{\mathcal{SFR}-\delta_{gal}}=0.15$ at only a slightly reduced significance ($\sim$8.5$\sigma$). 
Within the VUDS+ data there is a clear reversal of the $\mathcal{SFR}-$Density (hereafter $\mathcal{SFR}-\delta_{gal}$) relation.  

While it is tempting to interpret the result in the VUDS+ sample as an indication of the \emph{global} reversal of the $\mathcal{SFR}-\delta_{gal}$ relation at $z>2$, a few 
considerations are necessary to test the veracity of this result. As mentioned in \S\ref{represent}, the VUDS+ sample 
is nearly completely devoid of galaxies at the two extreme ends of the $\mathcal{SFR}$ spectrum: those that have ended their star-formation 
activity (i.e., quiescent) and those forming stars prodigiously (i.e., extremely obscured dusty star-forming galaxies). These galaxy populations are known to be associated with at least 
some overdensities at these redshifts (e.g., \citealt{kodama07, gobat11, connivingcasey15, taowang16, vernesa17, forrest20, wenjia20, federica21, shen21, shi21}). To incorporate the 
potential effects of such galaxies in the $\mathcal{SFR}-\delta_{gal}$ relation, we began by estimating the 
number of each of the two galaxies populations that should exist in our VUDS+ sample were it truly representative of the underlying photometric sample.

Starting with the COSMOS2015 catalog, objects were selected in the photometric
range $2 < z_{phot} < 5$ and, as in our final spectral sample, limited to $\log(\mathcal{M}_{\ast}/\mathcal{M}_{\odot})>9$ and [3.6]$<$25.3.
The COSMOS2015 catalog is chosen because it is the deepest of our three photometric catalogs. 
The percentage of quiescent galaxies in this sample was estimated by counting the number of galaxies that populated the quiescent region of rest-frame 
$NUVrJ$ color-color as defined by \cite{lem14b}:
\begin{equation}
\begin{aligned}
M_{NUV}-M_r > 3.45  \qquad\qquad\qquad      (M_r-M_J < 0.8)		\\
\\
M_{NUV}-M_r > 2.8(M_r-M_J)+1.21 \qquad  (M_r-M_J \ge 0.8).
\label{eqn:colorselectQ}
\end{aligned}
\end{equation}

\noindent For dusty star-forming galaxies, we estimated their fractional contribution by 
counting the number of objects that had rest-frame $NUVrJ$ colors that indicate an excess extinction of $A_{\rm{V}}\ge1.5$ relative to the average 
COSMOS2015 object at these redshifts, which is characterized by median $NUV-r$ and $r-J$ colors of 1.06 and 0.33, respectively. More specifically:
\begin{equation}
\begin{aligned}
M_{NUV}-M_r > 2 \; \wedge \; M_{NUV}-M_r > \frac{3.7}{M_r-M_J+0.3} \; \wedge \; \qquad \; \\
M_r-M_J > 0.7 \; \wedge \; M_{NUV}-M_r \le 3.45  \; \; \; (M_r-M_J < 0.8)  \; \; \; \\
\\
M_{NUV}-M_r \le 2.8(M_r-M_J)+1.21 \qquad  \; \; (M_r-M_J \ge 0.8). \; \; \; 
\label{eqn:colorselectDSFG}
\end{aligned}
\end{equation}
 
\noindent Using the cuts defined in Equations \ref{eqn:colorselectQ} and \ref{eqn:colorselectDSFG}, the fractional contribution of quiescent and dusty star-forming galaxies were found 
to be 0.8\% and 4.2\%, respectively. 

While these are modest fractions in both cases,
both populations lie at extreme values of $\mathcal{SFR}$, so their contribution is necessary to quantify. In addition, the detection 
image used for the COSMOS2015 catalog was chosen to be a $z^{++}YJHK$ $\chi^2$ image rather than the [3.6] image (see \citealt{laigle16} for details).
As a consequence, a small number of dusty, and therefore very red, highly star-forming galaxies at high redshift
may be missing from the COSMOS2015 catalog even if they satisfy our adopted limiting magnitude in the [3.6] band.
As such, their contribution may be underestimated here. We note, however, that that deep K-band observations are generally successful in 
detecting highly obscured dusty star forming galaxies at these redshifts (e.g., \citealt{romano20, wenjia20}), as is the COSMOS2015 catalog in general
(e.g., \citealt{gruppioni20}). Regardless, the results presented in this paper do not change meaningfully if we impose that the fractional 
contribution of such galaxies is increased by a factor of, for example, two, a number which roughly corresponds to the approximate 
incompleteness of the COSMOS2015 catalog for sources detected at far-infrared or radio wavelengths (\citealt{gruppioni20}, G. Zamorani, 
\emph{private communication}). Though we have no comparable estimate of the level of incompleteness of quiescent galaxies, as 
only a fraction of this population are bright in far-infrared/radio wavelengths, we find that our results are not meaningfully 
changed if the number of quiescent galaxies is doubled in the exercise described below.

Following the determination of the fractional contribution of underrepresented populations in our VUDS+ sample, a Monte-Carlo simulation was run in 
the following manner. For each of the 1000 iterations of the simulation, $\mathcal{SFR}$ and $\log(1+\delta_{gal})$ values were recalculated for each of the galaxies
in the VUDS+ sample by Gaussian sampling their associated uncertainties. To this, the appropriate number of quiescent and dusty star-forming galaxies
were added, $\sim$50 and $\sim$300, respectively, at random overdensity values taken from the VUDS+ $\log(1+\delta_{gal})$ distribution. Each quiescent galaxy
was assigned an $\mathcal{SFR}=0$. Each dusty star-forming galaxy was assigned a $\mathcal{SFR}$ statistically by sampling from a 
$\log(SFR/\mathcal{M}_{\odot}/\rm{yr})$ distribution with a mean of 3.25 and a $\sigma=0.12$. Such values are appropriate for this galaxy population at these 
redshifts (e.g., \citealt{lem14b, vernesa17}). For each of the 1000 iterations, the Spearman correlation coefficient was calculated as well as the significance 
of the rejection of the null hypothesis of no correlation. A weak ($\tilde{\rho}_{\mathcal{SFR}-\delta_{gal}}=0.08$) but still highly significant 
($\sim$8.5$\sigma$ on median) \emph{positive} correlation between $\mathcal{SFR}$ and $\log(1+\delta_{gal})$ was found over all iterations, with no iterations 
returning a significance of $<$6$\sigma$. 

As mentioned earlier, however, there are some hints 
that such galaxy populations are found to be associated with overdense environments at high redshift (e.g., \citealt{kubo13,taowang16, vernesa17, miller18, strazzullo18, shi19, shi20, ryley20}). 
As such, we repeated this exercise by sampling only from $\log(1+\delta_{gal})$ values from those VUDS+ galaxies found in non-field environments, in other words, either the Intermediate or Peak 
regions as defined earlier in this section. Since dusty star-forming galaxies outnumber quiescent galaxies by a factor of approximately six in this simulation, 
forcing these populations to reside in overdense regions only serves to increase both the strength and significance of the observed positive correlation
between $\mathcal{SFR}$ and $\delta_{gal}$.  As these tests account for the full range of the effect of measurement errors, and, in conjunction with those tests given 
in \S\ref{altmethod}, account for the effect of galaxy populations missing in our data, we conclude the reversal of the $\mathcal{SFR}-\delta_{gal}$ relation 
observed in the VUDS+ sample is a real feature of the galaxy population in the early universe. 

\begin{figure*}
\epsscale{1}
\plottwosuperspecial{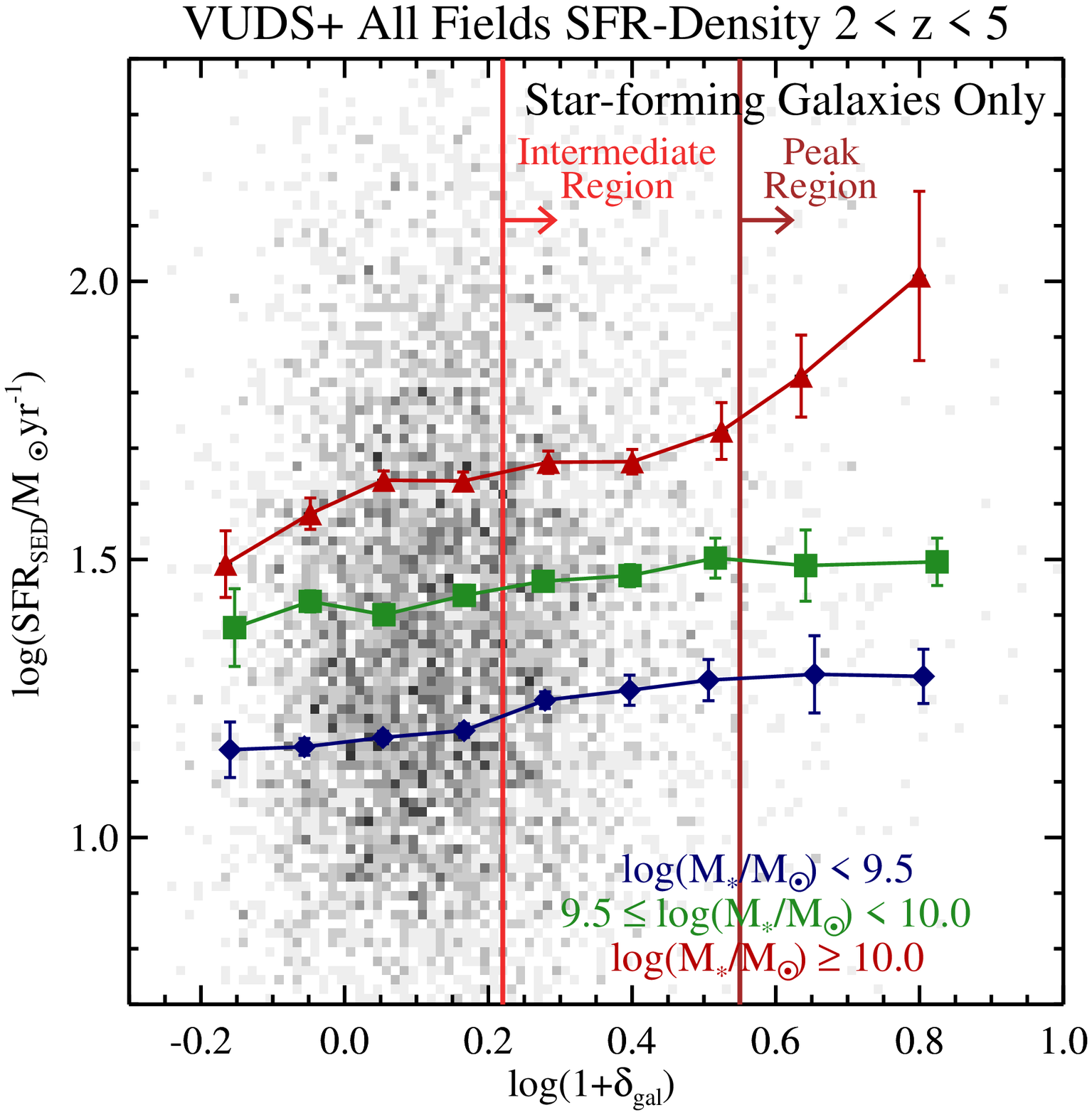}{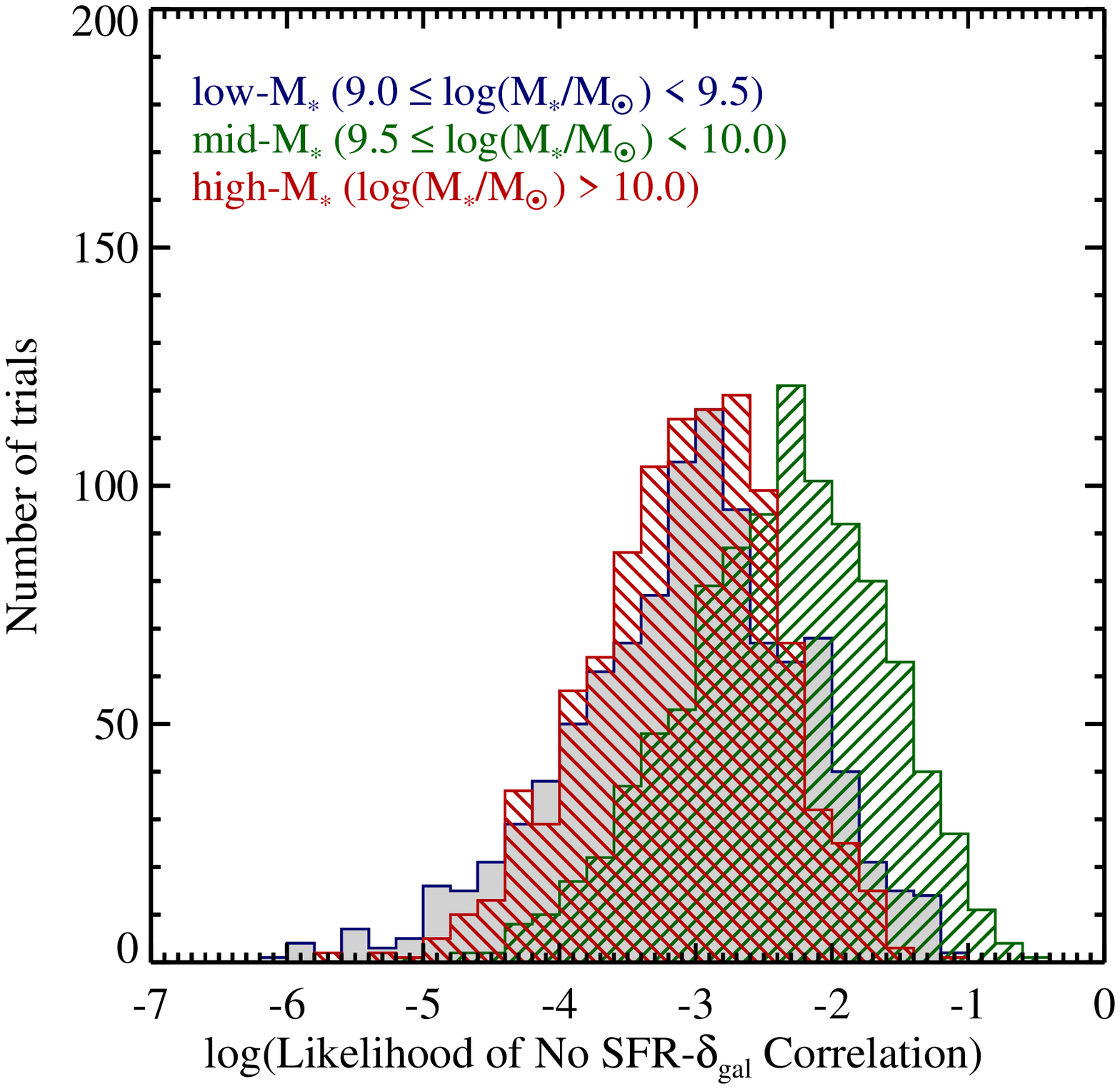}		
\caption{The $\mathcal{SFR}$$-$density relation in the VUDS+ sample as a function of stellar mass. \emph{Left:} As in the left panel of Figure \ref{fig:mainSFRnSMdens}, the 
$\mathcal{SFR}$-$\delta_{gal}$ relation for our final VUDS+ sample, but now with galaxies broken up 
into three approximately equally-sized samples binned in stellar mass using the limits indicated at the bottom right of the panel. The points and their associated uncertainties
have the same meaning as in Figure \ref{fig:mainSFRnSMdens}. The solid vertical lines and their meanings are
the same as in Figure \ref{fig:mainSFRnSMdens}. A weak but significant positive correlation between $\mathcal{SFR}$ and $\log(1+\delta_{gal})$ is seen in the two lower mass bins. 
A stronger and equally significant positive correlation is seen for the highest stellar mass galaxies and is especially apparent in the Peak Region.
\emph{Right:} Distribution of the logarithmic value of the likelihood of no correlation between $\mathcal{SFR}$ and $\log(1+\delta_{gal})$ of our 
final spectral sample for the three different stellar mass bins used in our analysis for the full redshift range $2 \le z \le 5$. This distribution is calculated from 1000 Monte 
Carlo realization in which we perturb each $\mathcal{SFR}$ and $\log(1+\delta_{gal})$ value by their associated errors. These histograms do not meaningfully  
change if we additionally include dusty star-forming galaxies with high rates of $\mathcal{SFR}$ and galaxies with no active star formation, both
of which are completely missed in our sample, at proportions appropriate for their number density at $2 \le z \le 5$ in our Monte Carlo simulations. For reference, 
a log(likelihood)$<$-2.5 corresponds to a positive correlation between the two parameters estimated at $\ge$3$\sigma$ confidence level.} 
\label{fig:SFRdensbymass}
\end{figure*}

\subsection{Stellar mass and redshift effects on the $\mathcal{SFR}-\delta_{gal}$ relation}
\label{SMnzeffects}

Plotted in the right panel of Figure \ref{fig:mainSFRnSMdens} is the equivalent relation between $\mathcal{M}_{\ast}$ and $\log(1+\delta_{gal})$. As in its $\mathcal{SFR}$
equivalent, a steady, nearly monotonic increase in $\mathcal{M}_{\ast}$ and $\log(1+\delta_{gal})$ is observed at about the same level ($\sim$0.3 dex) as that of the 
$\mathcal{SFR}$ over the same dynamic range in local environments. The Spearman's correlation coefficient is essentially equivalent to that calculated for 
$\mathcal{SFR}$, $\rho_{\mathcal{M}_{\ast}-\delta_{gal}}=0.15$, as is the significance of the positive correlation ($\sim$12$\sigma$). Because the star-forming properties of galaxies 
relate in a complex way to stellar mass, as well as redshift and environment (e.g., \citealt{peng10, wuyts11, tizzytasca15, AA16, AA19, lem19}), the level of concordance of 
the $\mathcal{M}_{\ast}-\delta_{gal}$ and $\mathcal{SFR}-\delta_{gal}$ relations seen in the VUDS+ sample suggests that stellar mass effects may be primarily responsible for driving the observed 
$\mathcal{SFR}-\delta_{gal}$ relation. In other words, higher stellar mass star-forming galaxies are preferentially situated in high-density environments in the VUDS+ 
sample. That these higher stellar mass star-forming galaxies, whether through an increased gas supply or a higher star-forming efficiency 
(e.g., \citealp{schinnerer16, smokynicky16, smokynicky17, taowang18}), have, on average, higher $\mathcal{SFR}$ values (e.g., \citealp{rodighiero14,tizzytasca15,AA16}), 
could explain the observed $\mathcal{SFR}-\delta_{gal}$ relation. Such an explanation is still wanting in
that the increased presence of higher $\mathcal{M}_{\ast}$ in higher-density environments has to be accounted for by some process acting at earlier times, but it at
least could potentially explain the increase in $\mathcal{SFR}$ with increasing $\log(1+\delta_{gal})$.

\begin{figure*}
\epsscale{1}
\plottwo{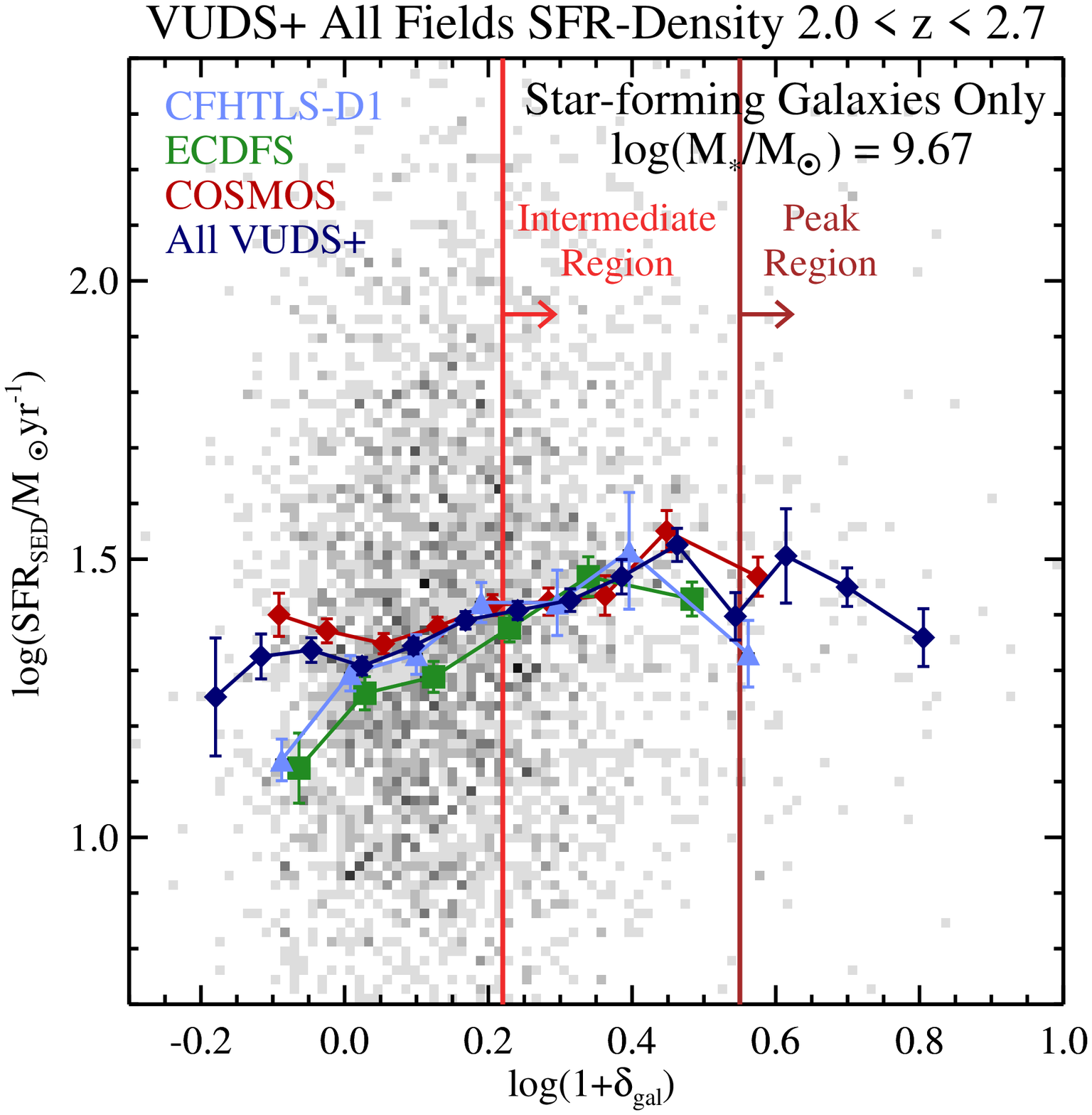}{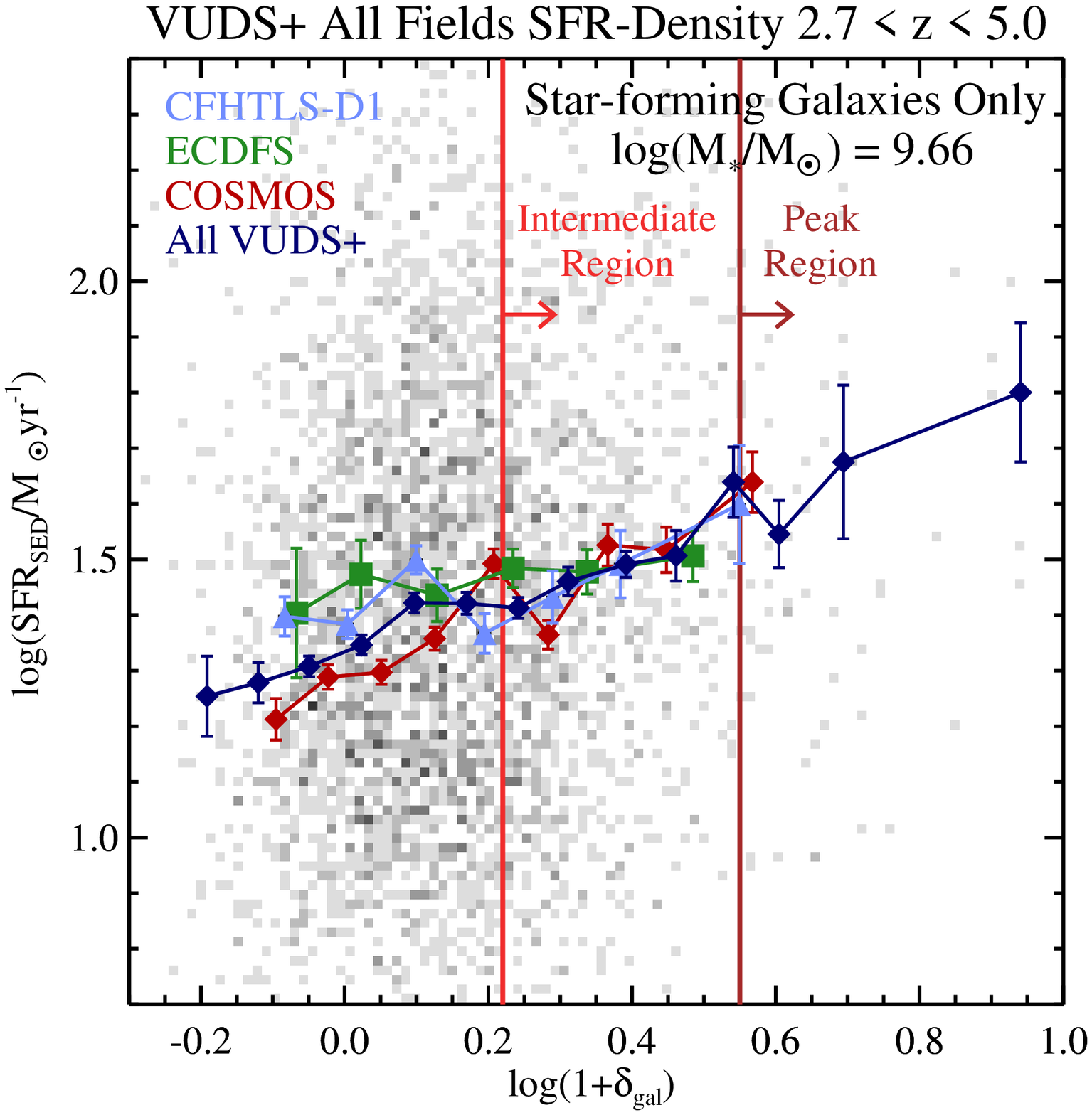}
\caption{The redshift dependence of the $\mathcal{SFR}-\delta_{gal}$ relation in the VUDS+ sample. The two panels are identical to the left panel of Figure \ref{fig:mainSFRnSMdens} except 
now the VUDS+ sample is broken up into two redshift bins with approximately equal numbers of 
galaxies. The meaning of the points and their associated uncertainties are identical to those in Figure \ref{fig:mainSFRnSMdens} as are the meanings of the vertical lines. 
As in Figure \ref{fig:mainSFRnSMdens}, the $\mathcal{SFR}-\delta_{gal}$ for all three VUDS+ fields are shown along with the combined sample. The \emph{left} panel shows 
the $\mathcal{SFR}-\delta_{gal}$ relation for the lower-redshift sample ($2.0 \le z \le 2.7$) and the \emph{right} panel shows the relation for the higher-redshift sample 
($2.7 < z \le 5.0$). The average stellar mass of each sample is given in the top-right portion of each panel and is effectively identical between the two samples. 
While a significant positive correlation between $\mathcal{SFR}-\delta_{gal}$ exists in the lower-redshift sample, this correlation disappears when considering
only galaxies in the Intermediate and Peak regions. In contrast, in the higher-redshift bin, the significant positive correlation is seen at all environments and increases in 
strength in the densest regions.}
\label{fig:SFRdensbyz}
\end{figure*}

\begin{figure}
\epsscale{1}
\plotone{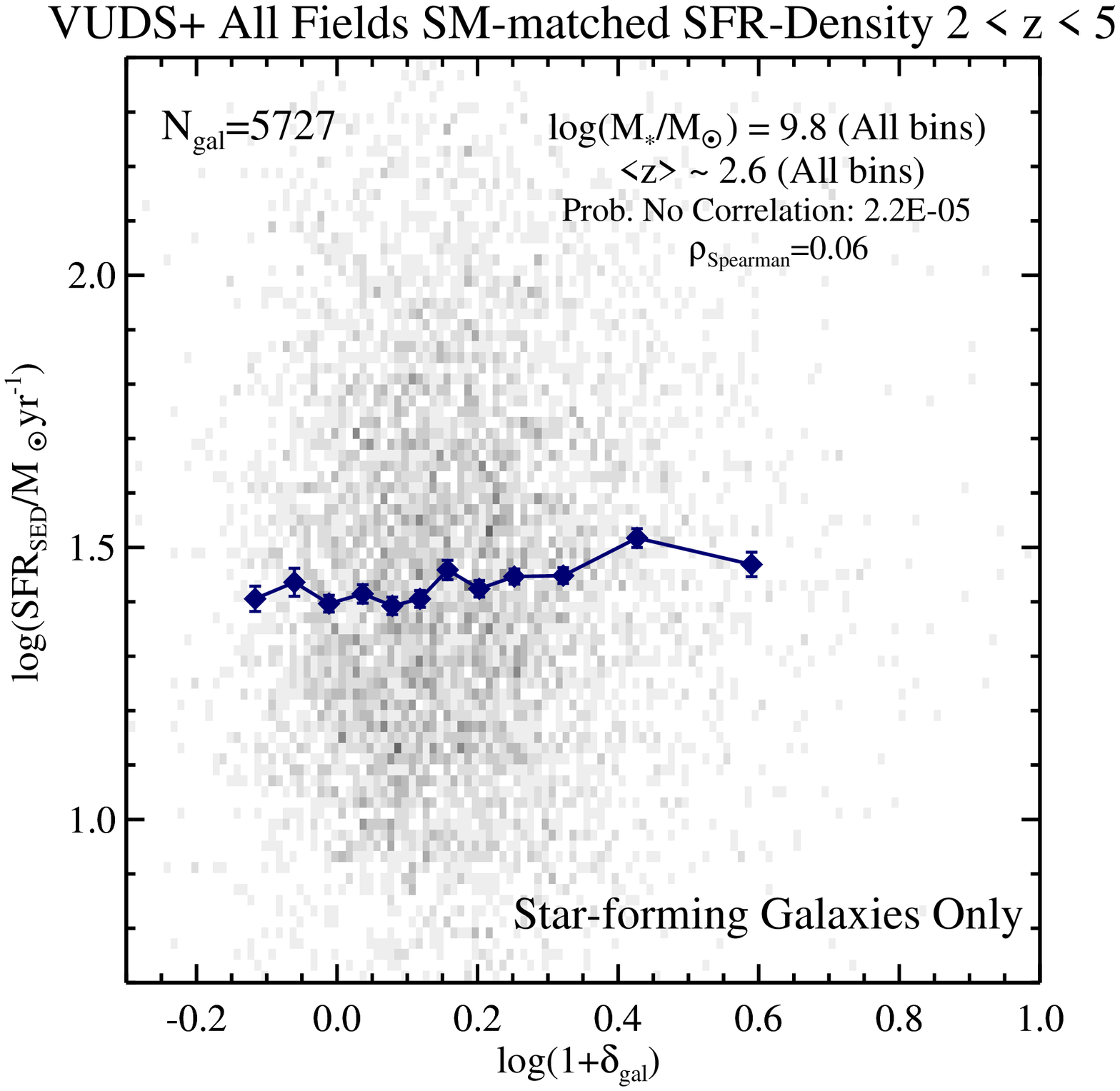}
\caption{Relationship between $\mathcal{SFR}$ and $\log(1+\delta_{gal})$ for a stellar-mass- and redshift-matched sample of galaxies. Each bin in $\log(1+\delta_{gal})$ 
originally contained 10\% of the final VUDS+ sample. Galaxies in each decile were clipped in order to have the same median redshift and stellar mass within the uncertainties with 
every other decile. The bins in the lower and upper decile of the $\log(1+\delta_{gal})$ distribution were cut in half for illustrative purposes.
The approximate median stellar mass and redshift of all bins is indicated in the top right and the number of galaxies remaining in the sample indicated in the top
left. The data points and their associated uncertainties were calculated in the same way as Figure \ref{fig:mainSFRnSMdens}. In the top right of the plot the Spearman's
correlation coefficient for this sample and the likelihood of no correlation are shown. Despite normalizing by stellar mass and redshift, a weak but significant positive
$\mathcal{SFR}-\delta_{gal}$ trend is observed.}
\label{fig:SFRdensbymassnz}
\end{figure}

In the left panel of Figure \ref{fig:SFRdensbymass} is plotted $\mathcal{SFR}$ versus $\log(1+\delta_{gal})$ in three different bins of stellar mass defined as low, mid,
and high $\mathcal{M}_{\ast}$ by the stellar mass ranges indicated in the bottom right of that plot. In all three cases
a significant positive correlation, as quantified by a Spearman's test, is observed, with $\rho_{\mathcal{SFR}-\delta_{gal}}$ ranging from 0.08-0.09 depending on the bin and a 
$>3.5\sigma$ rejection of the null hypothesis in each bin. This trend is relatively weak for the two lowest stellar mass bins, while slightly stronger for the galaxies in our 
sample that are the most massive in their stellar content. We note that these statements hold if we instead restrict the highest stellar mass bin to the stellar mass range 
$10 \le \log(\mathcal{M}_{\ast}/\mathcal{M}_{\odot})\le 10.5$ to remove any residual stellar mass-driven trends due to the wideness of that bin. As was done in 
\S\ref{generalSFRdens}, the reliability of these trends
were tested by running a Monte-Carlo simulation for each of the three samples. Because the stellar mass distribution of quiescent and dusty star-forming galaxies is 
not known, in this simulation we ignore their fractional contribution and only vary the $\mathcal{SFR}$ and $\log(1+\delta_{gal})$ of observed VUDS+ galaxies based on 
their formal uncertainties. However, this analysis is not meaningfully changed if we include their contribution equally across all $\mathcal{M}_{\ast}$ bins, or, since 
both galaxy populations are almost certainly some of the most massive at these redshifts (see, e.g., \citealt{connivingcasey14, AA14} and references therein), exclusively limit them 
to the higher-$\mathcal{M}_{\ast}$ bins. Plotted in the right panel of Figure \ref{fig:SFRdensbymass} is the distribution of the probability of the null hypothesis that $\mathcal{SFR}$ and 
$\log(1+\delta_{gal})$ are uncorrelated being correct for each of the three stellar mass samples across the 1000 Monte-Carlo iterations. For the vast majority 
of the iterations ($\sim80-90$\%) for the low- and high-$\mathcal{M}_{\ast}$ bins, a significant ($>3\sigma$, $\log(\rm{Likelihood})<-2.5$) positive correlation 
between $\mathcal{SFR}$ and $\log(1+\delta_{gal})$ is observed. For the mid-$\mathcal{M}_{\ast}$ bin, the trend is weaker, though $\sim$50\% of the iterations
still return a significantly positive $\mathcal{SFR}-\delta_{gal}$ correlation. It appears that even at fixed stellar mass, environmental effects serve to drive an 
increase in star formation in higher-density environments. Additionally, whatever this effect, it appears to be most effective on high-$\mathcal{M}_{\ast}$ galaxies,
as such galaxies increase their $\mathcal{SFR}$ by more than a factor of three from the lowest to the highest densities.

In order to investigate the possible redshift evolution of the $\mathcal{SFR}-\delta_{gal}$ relation, in the two panels of Figure \ref{fig:SFRdensbyz} we show the same 
plot as was shown in the left panel of Figure \ref{fig:mainSFRnSMdens} but now in two redshift bins. The lower-$z$ bin is defined as $2 \le z \le 2.7$ and the higher-$z$ bin as 
$2.7 < z \le 5$. These redshift ranges cut the sample approximately in half, allowing each bin to have considerable statistical power. Immediate differences can be seen between the 
two samples. In the lower-$z$ bin, while the median $\mathcal{SFR}$ is seen to gently increase with increasing $\log(1+\delta_{gal})$ in the Field and Intermediate regions, a turnover is 
observed in the Peak region that
is strong enough to decrease the $\mathcal{SFR}$ in the highest-$\log(1+\delta_{gal})$ bin to a value that is statistically indistinguishable (i.e., at a difference of $<1\sigma$) 
with that of the lowest-$\log(1+\delta_{gal})$ 
bin. Conversely, the increase in $\mathcal{SFR}$ with $\log(1+\delta_{gal})$ persists through the Peak region in the higher-$z$ bin. While the correlation coefficient is not appreciably different 
between the two redshift bins ($\rho_{\mathcal{SFR}-\delta_{gal}}$=0.13 and 0.14 in the lower- and higher-$z$ bins, respectively), the differences are significant when considering 
only the Intermediate and Peak regions. For galaxies with $\log(1+\delta_{gal})>0.22$,
no significant $\mathcal{SFR}-\delta_{gal}$ relation exists at $2.0 \le  z \le 2.7$ ($\rho_{\mathcal{SFR}-\delta_{gal}}=0.04$ with a rejection of the null hypothesis of no correlation
at $<2\sigma$). While a flat relation is still considerably different from the negative correlation observed at 
lower $z$ (i.e., $z<2$), galaxies in the higher-$z$ bin exhibit a dramatic and significant increase in their $\mathcal{SFR}$ with increasing $\log(1+\delta_{gal})$ in the Intermediate 
and Peak regions ($\rho_{\mathcal{SFR}-\delta_{gal}}=0.17$, $\sim$5$\sigma$). These results combined with the previous results binned by stellar mass imply that it is 
primarily higher stellar mass, higher-redshift 
galaxies that are driving the global reversal of the $\mathcal{SFR}-\delta_{gal}$ at these redshifts. 

Some caution in extrapolating these results is necessary, however. 
While galaxies in the Intermediate region of the higher-$z$ bin are drawn from a variety of different structures, a considerable fraction of the galaxies in the Peak region are 
members of PCl J1001+0220 at $z\sim4.57$ in the COSMOS field. Even so, this trend does persist at a similar significance and strength even after excising the members of PCl J1001+0220.
Additionally, it is clear from the right panel of Figure \ref{fig:SFRdensbyz} that this trend is also present in the CFHTLS-D1 and ECDFS fields, though at a less significant level 
($\sim$2-3$\sigma$ per field, $>$3$\sigma$ for the two fields combined). However, the number of galaxies in the 
peak region across all fields after excising the PCl J1001+0220 members is sufficiently small ($\sim$60 galaxies) to make sample variance a concern. Further data, both in PCl J1001+0220, and in 
structures at similar redshifts across the VUDS fields, will be necessary to see if such trends are general or isolated to a particular system or small set of proto-structures. 

Finally, we perform two tests in an attempt to further separate out the effects of stellar mass and redshift from those of environment in driving the $\mathcal{SFR}-\delta_{gal}$ 
relation. The first of these tests is a Spearman's partial rank correlation test, which is a formalism designed to estimate the strength and significance of the correlation 
between two variables after removing the contribution of other variables to the observed correlation. In this case, we are interested in computing the correlation between 
$\mathcal{SFR}$ and $\log(1+\delta_{gal})$ after removing the effects of $\mathcal{M}_{\ast}$ and $z$. The calculation that follows adopts the Spearman's partial rank 
correlation methodology described in \citet{macklin82}, in which it was shown that this test is efficient for the purpose of separating out the effect of different variables in
inducing observed correlations. The Spearman's partial 
rank coefficient, $\rho_{\rm{p}}$, for $\mathcal{SFR}$ and $\log(1+\delta_{gal})$ was calculated to be $\rho_{\rm{p},SFR-\delta_{gal}}=0.07$ with a significance of $\sim6\sigma$.
In other words, while the strength of the $\mathcal{SFR}$-$\delta_{gal}$ correlation has depreciated by approximately 50\% after taking into account the correlation between both 
$\mathcal{M}_{\ast}$ and $z$ and $\mathcal{SFR}$, though mostly the former, the correlation between $\mathcal{SFR}$-$\delta_{gal}$ still remains extremely significant. Incidentally,
the same is true if the sample is confined to only those galaxies in the COSMOS field, though at a significance of $\sim$4$\sigma$. 

The result of the second test is shown in 
Figure \ref{fig:SFRdensbymassnz}. In this test, galaxies are divided into ten equally-populated bins\footnote{The bins representing the lower and upper 10\% of 
the $\log(1+\delta_{gal})$ distribution were cut in half for illustrative purposes. This choice does not affect our analysis, which is performed on the individual galaxies in the 
clipped sample.} in $\log(1+\delta_{gal})$ and forced to have the same median stellar
mass and redshift across all bins to within the uncertainties of the median values in each bin. For each bin, galaxies at the extreme low end of the stellar mass distribution,		
but still always with $\log(\mathcal{M}_{\ast}/\mathcal{M}_{\odot})>9$, and the extreme high end of the redshift distribution were removed until the median values in
each bin converged, within their associated uncertainties, to the values shown in the top right of Figure \ref{fig:SFRdensbymassnz}. The median value of 
the $\mathcal{SFR}$ for each $\log(1+\delta_{gal})$ is overplotted against the backdrop of the individual measurements of the $\sim$5700 galaxies that remain in the 
$\mathcal{M}_{\ast}/z$-matched sample. Also shown in the top right of Figure \ref{fig:SFRdensbymassnz} is $\rho_{\mathcal{SFR}-\delta_{gal}}$ for the 
$\mathcal{M}_{\ast}/z$-matched sample and the probability of the null hypothesis of no correlation between $\mathcal{SFR}$ and $\log(1+\delta_{gal})$. It appears
that regardless of how the data are analyzed, there is a weak but highly significant positive correlation between $\mathcal{SFR}-\delta_{gal}$ and that at least some part 
of that trend is driven by environmental effects.  

\subsection{Effects of lack of spectral representativeness on the $\mathcal{SFR}-\delta_{gal}$ relation} 
\label{altmethod}

\begin{figure*}
\plottwo{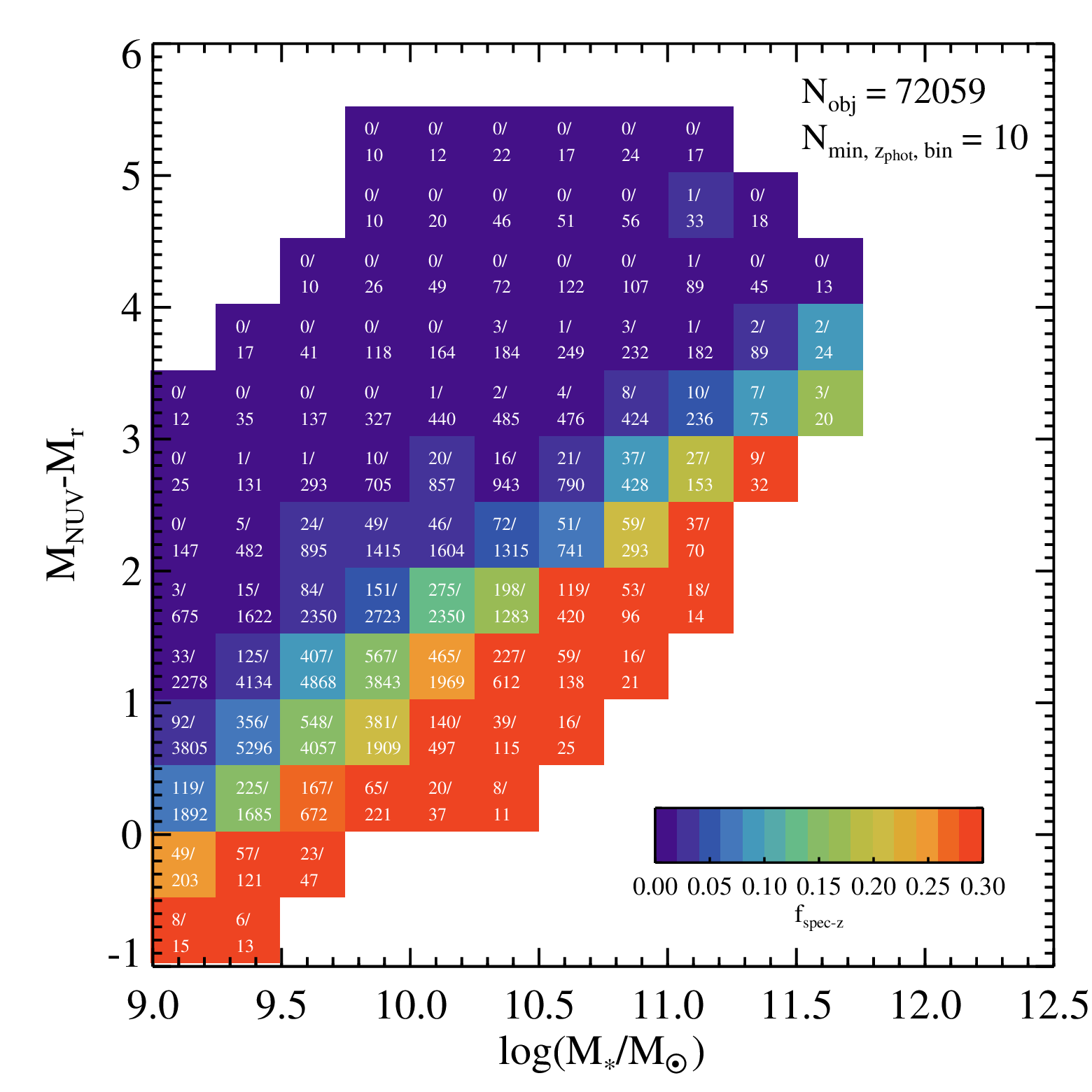}{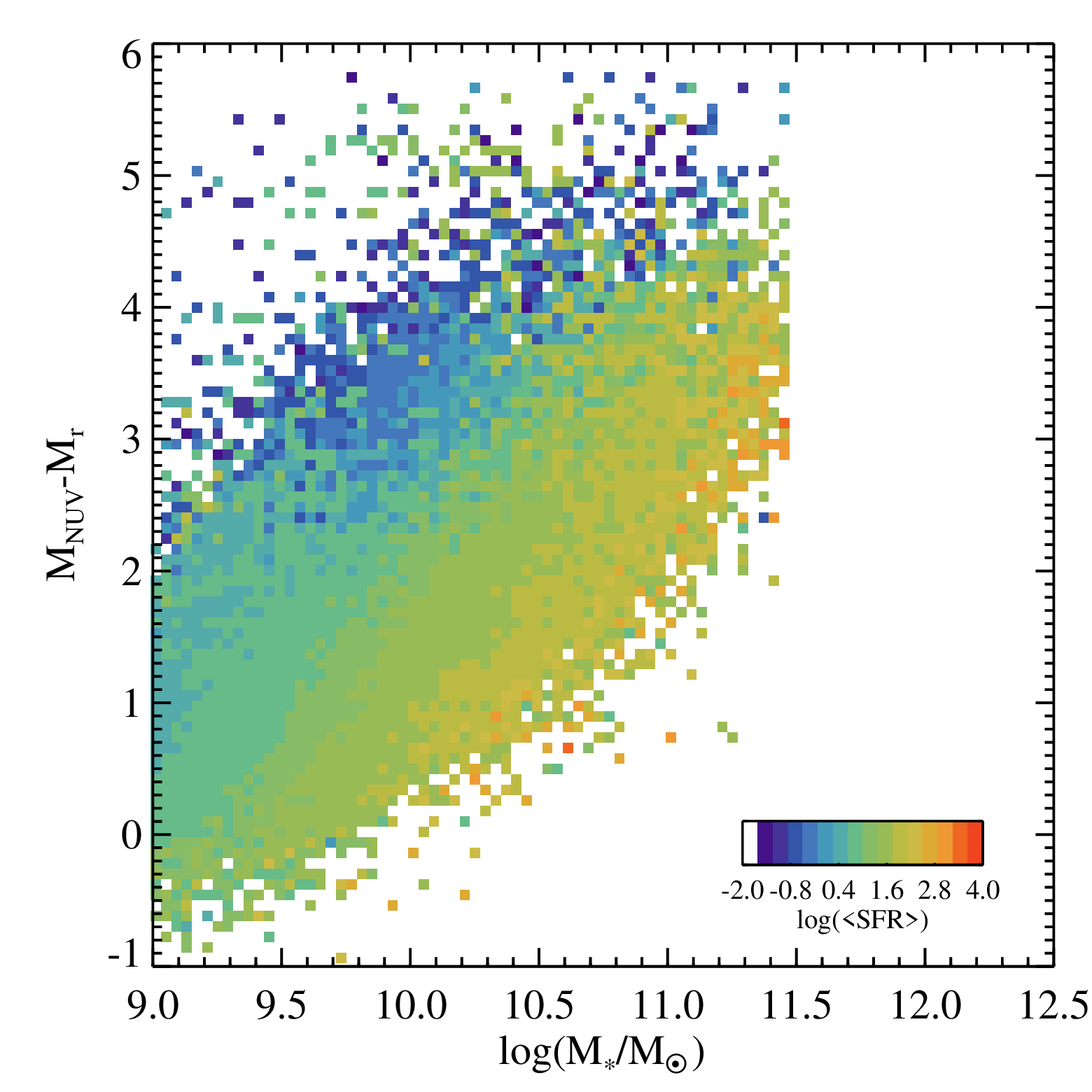}
\caption{Color, stellar mass, and $\mathcal{SFR}$ properties of the VUDS+ spectroscopic and parent photometric sample. \emph{Left:} Rest-frame color-stellar mass 
diagram of an observed-frame NIR-selected sample of $z_{phot}$ and $z_{spec}$ objects in across the three VUDS fields. The
fraction of $z_{phot}$ objects with a secure spectral redshift in each color/stellar mass bin is indicated by the color bar. In each color/stellar mass bin the number of
$z_{spec}$ and $z_{phot}$ objects are given by the upper and lower number in each box, respectively. Only those bins that contained ten or more $z_{phot}$ objects are plotted.
This plot is nearly identical to the right panel of Figure \ref{fig:representativeness},
but here the bins are coarser to reflect the size used to perform the weighted Monte Carlo $\mathcal{SFR}$-$\delta_{gal}$ analysis presented in this section.
\emph{Right:} Color-stellar mass diagram of the $z_{phot}$ sample plotted in the left panel with the color bar
now corresponding to the average SED-fit $\mathcal{SFR}$ of all objects in that bin. No color or stellar mass limits are imposed on the galaxies plotted here, and two-dimensional
bins are now plotted if they contain at least one object. The average $\mathcal{SFR}$ is seen to vary smoothly as a function of color and stellar mass over the range plotted.
While the clear preference of the VUDS+ spectral sampling toward bluer galaxies at fixed $\mathcal{M}_{\ast}$ appears to potentially bias the spectral sample to higher $\mathcal{SFR}$s,
from the analysis presented in this section we conclude that this potential bias does not drive the significant positive correlation between $\mathcal{SFR}$ and $\delta_{gal}$
seen in the full VUDS+ sample.}
\label{fig:weightedMC}
\end{figure*}

In \S\ref{represent} we discussed the lack of spectral representativeness of the VUDS+ sample. In particular, we noted that, at a given $\mathcal{M}_{\ast}$,
objects in the parent photometric sample at bluer colors in the range $2<z_{phot}<5$ generally appeared to have a higher percentage of spectroscopic
redshifts. This trend is shown again
in the left panel of Figure \ref{fig:weightedMC}, where we now plot the spectral fraction of objects in the NIR-selected photometric parent sample in coarser bins
of 0.5 and 0.25 in $M_{NUV}-M_{r}$ color and $\mathcal{M}_{\ast}$, respectively. In each two-dimensional bin, the number of galaxies with spectral redshifts within the
redshift range considered in this paper ($2 \le z \le 5$) are given as the top number. The number of objects that have $z_{phot}$ values within the redshift range considered
here that either do not have a spectral redshift or have a spectral redshift within our adopted redshift range is given as the bottom number in each bin. The color bar
indicates the $z_{spec}$ fraction in each bin.

It is again apparent from inspection of the left panel of Figure \ref{fig:weightedMC} that galaxies in the spectral sample preferentially have a secure spectral redshift
at the bluest colors at a given stellar mass, as $z_{spec}$ fractions appear at $\sim$25-75\% for the bluest colors and $\la$5\% for the reddest colors across nearly
the entire stellar mass range plotted. As mentioned in \S\ref{represent}, galaxies on the red extreme of the color distribution at a given stellar mass are subdominant, both in 
the $z_{phot}$ sample and, in terms of absolute numbers, in the spectral sample (see Figure \ref{fig:representativeness}). Therefore, the main channel for 
them to disrupt a general analysis
is if the average $\mathcal{SFR}$s are considerably higher than their counterparts at bluer colors\footnote{As a reminder, we are considering here only
star-forming galaxies with low to moderate dust contents. Extremely dusty star-forming galaxies and sub-millimeter populations are considered in \S\ref{generalSFRdens}.}. 
If this were the case, it would perhaps make our results more sensitive to small sample statistics and the sampling of such galaxies in each environment.

In the right panel of Figure \ref{fig:weightedMC} the average SED-derived $\mathcal{SFR}$ is shown in two-dimensional bins of color and stellar mass for the same observed-frame
NIR-limited sample. It
is immediately clear in this figure that objects appearing at redder colors generally show depreciated levels of star formation at fixed stellar mass relative to their bluer counterparts.
Further, the decrease in $\mathcal{SFR}$ is a slowly varying, smooth function of color and stellar mass for the populations plotted here. Because of the lack of stark change
in the average $\mathcal{SFR}$ for the bulk of this sample, the relative scarcity of redder galaxies at a given stellar mass, and the fact that we are preferentially sampling galaxies
at the high end of the $\mathcal{SFR}$ distribution at a given stellar mass (with the notable exception of extremely dusty starbursts), it is unlikely that the bias seen in the left panel of
Figure \ref{fig:weightedMC} meaningfully affects the main conclusions presented in this paper.

Despite the high likelihood of our results being invariant to this sampling effect, we formally tested the possible effect of this selection bias on our results in the 
following manner. We begun by looking at the spectral representativeness in each bin of color and stellar mass plotted in the left panel of Figure \ref{fig:weightedMC} by means of 
one-dimensional KS tests. For each bin in color and stellar mass, a KS test was run on the distribution of
the $z_{phot}$ objects and $z_{spec}$ galaxies in that bin. For            
nearly all (80\%) of the bins in stellar mass and color running $-0.5 \le M_{NUV}-M_r \le 4$ and $9 \le \log(\mathcal{M}_{\ast}/\mathcal{M}_{\odot}) \le 11.5$ for bins
that had sufficient numbers of both $z_{spec}$ and $z_{phot}$ objects, $N>3$ and $N>20$, respectively, the KS test did not return a significant rejection of the null hypothesis
($>3\sigma$) that the two distributions were drawn from the same sample.
We conclude from this exercise that the final spectral sample is broadly representative of the non-dusty star-forming galaxy population at these redshifts in each
individual color-$\mathcal{M}_{\ast}$ bin. This representativeness allows us to leverage the spectral sample in each individual color-$\mathcal{M}_{\ast}$ bin to proxy
as the underlying photometric sample.

Following this, we attempted to homogenize the $z_{spec}$ fraction using a Monte Carlo approach. We begun by identifying all color-$\mathcal{M}_{\ast}$ bins
in the left panel of Figure \ref{fig:weightedMC} that we consider well-sampled spectroscopically. For the purposes of this exercise, we defined this term as any bin whose
$z_{spec}$ fraction is $\ge3$\% and the number of galaxies with a secure $z_{spec}$ exceeds three. These cuts included the majority (70.0\%) of all photometric objects
within this redshift range and the results of this exercise do not meaningfully change if we slightly adjusted these limits. We then set a fiducial $z_{spec}$ fraction
that we will homogenize each color-$\mathcal{M}_{\ast}$ bin to for all Monte Carlo iterations. If the $z_{spec}$ fraction exceeded this fiducial value for a given bin, for
each Monte Carlo realization, we randomly sampled the $\mathcal{SFR}$ and $\log(1+\delta_{gal})$ values from the available $z_{spec}$ galaxies in that bin only enough
times to reduce the $z_{spec}$ fraction to the fiducial value. If the $z_{spec}$ fraction was below this fiducial value for a given bin, in addition to keeping all $z_{spec}$
galaxies in that bin, we again randomly sampled all $z_{spec}$ galaxies in this bin a sufficient number of times to raise the $z_{spec}$ fraction to the fiducial value.
At the end of this process, for each bin that is well-sampled spectroscopically, the $z_{spec}$ fraction is homogenized to the fiducial value and a unique set of
$\mathcal{SFR}$ and $\log(1+\delta_{gal})$ values were generated for that iteration. A fiducial $z_{spec}$ fraction of 0.2 was chosen here as an intermediate value
to all those shown in the left panel of Figure \ref{fig:weightedMC}, though the results of this exercise do not change meaningfully if this fraction is slightly varied.

At the end of each Monte Carlo iteration, we calculated the Spearman $\rho$ and significance of the rejection of the null hypothesis that the two parameters
are uncorrelated. Over 1000 Monte Carlo iterations we found a rejection of the null hypothesis at $>3\sigma$ for all iterations, with a median significance of 6.9$\sigma$
and a median $\rho_{\mathcal{SFR}-\delta_{gal}}=0.07$. While the strength of the correlation is slightly lower than that of the full spectral sample (see \S\ref{generalSFRdens}), this is 
expected given that the random sampling of the $\mathcal{SFR}$ and $\log(1+\delta_{gal})$ values in the Monte Carlo process necessarily dilutes the true signal when
averaging over a large number of realizations. Though we did not strictly impose the stellar mass and color limits used in this exercise (i.e.,
$-0.5 \le M_{NUV}-M_r \le 4$ and $9 \le \log(\mathcal{M}_{\ast}/\mathcal{M}_{\odot}) \le 11.5$) to define the final VUDS+ spectral
sample, these galaxies make a small contribution ($<10$\%) to the overall sample. Additionally, the main conclusions of this study do not change meaningfully if we instead
limit the final VUDS+ sample to the $>$6000 galaxies with $\log(\mathcal{M}_{\ast}/\mathcal{M}_{\odot})\ge9$ and $M_{NUV}-M_r\ge-1$.

\subsection{Comparison of the observed $\mathcal{SFR}-\delta_{gal}$ relation to other works}
\label{compare}

While a rigorous comparison of the observed VUDS+ $\mathcal{SFR}-\delta_{gal}$ relations derived in \S\ref{generalSFRdens} and \S\ref{SMnzeffects} and those predicted in simulations is deferred to future work, we take here a few points of comparison.
The first point of comparison relies on \citet{elbaz07}, who used light cones generated by \citet{kdubz07} to estimate the $\mathcal{SFR}-\delta_{gal}$ predicted at different redshifts
from the Millennium simulation \citep{springel05} as interpreted through the lens of the semi-analytic model of \citet{croton06}. This prediction showed an increase of $\sim$0.25 dex
at both $z\sim2$ and $z\sim3$ from the most rarefied to densest environments measured in the simulation\footnote{These light cones use the full Millennium simulation, which contains many 
structures massive enough to be considered as proto-groups or proto-clusters (see, e.g., \citealt{coolchiang17,arayaaraya21}), and thus span a similar dynamic range in environments 
as the VUDS+ sample. The same is true for the IllustrisTNG simulations discussed later in this section (see, e.g., \citealt{bose19}).}. While the $\mathcal{SFR}$s as measured in the 
VUDS+ sample are offset higher from these predictions by $\sim0.1-0.2$ dex across all environments, likely due to some combination of differing redshift ranges, 
ranges of $\mathcal{SFR}$s and $\mathcal{M}_{\ast}$ probed, and
differences in the assumed properties of the models used to infer physical properties, the magnitude of the increase in the average $\mathcal{SFR}$ across the full density range
measured in the VUDS+ sample is essentially identical to that predicted in the Millennium simulation. Such behavior was also predicted in the $N-$body simulations of \citet{delucia04}. 
Using semi-analytic prescriptions to estimate $\mathcal{SFR}$s of cluster and field galaxies, \citet{delucia04} predicted the reversal of the $\mathcal{SFR}-\delta_{gal}$ relation
at $z\sim2.5$, with the normalized average rate of star production of cluster galaxy outpacing field galaxies by $\sim$0.2 dex in the redshift range corresponding to the higher-redshift portion 
of our sample ($2.7 \le z \le 5$). More recently, similar behavior was also seen in \citet{hwang19}, where the predictions of the $\mathcal{SFR}-\delta_{gal}$ from the TNG300 run of 
the IllustrisTNG project \citep{nelson19} were investigated. While this investigation is only performed over the redshift range $0\le z \le 2$,
a range that just barely intersects with that of the VUDS+ sample, at $z=2$, a $\sim$0.3 dex increase in the average $\mathcal{SFR}$ is seen from the lowest to highest-density
environments for a sample of galaxies with comparable $\mathcal{M}_{\ast}$ values to those in the VUDS+ sample. Again, while the VUDS+ galaxies appear to be biased to slightly
higher $\mathcal{SFR}$s across all redshifts, the magnitude of the increase in the average $\mathcal{SFR}$ across the full dynamic range of environments in the VUDS+ sample across the 
full redshift range is essentially identical to that observed in the TNG300 simulations. However, when considering only the lower-redshift ($2.0 \le z \le 2.7$) VUDS+ sample, the 
sample that is most relevant for this comparison, a more modest increase is seen in our data of $\sim$0.1 dex. Such an increase is in moderate tension with the results presented in 
\citet{hwang19}, though the disparate environmental metrics in the VUDS+ and \citet{hwang19} samples preclude a direct comparison. Further work with these simulations using a common
environmental metric and spanning the full VUDS+ redshift range would help to illuminate if any real tension exists. 

Our results are also qualitatively consistent with predictions of updated semi-analytic models applied to the Millennium simulation
\citep{coolchiang17}, in which the volume-averaged star-formation activity within proto-cluster environments, in other words, those regions corresponding to our Intermediate and Peak regions,
is predicted to well outpace that of the field at $z>2$. Other simulations that suggest a strong reversal at later epochs ($z\sim1$, \citealt{tonneson14}) are qualitatively inconsistent
with our results, especially the flattening in the $\mathcal{SFR}-\delta_{gal}$ observed in the lower-redshift portion of our data (see \S\ref{SMnzeffects}). Such a reversal
globally persisting to $z\sim1$ is also ruled out by the observed anticorrelation between $\mathcal{SFR}-\delta_{gal}$ and the fractional increase of quiescent galaxies in dense
environments seen in large surveys of $z\sim1$ clusters and groups (e.g., \citealt{lem19, AA19}). Future work involving more comprehensive comparisons to these and other cosmological
simulations will be extremely helpful to inform the mechanisms that drive this reversal at $z\ga2$.

From an observational perspective, of note is the stark contrast of the observed $\mathcal{SFR}-\delta_{gal}$ relation in the VUDS+ sample to that observed in \citet{chartab20} the five CANDELS
fields\footnote{Two of the CANDELS fields, GOODS-S and COSMOS, are covered in the VUDS+ sample}. In that study, a clear negative correlation was observed between
$\mathcal{SFR}$ and $\delta_{gal}$ across the entire redshift range
running from $0.4\le z < 3.5$ over a similar range of local overdensities as was probed in our study. Though nearly identical methodologies were used to estimate $\mathcal{SFR}$s
in the two studies, the differences in the size of the sample, the metric of environment, the spectroscopic selection method, and the inclusion of photometrically selected
galaxies in their sample make the genesis for the different $\mathcal{SFR}-\delta_{gal}$ behavior difficult to isolate. It is possible that cosmic variance effects 
could explain at least some of the observed difference,
as the combined areal size of the CANDELS fields is approximately an order of magnitude smaller than area probed in our study. In order to test the likelihood 
that cosmic variance is the primary driver of the difference in behavior observed in the two samples, we broke up the VUDS+ sample into 100 different nonunique subsamples 
that probe the same volume as the sample in \citet{chartab20}. In all 100 samples we observed a significant ($>$3$\sigma$) positive correlation between $\mathcal{SFR}$ and 
$\delta_{gal}$, with an average $\rho_{\mathcal{SFR}-\delta_{gal}}$ of 0.15. Thus, it is likely that cosmic variance is not the primary source of the observed tension. The 
consideration that the strength of the observed $\mathcal{SFR}-\delta_{gal}$ correlation is strongest in the higher-redshift portion of our sample, a redshift range largely 
unprobed by  \citet{chartab20}, does serve to somewhat alleviate this tension.
 
\subsection{Interpreting the reversal of the $\mathcal{SFR}-\delta_{gal}$ relation}

We have shown in the previous sections that the reversal of the $\mathcal{SFR}-\delta_{gal}$ in the redshift range $2 < z < 5$ for VUDS+ galaxies is 
real and that these results are very likely applicable to the general population of $\log(\mathcal{M}_{\ast}/\mathcal{M}_{\odot})>9$ galaxies at
these redshifts. Despite the complicated relationship demonstrated between the average $\mathcal{SFR}$ of VUDS+ galaxies and stellar mass, redshift, and 
galaxy environment, we found increased star-formation activity in dense environments even after stellar mass and redshift effects were taken into account. 
Such definitive results were possible only by the immense spectroscopic data set compiled here as well as the supporting ancillary data
and methodologies that allowed for accurate and precise estimates of each of these parameters. Given the weakness of the trends observed in
the data and their large dispersion, a smaller sample size of $\sim$1000 galaxies with spectroscopic redshifts, if sampled representatively from the galaxy sample observed
here, would not yield sufficiently confident rejections ($>3\sigma$) of the null hypothesis of lack of correlation for many of the interdependences
seen here.

In this section we attempt to leverage these interdependences to speculate on the possible mechanisms that drive the 
measured decrease in the positive correlation between $\mathcal{SFR}-\delta_{gal}$ from the higher- to the lower-redshift bin. Several mechanisms present themselves 
as possible explanations for the observed behavior. For this discussion we will assume that the overdense structures probed and their constituent galaxies in the higher-redshift 
bins are possible progenitors of those observed in the lower-redshift bin. This may or may not be a valid assumption and will eventually need to be verified by the use of 
simulations (e.g., \citealp{superB14}) and semiempirical models (e.g., \citealp{AA17}). Given the large number of galaxies in the VUDS+ sample and 
the large number of structures probed, it is likely there is at least some level of truth to this assumption. Additionally, we again note that a considerable 
fraction ($\sim$50\%) of the high-$\delta_{gal}$, high-$\mathcal{M}_{\ast}$ galaxies in the higher-redshift bin belong to the $z\sim4.57$ PCl J1001+0220 proto-cluster,
and, thus, sample variance effects may be considerable. Future observations from the C3VO survey with Keck/DEIMOS and Keck/MOSFIRE of the most massive proto-clusters 
detected in VUDS across the full redshift range studied here will be crucial to determining the level of this effect. Additional future wide-field observations from instruments such 
as the Subaru Prime Focus Spectrograph (PFS; \citealt{sugai12}), the Multi Object Optical and Near-infrared Spectrograph (MOONS; \citealt{cirasuolo14}) on the VLT, or dedicated 
observations of specific proto-clusters with the \emph{James Webb Space Telescope} (\emph{JWST}, \citealt{gardner06}) will mitigate sample variance effects and immensely aid
in the calibration of various physical parameters.

Putting aside the above concerns, it appears that high-$\mathcal{M}_{\ast}$ galaxies in the densest environments in the higher-redshift bin are primarily driving
the reversal of the $\mathcal{SFR}-\delta_{gal}$ relation. This is in stark contrast to the demographics of higher-$\mathcal{M}_{\ast}$ group and cluster galaxies in the low-
and intermediate-redshift universe (i.e., $0\le z \le 1.5$) where such galaxies broadly exhibit higher quiescent fractions and have average $\mathcal{SFR}$s that are
lower than their lower-$\mathcal{M}_{\ast}$ counterparts (see, e.g., \citealt{lem19, AA19} and references therein). This dramatic change over the course of several
Gyr is likely the result of multiple processes present in high-density environments that serve to both enhance star-formation activity at higher redshift and quench it 
at lower redshifts. In what follows, we explore the possible effect of these processes and review potential constraints on these processes from our own data as well as 
various studies from the literature.

\subsubsection{Gas fraction and the $\mathcal{SFR}-\delta_{gal}$ relation}

It has been well established by observations of large samples of galaxies, primarily with the Atacama Large Millimeter/submillimeter Array (ALMA) and
the Plateau de Bure millimetre Interferometer (PdBI), that 
the fraction of gas mass to total baryonic mass ($\mu_{\rm{gas}}$) is a strong function of redshift, with galaxies exhibiting  
lower $\mu_{gas}$ values at fixed stellar mass with decreasing redshift (see, e.g., \citealt{genzel15, schinnerer16, smokynicky16, smokynicky17}). 
Adopting the formalism of \citet{smokynicky17}, $\mu_{\rm{gas}}$ is predicted to drop by $\sim$10\% in between the median redshift of our higher- and 
lower-redshift bins, with average rates of cosmological accretion \citep{dekel09a, dekel09b} decreasing by more than a factor of two. By the time the 
$\mathcal{SFR}-\delta_{gal}$ relation is firmly
seen as an anticorrelation. At $z\sim1$, gas fractions have decreased, on average, by more than 20\% at fixed stellar mass and expected gas accretion 
rates have declined by more than an order of magnitude relative to those at the median redshift of our higher-redshift sample. At fixed star-formation 
efficiency, such a decrease in gas mass will lead to an overall decrease in star formation across all environments.  

However, it is reasonable to assume that environment would modulate both the accretion rates and gas fraction of galaxies in varying ways across different redshifts,
potentially inducing excess accretion at some redshifts and stifling or ceasing accretion at other redshifts.
While such evidence is sparse due to the lack of observations of consistent samples of galaxies residing in average- and high-density environments at all redshifts, 
there have been a few hints from ALMA observations of the possible effect of environment at different redshifts. 

At the highest redshifts probed by our sample, such observations are still extremely limited. Recently, the ALMA Large Program to Investigate C$^+$ at Early times 
(ALPINE; \citealt{dong20, dirtyAF20, mbetz20}) survey, while primarily designed as a field survey, has begun to probe high-density environments at $4 < z < 5$. 
In ALPINE, several galaxies in and around the PCl J1001+0220 proto-cluster in the COSMOS field were observed (see, e.g., \citealt{jones19, ginolfi20b}). In these observations, 
massive, diffuse gas envelopes were observed to be associated with several ongoing merging events within the greater proto-cluster environment. If the ambient medium has not been 
sufficiently heated, likely a fair assumption at these redshifts (e.g., \citealt{overzier16, rhythm18b}), some of this diffuse gas will likely be (re-)captured
by the remnant galaxy. If most of this gas is retained, the resultant $\mu_{\rm{gas}}$ values would likely approach those observed in starburst galaxies (e.g., 
\citealt{smokynicky16, tacconi18} though see also \citealt{silverman18}), which could result in the fueling of 
virulent star-formation activity (see, e.g., the discussion in \S4.2 of \citealt{aoyama22} and references therein). There also appears evidence of cold, diffuse 
gas in several proto-clusters in our VUDS+ sample (e.g., 
\citealt{olga14, drew20}) that may accrete on to galaxies as they coalesce into the proto-cluster environment (see, e.g., \citealt{rhythm17,rhythm18b} for possible evidence
of such a phenomenon). Such a process would naturally be more effective in higher stellar mass galaxies due to their larger potential well. By contrast, 
in a study of the core of the exceptional SPT2349-56 proto-cluster at $z\sim4.3$, \citet{ryley21} found smaller $\mu_{\rm{gas}}$ 
and depletion timescales for proto-cluster members than those of comparable field galaxies. Though many of these galaxies were extremely massive, with approximately 
half of the sample exhibiting stellar masses of $\log(\mathcal{M}_{\ast}/\mathcal{M}_{\odot})>11$, well in excess of most of the galaxies in the VUDS+ sample, 
these results suggest considerable variation across structures and different galaxy populations even at these early epochs. 

At redshifts that correspond to the lower-redshift portion of our sample, $z\sim2-2.5$, the gas fraction at fixed stellar mass
appears to be somewhat anticorrelated with environment, with lower $\mu_{\rm{gas}}$ measured for at least some fraction of galaxies situated in denser environments 
(e.g., \citealt{coogan18, taowang18, zavala19}, though see references citing other trends below). One of the overdensities where this trend is observed 
\citep{taowang16,taowang18} is situated in the Hyperion proto-supercluster at 
$z\sim2.5$ \citep{olga18}, a structure that is probed as part of the sample presented here (though see \citealt{champagne21} for a different view). 
Recent observations of the most massive proto-cluster in Hyperion, Theia, 
using the NOrthern Extended Millimeter Array (NOEMA) also suggest a deficit of gas among the massive galaxies of that system, which implies this trend persists to some
level throughout Hyperion (Cucciati et al. \emph{in prep}). Such a deficit is also seen at lower redshifts ($z\sim0.7$; \citealt{betti19}) for galaxies in the highest-density
regions of massive group/cluster environments. If such a paucity is pervasive for galaxies in proto-cluster environments at $z\sim2-2.5$, it would help to explain the change in behavior of
the $\mathcal{SFR}-\delta_{gal}$ seen between our higher- and lower-redshift samples. However, such observations remain very sparse even at these redshifts, and the opposite trend 
(e.g., \citealt{gg19, tadaki19, aoyama22}) or a flat trend (e.g., \citealt{lee17, darvish18}) of $\mu_{\rm{gas}}$ with environment is also observed at these redshifts (also see
the discussion in \S3.4 of \citealt{tacconi20}). 
While it appears likely that \emph{in situ} gas depletion at $z\la2.5$ and gas enhancement at $z\ga2.5$ in high-density environments helps to drive the behavior 
of the $\mathcal{SFR}-\delta_{gal}$ relation seen here, future work with ALPINE and other observations from ALMA and PdBI at these redshifts are required in order to be able to make 
definitive claims. 
 
\subsubsection{Merging activity and the $\mathcal{SFR}-\delta_{gal}$ relation}

It has already been established that at least some of the galaxies in the high-redshift range of the VUDS+ sample are undergoing major merging events.
This evidence comes from detailed observations of individual merging systems (\citealt{jones19, ginolfi20b}) and from a general analysis of the PCl J1001+0220 
proto-cluster, in which the VUDS members are seen to potentially show elevated numbers of companions \citep{lem18}. Additionally, there is evidence, at least up to $z\sim2-4$, 
that the major merger rate broadly increases with increasing redshift (e.g., \citealt{carlos13, tizzytasca14}). Further, it has been shown that a large number of merging events 
in intermediate density environments, such as those in proto-clusters, are required at high redshift to explain the stellar mass function in group and cluster galaxies at $z\sim1$ 
\citep{AA17}. Given this evidence, it is possible merging activity is generally prevalent in our sample, though to a lesser extent in the lower-redshift portion of our 
sample ($z\sim2.5$). 

The drop in merging activity from higher to lower redshift observed in the field would likely be exacerbated in higher-density environments  
due to dynamical considerations. Specifically, for a given structure, the velocity dispersion of member galaxies will increase with decreasing redshift as a result of  
the concentration of mass in the forming cluster (see, e.g., Figure 4 of \citealt{olga14}), which will ultimately discourage merging activity in the cores of such structures
at lower redshifts. If such merging activity serves to enhance star formation (e.g., \citealp{hopkins08}), it is possible that changes in the merging rate could explain the increase in $\mathcal{SFR}$ 
seen in higher-density environments at higher redshift and the subsequent decline to lower redshifts as such activity begins to wane. 

Considering the high-$\mathcal{M}_{\ast}$ galaxies only, galaxies which are more likely to be strongly clustered at these redshifts than their lower-mass counterparts (see, e.g., \citealt{ania18}), 
the median overdensity in the Peak region in the highest-redshift bin for such galaxies is $\sim$5 gal/Mpc$^2$. This translates to an average inter-galaxy distance of $\sim$0.45 Mpc. 
For illustrative purposes, an 
average velocity offset $\Delta_v=400$ km s$^{-1}$ can be assumed for such galaxies, as this is a typical line-of-sight member velocity dispersion for member galaxies of 
massive proto-clusters at these redshifts (e.g., \citealt{olga14, olga18, dey16, tersetoshikawa20}). Incorporating the median stellar mass of the high-$\mathcal{M}_{\ast}$ 
sample at these redshifts, $\log(\tilde{\mathcal{M}_{\ast}}/\mathcal{M}_{\odot})={10.30}$, the resultant average dynamical friction timescale is $\sim$500 Myr 
\citep{binney87, burke13}. The difference in lookback times between the median redshift of galaxies in the two redshift 
bins used in this paper ($z=2.34$ to $z=3.13$) is $\sim$800 Myr in our adopted cosmology. This amount of time allows for galaxies that begin merging in the 
higher-redshift bin to coalesce, with several 100 megayears after coalescence to begin the transition to lower levels of $\mathcal{SFR}$. 

Following this line of thought, we consider two toy scenarios. In the first we assume galaxies in the higher-mass bin have a constant $\mathcal{SFR}$ of 
$\log(\mathcal{SFR}/\mathcal{M}_{\odot}/\rm{yr})=1.8$, 
a value intermediate to the average $\mathcal{SFR}$ of high-$\mathcal{M}_{\ast}$ galaxies in the lower- and higher-redshift samples, and that such activity persists over 
the time period elapsed between the two redshift bins. Adding the resultant newly created stellar mass to the average stellar mass of the high-$\mathcal{M}_{\ast}$ galaxies in the higher-z bin, 
the average merger remnant formed through merging processes that began in the higher-redshift bin approaches $\log(\mathcal{M}_{\ast}/\mathcal{M}_{\odot})=11$ by the redshift range
of our lower-redshift bin. This stellar mass rivals some of the most massive galaxies observed at these redshifts (e.g., \citealt{AA14, forrest20, shen21}) and begins to approach the stellar 
mass of the most massive cluster galaxies observed at intermediate and low redshift (see \citealt{superB14} and references therein). Many such galaxies are still observed to 
be star forming at $z=1-2$, though typically at modest levels, making this a plausible scenario for the formation of such galaxies.  

If, conversely, merging galaxies in the highest-redshift bin have 
already coalesced and are experiencing relatively short-lived elevated star-formation activity as a result of this coalescence, they would have $\sim$800 Myr
to begin to transition to a more quiescent state as the result of secular or larger scale environmental processes. Such a timescale is sufficient to begin to 
transition to quiescence for those galaxies whose SFHs are governed by an exponential decline with a relatively small $\tau$ value
(e.g., \citealp{thibaud16}). However, the lack of an anticorrelated $\mathcal{SFR}-\delta_{gal}$ trend for high-$\mathcal{M}_{\ast}$ galaxies in the lower-redshift 
end of the VUDS+ sample precludes this $\tau$ from being too small unless they quench so rapidly as to be outside the star-forming selection criteria of the VUDS+
spectral sample. In this case, such a small $\tau$ would require a large number of truly quiescent galaxies to be formed in such environments. 

These two scenarios should lead to significantly different behavior in the stellar mass function for quiescent and star-forming galaxies as a function of both redshift and environment. 
Future observations from NIR spectrographs in these environments aimed at constraining the pervasiveness of high-$\mathcal{M}_{\ast}$ quiescent galaxies will,
when paired with data sets such as the one presented here, allow for robust constraints on the timescale and manner of quenching (as done in, e.g., 
\citealt{balogh16, foltz18, lem19, AA19} at intermediate redshifts) and the amount of merging activity in different environments (as done in, e.g., \citealt{AA17}). 
Additional future work on constraining the recent SFHs of galaxies (e.g., \citealt{cassara16, carnall19, leja19}) in higher-density environments, especially 
those at high $\mathcal{M}_{\ast}$, would be extremely useful to evaluate the likelihood of such scenarios.  

We note, though, that the value quoted for the dynamical friction timescale, a value upon which these scenarios are predicated, should be considered a soft lower limit to the true average 
dynamical friction timescale as this estimate ought to be based on the true volume density of such galaxies rather than the surface density. We do not attempt a conversion here,
as the true three dimensional structure of overdense regions is masked by peculiar motions. However, if the mean separation between galaxies in the Peak region is, in reality, 
even only $\sim$50\% larger, there is little chance that galaxies, which have just begun to merge in the highest-redshift bin, have completed this process by the lower-redshift
bin. In such a case, while star-formation activity could naturally be enhanced by merging processes in the higher-redshift bin, there is no natural explanation as to why such activity would 
begin to depreciate in our lower-redshift bin unless larger-scale processes begin to act while merging is still ongoing.  
Future work characterizing the frequency of merging signatures in the VUDS+ sample across different environments at fixed $\mathcal{M}_{\ast}$ for those 
galaxies imaged with the requisite resolution and depth to detect such signatures will be crucial to assess the veracity of such scenarios. 

\subsubsection{Large scale processes and the $\mathcal{SFR}-\delta_{gal}$ relation} 

It is a distinct possibility, however, that larger-scale processes may be preferentially acting on galaxies in the lower-redshift portion of our sample. 
Observational evidence exists for at least one overdensity at $z\sim2.5$ to contain a hot intracluster medium (ICM), an overdensity that is in our sample 
(\citealt{taowang16}, though again see \citealt{champagne21} for a different possibility). Indeed, as mentioned earlier, in this particular overdensity, massive galaxies toward 
the center of the structure possibly appear to be 
deficient in their gas content relative to galaxies near the outskirts, a deficiency that may be due to stripping processes associated with the dense environment 
\citep{taowang18}. Additionally, it appears that it might be a general characteristic, at least of massive proto-clusters
to begin to transition from a relatively cold ICM marked by strong absorption of neutral hydrogen (e.g., \citealp{lee16, drew20}) to a gravitationally heated ICM
at redshifts that mark the boundary between our two redshift bins \citep{overzier16}. Other case studies that probe representative galaxy
populations at similar redshifts also show hints that star-formation activity in the densest regions is beginning to slow, potentially due 
to the presence of a hot ICM (e.g., \citealt{rhythm18b}). Thus, it is natural to suspect that the development and heating of
this ICM may be at least partially responsible for the observed reduction in the average $\mathcal{SFR}$ of galaxies in the Peak regions from $z\sim3.1$ to
$z\sim2.3$, and that such a drop could occur without the need to invoke a widespread merging scenario. 

The dynamical timescale for galaxies in the highest-redshift bin that are accreting into proto-cluster environments, a quantity that is defined as the time from 
infall at $R_{vir}$ to first pericentric passage, is $\sim0.3$ Gyr adopting the formalism of \citet{wetzel11} and \citet{wetzel13}. Such a timescale implies that a galaxy
situated deep in the core region\footnote{We note that $R_{vir}$ or its counterpart $R_{200}$ is very small at these redshifts, with $R_{200}\sim$225 kpc at $z=3.13$ 
for a proto-cluster with $\mathcal{M}_{200}=5\times10^{13}$ $\mathcal{M}_{\odot}$.}
is able to make at least one or two passages through the proto-cluster core in the time elapsed between the two redshift bins in our sample. Such passages 
may have several consequences including decreasing the merging time of any ongoing mergers through tidal torquing, tidal stripping of the \emph{in situ} or circumgalactic \ion{H}{I} gas, 
and, if a hot ICM begins to develop, ram pressure stripping (see, e.g., \citealt{cen14, ciao16, steinhauser16, gavazzi18, moretti18}). Stellar feedback induced through merging activity or 
through elevated levels of star formation, as well as any associated active galactic nuclei (AGN) activity (see, e.g., \citealt{rhythm18}), would likely only serve to increase the effectiveness of 
such stripping \citep{bahe15, lem17} as otherwise usable gas is pushed to larger galactocentric radii to eventually enhance and enrich the ICM. Such feedback processes are 
likely more common and more severe in our higher-$\mathcal{M}_{\ast}$ galaxies. This sub-sample contains the galaxies that are observed with the highest average $\mathcal{SFR}$s 
of all galaxies in VUDS+. Such
elevated $\mathcal{SFR}$s serve to increase the potency of stellar feedback processes (e.g., \citealt{heckman15,cicone16, ginolfi20a}) and, though perhaps not causally, 
increase likelihood of various types of AGN
activity (\citealt{juneau13, lem14b, aird19}). It is indeed these galaxies that are inferred to be preferentially quenching from the higher- to the lower-redshift bin 
in our own data. 

While such a scenario 
appears to satisfy at least some trends observed here and in the literature, considerable effort will need to be made to test scenarios such as these through the 
use of simulations and models that encode merging activity and the enrichment 
of the ICM. Additionally, future work will be done as part of the C3VO survey to characterize the entire population of VUDS proto-clusters and groups in the same manner as was done at $z\sim1$ 
in \citet{denise20, denise21}. In concert with the exquisite X-ray and other ancillary data available in the VUDS fields, the large number of overdensities that will be identified through 
our galaxy density mapping and overdensity detection approach will enable
a comprehensive search for evidence of a hot ICM both in individual cases and through stacked analysis. Such a search will ultimately allow us to tie the presence 
or absence of a hot ICM in different structures to the properties of the galaxies contained within those structures. Through the use of simulations and semiempirical models, 
such evolution can eventually be connected to galaxy populations at low to intermediate redshifts ($0 \le z \la 1.5$) 
where a clear anticorrelation between star-formation activity and environment is observed. 


 
\section{Summary and conclusions}
\label{conclusions}

In this paper we use a sample of 6730 star-forming galaxies spectroscopically confirmed at $2 \le z_{spec} \le 5$ in three extragalactic legacy fields 
(COSMOS, ECDFS, and CFHTLS-D1) to investigate the relationship between the average $\mathcal{SFR}$ of galaxies, as measured by spectral energy distribution fitting, 
and their local environment ($\delta_{gal}$). These 
spectroscopic data were drawn primarily from the VIMOS Ultra Deep Survey and supplemented by a variety of other spectral surveys taken across the three fields,
including new observations taken with Keck/DEIMOS as part of the C3VO survey. These data were supplemented by high-quality photometric redshift measurements, which
were used in conjunction with the spectroscopic data to estimate the galaxy density field at these redshifts through a technique known as Voronoi Monte Carlo 
mapping. While at low to intermediate redshift ($z\la1.5$) it has been shown that $\mathcal{SFR}$ and $\delta_{gal}$ are generally anticorrelated, at 
$2\le z \le 5$ we observe a reversal of the $\mathcal{SFR}-\delta_{gal}$ relation, with the two quantities showing a weak but significant \emph{positive} correlation
over more than an order of magnitude in local density. We list our main conclusions below:

\begin{itemize}

\renewcommand{\labelitemi}{$\bullet$}

\item The full spectral sample, which probes from rarefied field environments to the core of massive proto-clusters, exhibits a nearly monotonic rise in $\mathcal{SFR}$ 
over more than an order of magnitude in local overdensity. Similar behavior is observed between stellar mass ($\mathcal{M}_{\ast}$) and local overdensity. 

\vskip 6pt

\item The reversal of the $\mathcal{SFR}-\delta_{gal}$ relation is observed across all three fields and across all redshifts probed by our sample with varying degrees 
of strength and significance. The reversal remains significant after taking into account all measurement uncertainties and selection effects, including the possible
contribution of extremely dusty star-forming and quiescent galaxies that are largely absent from our sample.

\vskip 6pt

\item A positive $\mathcal{SFR}-\delta_{gal}$ relation is also observed at fixed $\mathcal{M}_{\ast}$, with the strongest and most significant positive correlation seen among 
the most massive galaxies in our sample ($\log(\mathcal{M}_{\ast}/\mathcal{M}_{\odot})\ge10$). After accounting for and removing both redshift and stellar mass effects 
from the $\mathcal{SFR}-\delta_{gal}$ relation, a weaker, but still highly significant positive correlation is observed.

\vskip 6pt

\item The $\mathcal{SFR}-\delta_{gal}$ relation in the higher-redshift portion of the sample ($2.7 < z \le 5$) is observed to be considerably stronger than in
the lower-redshift portion of our sample ($2 \le z \le 2.7$). For environments that correspond to the outskirts and core of proto-clusters, termed ``Intermediate''
and ``Peak'' regions, no significant correlation is observed between $\mathcal{SFR}-\delta_{gal}$ at $2 \le z \le 2.7$, with galaxies in the field and the 
densest proto-cluster environments showing statistically consistent $\mathcal{SFR}$s. Conversely, in the higher-redshift portion of our sample, a highly significant
positive $\mathcal{SFR}-\delta_{gal}$ relation is found, with galaxies in the densest proto-cluster environments exhibiting average $\mathcal{SFR}$s $>$0.5 dex higher 
than those in the coeval field.

\vskip 6pt

\item The observed $\mathcal{SFR}-\delta_{gal}$ is generally consistent with investigations in other works of this relation in the Millennium, IllustrisTNG, and other simulations, 
with the magnitude of the increase in the average $\mathcal{SFR}$ observed to be broadly similar across the full density range probed in
simulations with a few exceptions. 

\vskip 6pt 

\item We explore various scenarios that could cause the increased $\mathcal{SFR}$ activity seen in the high-density environments at the highest 
redshifts probed in our sample and the subsequent decrease of this activity in such environments to $z\sim2$. We conclude that accelerated gas depletion, 
merging activity, and processes related to the large-scale environment, which include those originating from the development of a hot ICM, are all plausible 
mechanisms to explain the observed behavior. 

\end{itemize}

This work marks an important step forward in establishing the general relationship between galaxies and their environment at high redshift. However, considerable 
work, both with the VUDS+ sample and with larger samples in the future from Subaru/PFS and VLT/MOONS, will be necessary to begin to quantify the average contribution 
of the various processes in shaping the $\mathcal{SFR}-\delta_{gal}$ at high redshift. Additional work will also be needed to identify 
systems in such surveys that are in special stages of formation at these epochs, such as those with fractionally large populations of strongly starbursting galaxies or massive 
quiescent galaxies, systems that are generally more difficult or impossible to detect from an observed-frame optical/NIR perspective. Ongoing, dedicated follow up at optical, 
near-infrared, and sub-millimeter wavelengths of several exceptional structures in the VUDS+ sample, as well as systems that appear to be in a more typical stag of evolution, will 
be eventually used to understand the various pathways through which various processes might incite and suppress star-formation activity in high-density environments in the 
early universe. 

\begin{acknowledgements}
Based on data obtained with the European Southern Observatory Very Large Telescope, Paranal, Chile, under Large Programs 175.A-0839, 177.A-0837, and
185.A-0791. Some of the material presented in this paper is based upon work supported by the National Science Foundation under Grant No. 1908422. 
This work was additionally supported by the France-Berkeley Fund, a joint venture between UC Berkeley, UC Davis, and le Centre National 
de la Recherche Scientifique de France promoting lasting institutional and intellectual cooperation between France and the United States. 
This work was supported by funding from the European Research Council Advanced Grant ERC-2010-AdG-268107-EARLY and by INAF Grants PRIN 2010, 
PRIN 2012 and PICS 2013. RA acknowledges support from FONDECYT Regular Grant 1202007. 
BCL thanks Maru\v{s}a Brada\v{c} for providing the time and space necessary to finish this work. BCL additionally
thanks Helmut Dannerbaur and Ga\"{e}l Noirot for useful discussions and the organizers of the Protoclusters: Galaxy Evolution in Confinement workshop, a workshop 
which stimulated discussions on this topic that improved the manuscript.
We also thank the anonymous referee for the careful reading of the manuscript.
This work is partially based on observations obtained with MegaPrime/MegaCam, a joint project of CFHT and CEA/IRFU, at the Canada-France-Hawaii Telescope (CFHT) which is operated by the 
National Research Council (NRC) of Canada, the Institut National des Science de l'Univers of the Centre National de la Recherche Scientifique (CNRS) of France, and the 
University of Hawaii. This work is based in part on data products produced at Terapix available at the Canadian Astronomy Data Centre as part of the Canada-France-Hawaii 
Telescope Legacy Survey, a collaborative project of NRC and CNRS. This work is based, in part, on observations made with the Spitzer Space Telescope, which is operated by the 
Jet Propulsion Laboratory, California Institute of Technology under a contract with NASA. We thank ESO staff for their support for the VUDS survey, particularly the Paranal staff 
conducting the observations and Marina Rejkuba and the ESO user support group in Garching. 
Some of the spectrographic data presented herein were obtained at the W.M. Keck Observatory, which is operated as a scientific partnership among
the California Institute of Technology, the University of California, and the National Aeronautics and Space Administration. The Observatory was 
made possible by the generous financial support of the W.M. Keck Foundation. We thank the indigenous Hawaiian community for allowing us to be guests 
on their sacred mountain, a privilege, without which, this work would not have been possible. We are most fortunate to be able to conduct observations from this site.

\end{acknowledgements}


\bibliographystyle{aa} 
\bibliography{vuds_SFRdensity_reversal_v12_FINAL.bib} 

\newcommand{\noop}[1]{}
\begin{thebibliography}{235}
\expandafter\ifx\csname natexlab\endcsname\relax\def\natexlab#1{#1}\fi

\bibitem[{{Aird} {et~al.}(2019){Aird}, {Coil}, \& {Georgakakis}}]{aird19}
{Aird}, J., {Coil}, A.~L., \& {Georgakakis}, A. 2019, \mnras, 484, 4360

\bibitem[{{Aoyama} {et~al.}(2022){Aoyama}, {Kodama}, {Suzuki}, {Tadaki},
  {Shimakawa}, {Hayashi}, {Koyama}, \& {P{\'e}rez-Mart{\'\i}nez}}]{aoyama22}
{Aoyama}, K., {Kodama}, T., {Suzuki}, T.~L., {et~al.} 2022, \apj, 924, 74

\bibitem[{{Araya-Araya} {et~al.}(2021){Araya-Araya}, {Vicentin}, {Sodr{\'e}},
  {Overzier}, \& {Cuevas}}]{arayaaraya21}
{Araya-Araya}, P., {Vicentin}, M.~C., {Sodr{\'e}}, Laerte, J., {Overzier},
  R.~A., \& {Cuevas}, H. 2021, \mnras, 504, 5054

\bibitem[{{Arnouts} {et~al.}(1999){Arnouts}, {Cristiani}, {Moscardini},
  {Matarrese}, {Lucchin}, {Fontana}, \& {Giallongo}}]{stephane99}
{Arnouts}, S., {Cristiani}, S., {Moscardini}, L., {et~al.} 1999, \mnras, 310,
  540

\bibitem[{{Arnouts} {et~al.}(2013){Arnouts}, {Le Floc'h}, {Chevallard},
  {Johnson}, {Ilbert}, {Treyer}, {Aussel}, {Capak}, {Sanders}, {Scoville},
  {McCracken}, {Milliard}, {Pozzetti}, \& {Salvato}}]{stephane13}
{Arnouts}, S., {Le Floc'h}, E., {Chevallard}, J., {et~al.} 2013, \aap, 558, A67

\bibitem[{{Ascaso} {et~al.}(2014){Ascaso}, {Lemaux}, {Lubin}, {Gal},
  {Kocevski}, {Rumbaugh}, \& {Squires}}]{superB14}
{Ascaso}, B., {Lemaux}, B.~C., {Lubin}, L.~M., {et~al.} 2014, \mnras, 442, 589

\bibitem[{{Bah{\'e}} \& {McCarthy}(2015)}]{bahe15}
{Bah{\'e}}, Y.~M. \& {McCarthy}, I.~G. 2015, \mnras, 447, 969

\bibitem[{{Balestra} {et~al.}(2010){Balestra}, {Mainieri}, {Popesso},
  {Dickinson}, {Nonino}, {Rosati}, {Teimoorinia}, {Vanzella}, {Cristiani}, \&
  {Cesarsky}}]{balestra10}
{Balestra}, I., {Mainieri}, V., {Popesso}, P., {et~al.} 2010, \aap, 512, A12

\bibitem[{{Balogh} {et~al.}(2017){Balogh}, {Gilbank}, {Muzzin}, {Rudnick},
  {Cooper}, {Lidman}, {Biviano}, {Demarco}, {McGee}, {Nantais}, {Noble}, {Old},
  {Wilson}, {Yee}, {Bellhouse}, {Cerulo}, {Chan}, {Pintos-Castro}, {Simpson},
  {van der Burg}, {Zaritsky}, {Ziparo}, {Alonso}, {Bower}, {De Lucia},
  {Finoguenov}, {Lambas}, {Muriel}, {Parker}, {Rettura}, {Valotto}, \&
  {Wetzel}}]{balogh17}
{Balogh}, M.~L., {Gilbank}, D.~G., {Muzzin}, A., {et~al.} 2017, \mnras, 470,
  4168

\bibitem[{{Balogh} {et~al.}(2016){Balogh}, {McGee}, {Mok}, {Muzzin}, {van der
  Burg}, {Bower}, {Finoguenov}, {Hoekstra}, {Lidman}, {Mulchaey}, {Noble},
  {Parker}, {Tanaka}, {Wilman}, {Webb}, {Wilson}, \& {Yee}}]{balogh16}
{Balogh}, M.~L., {McGee}, S.~L., {Mok}, A., {et~al.} 2016, \mnras, 456, 4364

\bibitem[{{Balogh} {et~al.}(2021){Balogh}, {van der Burg}, {Muzzin}, {Rudnick},
  {Wilson}, {Webb}, {Biviano}, {Boak}, {Cerulo}, {Chan}, {Cooper}, {Gilbank},
  {Gwyn}, {Lidman}, {Matharu}, {McGee}, {Old}, {Pintos-Castro}, {Reeves},
  {Shipley}, {Vulcani}, {Yee}, {Alonso}, {Bellhouse}, {Cooke}, {Davidson}, {De
  Lucia}, {Demarco}, {Drakos}, {Fillingham}, {Finoguenov}, {Forrest},
  {Golledge}, {Jablonka}, {Lambas Garcia}, {McNab}, {Muriel}, {Nantais},
  {Noble}, {Parker}, {Petter}, {Poggianti}, {Townsend}, {Valotto}, {Webb}, \&
  {Zaritsky}}]{balogh21}
{Balogh}, M.~L., {van der Burg}, R. F.~J., {Muzzin}, A., {et~al.} 2021, \mnras,
  500, 358

\bibitem[{{B{\'e}thermin} {et~al.}(2020){B{\'e}thermin}, {Fudamoto}, {Ginolfi},
  {Loiacono}, {Khusanova}, {Capak}, {Cassata}, {Faisst}, {Le F{\`e}vre},
  {Schaerer}, {Silverman}, {Yan}, {Amorin}, {Bardelli}, {Boquien}, {Cimatti},
  {Davidzon}, {Dessauges-Zavadsky}, {Fujimoto}, {Gruppioni}, {Hathi}, {Ibar},
  {Jones}, {Koekemoer}, {Lagache}, {Lemaux}, {Moreau}, {Oesch}, {Pozzi},
  {Riechers}, {Talia}, {Toft}, {Vallini}, {Vergani}, {Zamorani}, \&
  {Zucca}}]{mbetz20}
{B{\'e}thermin}, M., {Fudamoto}, Y., {Ginolfi}, M., {et~al.} 2020, \aap, 643,
  A2

\bibitem[{{Betti} {et~al.}(2019){Betti}, {Pope}, {Scoville}, {Yun}, {Aussel},
  {Kartaltepe}, \& {Sheth}}]{betti19}
{Betti}, S.~K., {Pope}, A., {Scoville}, N., {et~al.} 2019, \apj, 874, 53

\bibitem[{{Bielby} {et~al.}(2012){Bielby}, {Hudelot}, {McCracken}, {Ilbert},
  {Daddi}, {Le F{\`e}vre}, {Gonzalez-Perez}, {Kneib}, {Marmo}, {Mellier},
  {Salvato}, {Sanders}, \& {Willott}}]{bielby12}
{Bielby}, R., {Hudelot}, P., {McCracken}, H.~J., {et~al.} 2012, \aap, 545, A23

\bibitem[{{Binney} \& {Tremaine}(1987)}]{binney87}
{Binney}, J. \& {Tremaine}, S. 1987, {Galactic dynamics}

\bibitem[{{Boquien} {et~al.}(2019){Boquien}, {Burgarella}, {Roehlly}, {Buat},
  {Ciesla}, {Corre}, {Inoue}, \& {Salas}}]{mederic19}
{Boquien}, M., {Burgarella}, D., {Roehlly}, Y., {et~al.} 2019, \aap, 622, A103

\bibitem[{{Bose} {et~al.}(2019){Bose}, {Eisenstein}, {Hernquist}, {Pillepich},
  {Nelson}, {Marinacci}, {Springel}, \& {Vogelsberger}}]{bose19}
{Bose}, S., {Eisenstein}, D.~J., {Hernquist}, L., {et~al.} 2019, \mnras, 490,
  5693

\bibitem[{{Boselli} {et~al.}(2016){Boselli}, {Roehlly}, {Fossati}, {Buat},
  {Boissier}, {Boquien}, {Burgarella}, {Ciesla}, {Gavazzi}, \&
  {Serra}}]{ciao16}
{Boselli}, A., {Roehlly}, Y., {Fossati}, M., {et~al.} 2016, \aap, 596, A11

\bibitem[{{Boulade} {et~al.}(2003){Boulade}, {Charlot}, {Abbon}, {Aune},
  {Borgeaud}, {Carton}, {Carty}, {Da Costa}, {Deschamps}, {Desforge},
  {Eppell{\'e}}, {Gallais}, {Gosset}, {Granelli}, {Gros}, {de Kat}, {Loiseau},
  {Ritou}, {Rouss{\'e}}, {Starzynski}, {Vignal}, \& {Vigroux}}]{boulade03}
{Boulade}, O., {Charlot}, X., {Abbon}, P., {et~al.} 2003, in \procspie, Vol.
  4841, Instrument Design and Performance for Optical/Infrared Ground-based
  Telescopes, ed. M.~{Iye} \& A.~F.~M. {Moorwood}, 72--81

\bibitem[{{Bruzual} \& {Charlot}(2003)}]{BC03}
{Bruzual}, G. \& {Charlot}, S. 2003, \mnras, 344, 1000

\bibitem[{{Burgarella} {et~al.}(2005){Burgarella}, {Buat}, \&
  {Iglesias-P{\'a}ramo}}]{denis05}
{Burgarella}, D., {Buat}, V., \& {Iglesias-P{\'a}ramo}, J. 2005, \mnras, 360,
  1413

\bibitem[{{Burke} \& {Collins}(2013)}]{burke13}
{Burke}, C. \& {Collins}, C.~A. 2013, \mnras, 434, 2856

\bibitem[{{Calzetti} {et~al.}(2000){Calzetti}, {Armus}, {Bohlin}, {Kinney},
  {Koornneef}, \& {Storchi-Bergmann}}]{calzetti00}
{Calzetti}, D., {Armus}, L., {Bohlin}, R.~C., {et~al.} 2000, \apj, 533, 682

\bibitem[{{Capak} {et~al.}(2007){Capak}, {Aussel}, {Ajiki}, {McCracken},
  {Mobasher}, {Scoville}, {Shopbell}, {Taniguchi}, {Thompson}, {Tribiano},
  {Sasaki}, {Blain}, {Brusa}, {Carilli}, {Comastri}, {Carollo}, {Cassata},
  {Colbert}, {Ellis}, {Elvis}, {Giavalisco}, {Green}, {Guzzo}, {Hasinger},
  {Ilbert}, {Impey}, {Jahnke}, {Kartaltepe}, {Kneib}, {Koda}, {Koekemoer},
  {Komiyama}, {Leauthaud}, {Le Fevre}, {Lilly}, {Liu}, {Massey}, {Miyazaki},
  {Murayama}, {Nagao}, {Peacock}, {Pickles}, {Porciani}, {Renzini}, {Rhodes},
  {Rich}, {Salvato}, {Sanders}, {Scarlata}, {Schiminovich}, {Schinnerer},
  {Scodeggio}, {Sheth}, {Shioya}, {Tasca}, {Taylor}, {Yan}, \&
  {Zamorani}}]{capak07}
{Capak}, P., {Aussel}, H., {Ajiki}, M., {et~al.} 2007, \apjs, 172, 99

\bibitem[{{Cardamone} {et~al.}(2010){Cardamone}, {van Dokkum}, {Urry},
  {Taniguchi}, {Gawiser}, {Brammer}, {Taylor}, {Damen}, {Treister}, {Cobb},
  {Bond}, {Schawinski}, {Lira}, {Murayama}, {Saito}, \& {Sumikawa}}]{car10}
{Cardamone}, C.~N., {van Dokkum}, P.~G., {Urry}, C.~M., {et~al.} 2010, \apjs,
  189, 270

\bibitem[{{Carnall} {et~al.}(2019){Carnall}, {Leja}, {Johnson}, {McLure},
  {Dunlop}, \& {Conroy}}]{carnall19}
{Carnall}, A.~C., {Leja}, J., {Johnson}, B.~D., {et~al.} 2019, \apj, 873, 44

\bibitem[{{Casey} {et~al.}(2015){Casey}, {Cooray}, {Capak}, {Fu}, {Kovac},
  {Lilly}, {Sanders}, {Scoville}, \& {Treister}}]{connivingcasey15}
{Casey}, C.~M., {Cooray}, A., {Capak}, P., {et~al.} 2015, \apjl, 808, L33

\bibitem[{{Casey} {et~al.}(2014){Casey}, {Narayanan}, \&
  {Cooray}}]{connivingcasey14}
{Casey}, C.~M., {Narayanan}, D., \& {Cooray}, A. 2014, \physrep, 541, 45

\bibitem[{{Cassar{\`a}} {et~al.}(2016){Cassar{\`a}}, {Maccagni}, {Garilli},
  {Scodeggio}, {Thomas}, {Le F{\`e}vre}, {Zamorani}, {Schaerer}, {Lemaux},
  {Cassata}, {Le Brun}, {Pentericci}, {Tasca}, {Vanzella}, {Zucca},
  {Amor{\'\i}n}, {Bardelli}, {Castellano}, {Cimatti}, {Cucciati}, {Durkalec},
  {Fontana}, {Giavalisco}, {Grazian}, {Hathi}, {Ilbert}, {Paltani}, {Ribeiro},
  {Sommariva}, {Talia}, {Tresse}, {Vergani}, {Capak}, {Charlot}, {Contini}, {de
  la Torre}, {Dunlop}, {Fotopoulou}, {Guaita}, {Koekemoer},
  {L{\'o}pez-Sanjuan}, {Mellier}, {Pforr}, {Salvato}, {Scoville}, {Taniguchi},
  \& {Wang}}]{cassara16}
{Cassar{\`a}}, L.~P., {Maccagni}, D., {Garilli}, B., {et~al.} 2016, \aap, 593,
  A9

\bibitem[{{Cassata} {et~al.}(2013){Cassata}, {Giavalisco}, {Williams}, {Guo},
  {Lee}, {Renzini}, {Ferguson}, {Faber}, {Barro}, {McIntosh}, {Lu}, {Bell},
  {Koo}, {Papovich}, {Ryan}, {Conselice}, {Grogin}, {Koekemoer}, \&
  {Hathi}}]{paolo13}
{Cassata}, P., {Giavalisco}, M., {Williams}, C.~C., {et~al.} 2013, \apj, 775,
  106

\bibitem[{{Cassata} {et~al.}(2015){Cassata}, {Tasca}, {Le F{\`e}vre}, {Lemaux},
  {Garilli}, {Le Brun}, {Maccagni}, {Pentericci}, {Thomas}, {Vanzella},
  {Zamorani}, {Zucca}, {Amorin}, {Bardelli}, {Capak}, {Cassar{\`a}},
  {Castellano}, {Cimatti}, {Cuby}, {Cucciati}, {de la Torre}, {Durkalec},
  {Fontana}, {Giavalisco}, {Grazian}, {Hathi}, {Ilbert}, {Moreau}, {Paltani},
  {Ribeiro}, {Salvato}, {Schaerer}, {Scodeggio}, {Sommariva}, {Talia},
  {Taniguchi}, {Tresse}, {Vergani}, {Wang}, {Charlot}, {Contini}, {Fotopoulou},
  {Koekemoer}, {L{\'o}pez-Sanjuan}, {Mellier}, \& {Scoville}}]{paolo15}
{Cassata}, P., {Tasca}, L.~A.~M., {Le F{\`e}vre}, O., {et~al.} 2015, \aap, 573,
  A24

\bibitem[{{Cen} {et~al.}(2014){Cen}, {Roxana Pop}, \& {Bahcall}}]{cen14}
{Cen}, R., {Roxana Pop}, A., \& {Bahcall}, N.~A. 2014, Proceedings of the
  National Academy of Science, 111, 7914

\bibitem[{{Chabrier}(2003)}]{chab03}
{Chabrier}, G. 2003, \pasp, 115, 763

\bibitem[{{Champagne} {et~al.}(2021){Champagne}, {Casey}, {Zavala}, {Cooray},
  {Dannerbauer}, {Fabian}, {Hayward}, {Long}, \& {Spilker}}]{champagne21}
{Champagne}, J.~B., {Casey}, C.~M., {Zavala}, J.~A., {et~al.} 2021, \apj, 913,
  110

\bibitem[{{Chartab} {et~al.}(2020){Chartab}, {Mobasher}, {Darvish},
  {Finkelstein}, {Guo}, {Kodra}, {Lee}, {Newman}, {Pacifici}, {Papovich},
  {Sattari}, {Shahidi}, {Dickinson}, {Faber}, {Ferguson}, {Giavalisco}, \&
  {Jafariyazani}}]{chartab20}
{Chartab}, N., {Mobasher}, B., {Darvish}, B., {et~al.} 2020, \apj, 890, 7

\bibitem[{{Cheng} {et~al.}(2020){Cheng}, {Clements}, {Greenslade}, {Cairns},
  {Andreani}, {Bremer}, {Conversi}, {Cooray}, {Dannerbauer}, {De Zotti},
  {Eales}, {Gonz{\'a}lez-Nuevo}, {Ibar}, {Leeuw}, {Ma}, {Micha{\l}owski},
  {Nayyeri}, {Riechers}, {Scott}, {Temi}, {Vaccari}, {Valtchanov}, {van
  Kampen}, \& {Wang}}]{cheng20}
{Cheng}, T., {Clements}, D.~L., {Greenslade}, J., {et~al.} 2020, \mnras, 494,
  5985

\bibitem[{{Cheng} {et~al.}(2019){Cheng}, {Clements}, {Greenslade}, {Cairns},
  {Andreani}, {Bremer}, {Conversi}, {Cooray}, {Dannerbauer}, {De Zotti},
  {Eales}, {Gonz{\'a}lez-Nuevo}, {Ibar}, {Leeuw}, {Ma}, {Micha{\l}owski},
  {Nayyeri}, {Riechers}, {Scott}, {Temi}, {Vaccari}, {Valtchanov}, {van
  Kampen}, \& {Wang}}]{cheng19}
{Cheng}, T., {Clements}, D.~L., {Greenslade}, J., {et~al.} 2019, \mnras, 490,
  3840

\bibitem[{{Chiang} {et~al.}(2014){Chiang}, {Overzier}, \&
  {Gebhardt}}]{coolchiang14}
{Chiang}, Y.-K., {Overzier}, R., \& {Gebhardt}, K. 2014, \apjl, 782, L3

\bibitem[{{Chiang} {et~al.}(2015){Chiang}, {Overzier}, {Gebhardt},
  {Finkelstein}, {Chiang}, {Hill}, {Blanc}, {Drory}, {Chonis}, {Zeimann},
  {Hagen}, {Schneider}, {Jogee}, {Ciardullo}, \& {Gronwall}}]{coolchiang15}
{Chiang}, Y.-K., {Overzier}, R.~A., {Gebhardt}, K., {et~al.} 2015, \apj, 808,
  37

\bibitem[{{Chiang} {et~al.}(2017){Chiang}, {Overzier}, {Gebhardt}, \&
  {Henriques}}]{coolchiang17}
{Chiang}, Y.-K., {Overzier}, R.~A., {Gebhardt}, K., \& {Henriques}, B. 2017,
  \apjl, 844, L23

\bibitem[{{Cicone} {et~al.}(2016){Cicone}, {Maiolino}, \& {Marconi}}]{cicone16}
{Cicone}, C., {Maiolino}, R., \& {Marconi}, A. 2016, \aap, 588, A41

\bibitem[{{Cirasuolo} {et~al.}(2014){Cirasuolo}, {Afonso}, {Carollo}, {Flores},
  {Maiolino}, {Oliva}, {Paltani}, {Vanzi}, {Evans}, {Abreu}, {Atkinson},
  {Babusiaux}, {Beard}, {Bauer}, {Bellazzini}, {Bender}, {Best}, {Bezawada},
  {Bonifacio}, {Bragaglia}, {Bryson}, {Busher}, {Cabral}, {Caputi}, {Centrone},
  {Chemla}, {Cimatti}, {Cioni}, {Clementini}, {Coelho}, {Crnojevic}, {Daddi},
  {Dunlop}, {Eales}, {Feltzing}, {Ferguson}, {Fisher}, {Fontana}, {Fynbo},
  {Garilli}, {Gilmore}, {Glauser}, {Guinouard}, {Hammer}, {Hastings}, {Hess},
  {Ivison}, {Jagourel}, {Jarvis}, {Kaper}, {Kauffman}, {Kitching}, {Lawrence},
  {Lee}, {Lemasle}, {Licausi}, {Lilly}, {Lorenzetti}, {Lunney}, {Maiolino},
  {Mannucci}, {McLure}, {Minniti}, {Montgomery}, {Muschielok}, {Nandra},
  {Navarro}, {Norberg}, {Oliver}, {Origlia}, {Padilla}, {Peacock}, {Pedichini},
  {Peng}, {Pentericci}, {Pragt}, {Puech}, {Randich}, {Rees}, {Renzini}, {Ryde},
  {Rodrigues}, {Roseboom}, {Royer}, {Saglia}, {Sanchez}, {Schiavon},
  {Schnetler}, {Sobral}, {Speziali}, {Sun}, {Stuik}, {Taylor}, {Taylor},
  {Todd}, {Tolstoy}, {Torres}, {Tosi}, {Vanzella}, {Venema}, {Vitali},
  {Wegner}, {Wells}, {Wild}, {Wright}, {Zamorani}, \& {Zoccali}}]{cirasuolo14}
{Cirasuolo}, M., {Afonso}, J., {Carollo}, M., {et~al.} 2014, in Society of
  Photo-Optical Instrumentation Engineers (SPIE) Conference Series, Vol. 9147,
  Ground-based and Airborne Instrumentation for Astronomy V, 91470N

\bibitem[{{Coogan} {et~al.}(2018){Coogan}, {Daddi}, {Sargent}, {Strazzullo},
  {Valentino}, {Gobat}, {Magdis}, {Bethermin}, {Pannella}, {Onodera}, {Liu},
  {Cimatti}, {Dannerbauer}, {Carollo}, {Renzini}, \& {Tremou}}]{coogan18}
{Coogan}, R.~T., {Daddi}, E., {Sargent}, M.~T., {et~al.} 2018, \mnras, 479, 703

\bibitem[{{Cooper} {et~al.}(2010){Cooper}, {Coil}, {Gerke}, {Newman}, {Bundy},
  {Conselice}, {Croton}, {Davis}, {Faber}, {Guhathakurta}, {Koo}, {Lin},
  {Weiner}, {Willmer}, \& {Yan}}]{mcoopz10}
{Cooper}, M.~C., {Coil}, A.~L., {Gerke}, B.~F., {et~al.} 2010, \mnras, 409, 337

\bibitem[{{Cooper} {et~al.}(2007){Cooper}, {Newman}, {Coil}, {Croton}, {Gerke},
  {Yan}, {Davis}, {Faber}, {Guhathakurta}, {Koo}, {Weiner}, \&
  {Willmer}}]{mcoopz07}
{Cooper}, M.~C., {Newman}, J.~A., {Coil}, A.~L., {et~al.} 2007, \mnras, 376,
  1445

\bibitem[{{Cooper} {et~al.}(2008){Cooper}, {Newman}, {Weiner}, {Yan},
  {Willmer}, {Bundy}, {Coil}, {Conselice}, {Davis}, {Faber}, {Gerke},
  {Guhathakurta}, {Koo}, \& {Noeske}}]{mcoopz08}
{Cooper}, M.~C., {Newman}, J.~A., {Weiner}, B.~J., {et~al.} 2008, \mnras, 383,
  1058

\bibitem[{{Cooper} {et~al.}(2012){Cooper}, {Yan}, {Dickinson}, {Juneau},
  {Lotz}, {Newman}, {Papovich}, {Salim}, {Walth}, \& {Weiner}}]{cooper12}
{Cooper}, M.~C., {Yan}, R., {Dickinson}, M., {et~al.} 2012, \mnras, 425, 2116

\bibitem[{{Croton} {et~al.}(2006){Croton}, {Springel}, {White}, {De Lucia},
  {Frenk}, {Gao}, {Jenkins}, {Kauffmann}, {Navarro}, \& {Yoshida}}]{croton06}
{Croton}, D.~J., {Springel}, V., {White}, S. D.~M., {et~al.} 2006, \mnras, 365,
  11

\bibitem[{{Cucciati} {et~al.}(2017){Cucciati}, {Davidzon}, {Bolzonella},
  {Granett}, {De Lucia}, {Branchini}, {Zamorani}, {Iovino}, {Garilli}, {Guzzo},
  {Scodeggio}, {de la Torre}, {Abbas}, {Adami}, {Arnouts}, {Bottini}, {Cappi},
  {Franzetti}, {Fritz}, {Krywult}, {Le Brun}, {Le F{\`e}vre}, {Maccagni},
  {Ma{\l}ek}, {Marulli}, {Moutard}, {Polletta}, {Pollo}, {Tasca}, {Tojeiro},
  {Vergani}, {Zanichelli}, {Bel}, {Blaizot}, {Coupon}, {Hawken}, {Ilbert},
  {Moscardini}, {Peacock}, \& {Gargiulo}}]{olga17}
{Cucciati}, O., {Davidzon}, I., {Bolzonella}, M., {et~al.} 2017, \aap, 602, A15

\bibitem[{{Cucciati} {et~al.}(2006){Cucciati}, {Iovino}, {Marinoni}, {Ilbert},
  {Bardelli}, {Franzetti}, {Le F{\`e}vre}, {Pollo}, {Zamorani}, {Cappi},
  {Guzzo}, {McCracken}, {Meneux}, {Scaramella}, {Scodeggio}, {Tresse}, {Zucca},
  {Bottini}, {Garilli}, {Le Brun}, {Maccagni}, {Picat}, {Vettolani},
  {Zanichelli}, {Adami}, {Arnaboldi}, {Arnouts}, {Bolzonella}, {Charlot},
  {Ciliegi}, {Contini}, {Foucaud}, {Gavignaud}, {Marano}, {Mazure}, {Merighi},
  {Paltani}, {Pell{\`o}}, {Pozzetti}, {Radovich}, {Bondi}, {Bongiorno},
  {Busarello}, {de la Torre}, {Gregorini}, {Lamareille}, {Mathez}, {Mellier},
  {Merluzzi}, {Ripepi}, {Rizzo}, {Temporin}, \& {Vergani}}]{olga06}
{Cucciati}, O., {Iovino}, A., {Marinoni}, C., {et~al.} 2006, \aap, 458, 39

\bibitem[{{Cucciati} {et~al.}(2018){Cucciati}, {Lemaux}, {Zamorani}, {Le
  F{\`e}vre}, {Tasca}, {Hathi}, {Lee}, {Bardelli}, {Cassata}, {Garilli}, {Le
  Brun}, {Maccagni}, {Pentericci}, {Thomas}, {Vanzella}, {Zucca}, {Lubin},
  {Amorin}, {Cassar{\`a}}, {Cimatti}, {Talia}, {Vergani}, {Koekemoer}, {Pforr},
  \& {Salvato}}]{olga18}
{Cucciati}, O., {Lemaux}, B.~C., {Zamorani}, G., {et~al.} 2018, \aap, 619, A49

\bibitem[{{Cucciati} {et~al.}(2014){Cucciati}, {Zamorani}, {Lemaux},
  {Bardelli}, {Cimatti}, {Le F{\`e}vre}, {Cassata}, {Garilli}, {Le Brun},
  {Maccagni}, {Pentericci}, {Tasca}, {Thomas}, {Vanzella}, {Zucca}, {Amorin},
  {Capak}, {Cassar{\`a}}, {Castellano}, {Cuby}, {de la Torre}, {Durkalec},
  {Fontana}, {Giavalisco}, {Grazian}, {Hathi}, {Ilbert}, {Moreau}, {Paltani},
  {Ribeiro}, {Salvato}, {Schaerer}, {Scodeggio}, {Sommariva}, {Talia},
  {Taniguchi}, {Tresse}, {Vergani}, {Wang}, {Charlot}, {Contini}, {Fotopoulou},
  {L{\'o}pez-Sanjuan}, {Mellier}, \& {Scoville}}]{olga14}
{Cucciati}, O., {Zamorani}, G., {Lemaux}, B.~C., {et~al.} 2014, \aap, 570, A16

\bibitem[{{Damen} {et~al.}(2011){Damen}, {Labb{\'e}}, {van Dokkum}, {Franx},
  {Taylor}, {Brandt}, {Dickinson}, {Gawiser}, {Illingworth}, {Kriek},
  {Marchesini}, {Muzzin}, {Papovich}, \& {Rix}}]{damen11}
{Damen}, M., {Labb{\'e}}, I., {van Dokkum}, P.~G., {et~al.} 2011, \apj, 727, 1

\bibitem[{{Darvish} {et~al.}(2015){Darvish}, {Mobasher}, {Sobral}, {Scoville},
  \& {Aragon-Calvo}}]{darvish15}
{Darvish}, B., {Mobasher}, B., {Sobral}, D., {Scoville}, N., \& {Aragon-Calvo},
  M. 2015, \apj, 805, 121

\bibitem[{{Darvish} {et~al.}(2018){Darvish}, {Scoville}, {Martin}, {Mobasher},
  {Diaz-Santos}, \& {Shen}}]{darvish18}
{Darvish}, B., {Scoville}, N.~Z., {Martin}, C., {et~al.} 2018, \apj, 860, 111

\bibitem[{{Davis} {et~al.}(2003){Davis}, {Faber}, {Newman}, {Phillips},
  {Ellis}, {Steidel}, {Conselice}, {Coil}, {Finkbeiner}, {Koo}, {Guhathakurta},
  {Weiner}, {Schiavon}, {Willmer}, {Kaiser}, {Luppino}, {Wirth}, {Connolly},
  {Eisenhardt}, {Cooper}, \& {Gerke}}]{davis03}
{Davis}, M., {Faber}, S.~M., {Newman}, J., {et~al.} 2003, in \procspie, Vol.
  4834, Discoveries and Research Prospects from 6- to 10-Meter-Class Telescopes
  II, ed. P.~{Guhathakurta}, 161--172

\bibitem[{{De Lucia} {et~al.}(2004){De Lucia}, {Kauffmann}, \&
  {White}}]{delucia04}
{De Lucia}, G., {Kauffmann}, G., \& {White}, S. D.~M. 2004, \mnras, 349, 1101

\bibitem[{{Dekel} {et~al.}(2009{\natexlab{a}}){Dekel}, {Birnboim}, {Engel},
  {Freundlich}, {Goerdt}, {Mumcuoglu}, {Neistein}, {Pichon}, {Teyssier}, \&
  {Zinger}}]{dekel09a}
{Dekel}, A., {Birnboim}, Y., {Engel}, G., {et~al.} 2009{\natexlab{a}}, \nat,
  457, 451

\bibitem[{{Dekel} {et~al.}(2009{\natexlab{b}}){Dekel}, {Sari}, \&
  {Ceverino}}]{dekel09b}
{Dekel}, A., {Sari}, R., \& {Ceverino}, D. 2009{\natexlab{b}}, \apj, 703, 785

\bibitem[{{Dey} {et~al.}(2016){Dey}, {Lee}, {Reddy}, {Cooper}, {Inami}, {Hong},
  {Gonzalez}, \& {Jannuzi}}]{dey16}
{Dey}, A., {Lee}, K.-S., {Reddy}, N., {et~al.} 2016, \apj, 823, 11

\bibitem[{{Diener} {et~al.}(2013){Diener}, {Lilly}, {Knobel}, {Zamorani},
  {Lemson}, {Kampczyk}, {Scoville}, {Carollo}, {Contini}, {Kneib}, {Le Fevre},
  {Mainieri}, {Renzini}, {Scodeggio}, {Bardelli}, {Bolzonella}, {Bongiorno},
  {Caputi}, {Cucciati}, {de la Torre}, {de Ravel}, {Franzetti}, {Garilli},
  {Iovino}, {Kova{\v c}}, {Lamareille}, {Le Borgne}, {Le Brun}, {Maier},
  {Mignoli}, {Pello}, {Peng}, {Perez Montero}, {Presotto}, {Silverman},
  {Tanaka}, {Tasca}, {Tresse}, {Vergani}, {Zucca}, {Bordoloi}, {Cappi},
  {Cimatti}, {Coppa}, {Koekemoer}, {L{\'o}pez-Sanjuan}, {McCracken}, {Moresco},
  {Nair}, {Pozzetti}, \& {Welikala}}]{dutifuldiener13}
{Diener}, C., {Lilly}, S.~J., {Knobel}, C., {et~al.} 2013, \apj, 765, 109

\bibitem[{{Diener} {et~al.}(2015){Diener}, {Lilly}, {Ledoux}, {Zamorani},
  {Bolzonella}, {Murphy}, {Capak}, {Ilbert}, \& {McCracken}}]{dutifuldiener15}
{Diener}, C., {Lilly}, S.~J., {Ledoux}, C., {et~al.} 2015, \apj, 802, 31

\bibitem[{{Durkalec} {et~al.}(2018){Durkalec}, {Le F{\`e}vre}, {Pollo},
  {Zamorani}, {Lemaux}, {Garilli}, {Bardelli}, {Hathi}, {Koekemoer}, {Pforr},
  \& {Zucca}}]{ania18}
{Durkalec}, A., {Le F{\`e}vre}, O., {Pollo}, A., {et~al.} 2018, \aap, 612, A42

\bibitem[{{Elbaz} {et~al.}(2007){Elbaz}, {Daddi}, {Le Borgne}, {Dickinson},
  {Alexander}, {Chary}, {Starck}, {Brand t}, {Kitzbichler}, {MacDonald},
  {Nonino}, {Popesso}, {Stern}, \& {Vanzella}}]{elbaz07}
{Elbaz}, D., {Daddi}, E., {Le Borgne}, D., {et~al.} 2007, \aap, 468, 33

\bibitem[{{Faber} {et~al.}(2003){Faber}, {Phillips}, {Kibrick}, {Alcott},
  {Allen}, {Burrous}, {Cantrall}, {Clarke}, {Coil}, {Cowley}, {Davis}, {Deich},
  {Dietsch}, {Gilmore}, {Harper}, {Hilyard}, {Lewis}, {McVeigh}, {Newman},
  {Osborne}, {Schiavon}, {Stover}, {Tucker}, {Wallace}, {Wei}, {Wirth}, \&
  {Wright}}]{fab03}
{Faber}, S.~M., {Phillips}, A.~C., {Kibrick}, R.~I., {et~al.} 2003, in Society
  of Photo-Optical Instrumentation Engineers (SPIE) Conference Series, Vol.
  4841, Instrument Design and Performance for Optical/Infrared Ground-based
  Telescopes, ed. M.~{Iye} \& A.~F.~M. {Moorwood}, 1657--1669

\bibitem[{{Faisst} {et~al.}(2020){Faisst}, {Schaerer}, {Lemaux}, {Oesch},
  {Fudamoto}, {Cassata}, {B{\'e}thermin}, {Capak}, {Le F{\`e}vre}, {Silverman},
  {Yan}, {Ginolfi}, {Koekemoer}, {Morselli}, {Amor{\'\i}n}, {Bardelli},
  {Boquien}, {Brammer}, {Cimatti}, {Dessauges-Zavadsky}, {Fujimoto},
  {Gruppioni}, {Hathi}, {Hemmati}, {Ibar}, {Jones}, {Khusanova}, {Loiacono},
  {Pozzi}, {Talia}, {Tasca}, {Riechers}, {Rodighiero}, {Romano}, {Scoville},
  {Toft}, {Vallini}, {Vergani}, {Zamorani}, \& {Zucca}}]{dirtyAF20}
{Faisst}, A.~L., {Schaerer}, D., {Lemaux}, B.~C., {et~al.} 2020, \apjs, 247, 61

\bibitem[{{Fazio} {et~al.}(2004){Fazio}, {Hora}, {Allen}, {Ashby}, {Barmby},
  {Deutsch}, {Huang}, {Kleiner}, {Marengo}, {Megeath}, {Melnick}, {Pahre},
  {Patten}, {Polizotti}, {Smith}, {Taylor}, {Wang}, {Willner}, {Hoffmann},
  {Pipher}, {Forrest}, {McMurty}, {McCreight}, {McKelvey}, {McMurray}, {Koch},
  {Moseley}, {Arendt}, {Mentzell}, {Marx}, {Losch}, {Mayman}, {Eichhorn},
  {Krebs}, {Jhabvala}, {Gezari}, {Fixsen}, {Flores}, {Shakoorzadeh}, {Jungo},
  {Hakun}, {Workman}, {Karpati}, {Kichak}, {Whitley}, {Mann}, {Tollestrup},
  {Eisenhardt}, {Stern}, {Gorjian}, {Bhattacharya}, {Carey}, {Nelson},
  {Glaccum}, {Lacy}, {Lowrance}, {Laine}, {Reach}, {Stauffer}, {Surace},
  {Wilson}, {Wright}, {Hoffman}, {Domingo}, \& {Cohen}}]{fazio04}
{Fazio}, G.~G., {Hora}, J.~L., {Allen}, L.~E., {et~al.} 2004, \apjs, 154, 10

\bibitem[{{Foltz} {et~al.}(2018){Foltz}, {Wilson}, {Muzzin}, {Cooper},
  {Nantais}, {van der Burg}, {Cerulo}, {Chan}, {Fillingham}, {Surace}, {Webb},
  {Noble}, {Lacy}, {McDonald}, {Rudnick}, {Lidman}, {Demarco},
  {Hlavacek-Larrondo}, {Yee}, {Perlmutter}, \& {Hayden}}]{foltz18}
{Foltz}, R., {Wilson}, G., {Muzzin}, A., {et~al.} 2018, \apj, 866, 136

\bibitem[{{Forrest} {et~al.}(2020){Forrest}, {Marsan}, {Annunziatella},
  {Wilson}, {Muzzin}, {Marchesini}, {Cooper}, {Chan}, {McConachie}, {Gomez},
  {Kado-Fong}, {Barbera}, {Lange-Vagle}, {Nantais}, {Nonino}, {Saracco},
  {Stefanon}, \& {van der Burg}}]{forrest20}
{Forrest}, B., {Marsan}, Z.~C., {Annunziatella}, M., {et~al.} 2020, \apj, 903,
  47

\bibitem[{{Fukugita} {et~al.}(1996){Fukugita}, {Ichikawa}, {Gunn}, {Doi},
  {Shimasaku}, \& {Schneider}}]{fukugita96}
{Fukugita}, M., {Ichikawa}, T., {Gunn}, J.~E., {et~al.} 1996, \aj, 111, 1748

\bibitem[{{Fumagalli} {et~al.}(2016){Fumagalli}, {Franx}, {van Dokkum},
  {Whitaker}, {Skelton}, {Brammer}, {Nelson}, {Maseda}, {Momcheva}, {Kriek},
  {Labb{\'e}}, {Lundgren}, \& {Rix}}]{fumagalli16}
{Fumagalli}, M., {Franx}, M., {van Dokkum}, P., {et~al.} 2016, \apj, 822, 1

\bibitem[{{Gardner} {et~al.}(2006){Gardner}, {Mather}, {Clampin}, {Doyon},
  {Greenhouse}, {Hammel}, {Hutchings}, {Jakobsen}, {Lilly}, {Long}, {Lunine},
  {McCaughrean}, {Mountain}, {Nella}, {Rieke}, {Rieke}, {Rix}, {Smith},
  {Sonneborn}, {Stiavelli}, {Stockman}, {Windhorst}, \& {Wright}}]{gardner06}
{Gardner}, J.~P., {Mather}, J.~C., {Clampin}, M., {et~al.} 2006, \ssr, 123, 485

\bibitem[{{Gavazzi} {et~al.}(2018){Gavazzi}, {Consolandi}, {Gutierrez},
  {Boselli}, \& {Yoshida}}]{gavazzi18}
{Gavazzi}, G., {Consolandi}, G., {Gutierrez}, M.~L., {Boselli}, A., \&
  {Yoshida}, M. 2018, \aap, 618, A130

\bibitem[{{Gawiser} {et~al.}(2006){Gawiser}, {van Dokkum}, {Herrera}, {Maza},
  {Castander}, {Infante}, {Lira}, {Quadri}, {Toner}, {Treister}, {Urry},
  {Altmann}, {Assef}, {Christlein}, {Coppi}, {Dur{\'a}n}, {Franx}, {Galaz},
  {Huerta}, {Liu}, {L{\'o}pez}, {M{\'e}ndez}, {Moore}, {Rubio}, {Ruiz}, {Toft},
  \& {Yi}}]{fukinericgawiser06}
{Gawiser}, E., {van Dokkum}, P.~G., {Herrera}, D., {et~al.} 2006, \apjs, 162, 1

\bibitem[{{Genzel} {et~al.}(2015){Genzel}, {Tacconi}, {Lutz}, {Saintonge},
  {Berta}, {Magnelli}, {Combes}, {Garc{\'\i}a-Burillo}, {Neri}, {Bolatto},
  {Contini}, {Lilly}, {Boissier}, {Boone}, {Bouch{\'e}}, {Bournaud}, {Burkert},
  {Carollo}, {Colina}, {Cooper}, {Cox}, {Feruglio}, {F{\"o}rster Schreiber},
  {Freundlich}, {Gracia-Carpio}, {Juneau}, {Kovac}, {Lippa}, {Naab}, {Salome},
  {Renzini}, {Sternberg}, {Walter}, {Weiner}, {Weiss}, \& {Wuyts}}]{genzel15}
{Genzel}, R., {Tacconi}, L.~J., {Lutz}, D., {et~al.} 2015, \apj, 800, 20

\bibitem[{{Ginolfi} {et~al.}(2020{\natexlab{a}}){Ginolfi}, {Jones},
  {B{\'e}thermin}, {Faisst}, {Lemaux}, {Schaerer}, {Fudamoto}, {Oesch},
  {Dessauges-Zavadsky}, {Fujimoto}, {Carniani}, {Le F{\`e}vre}, {Cassata},
  {Silverman}, {Capak}, {Yan}, {Bardelli}, {Cucciati}, {Gal}, {Gruppioni},
  {Hathi}, {Lubin}, {Maiolino}, {Morselli}, {Pelliccia}, {Talia}, {Vergani}, \&
  {Zamorani}}]{ginolfi20b}
{Ginolfi}, M., {Jones}, G.~C., {B{\'e}thermin}, M., {et~al.}
  2020{\natexlab{a}}, \aap, 643, A7

\bibitem[{{Ginolfi} {et~al.}(2020{\natexlab{b}}){Ginolfi}, {Jones},
  {B{\'e}thermin}, {Fudamoto}, {Loiacono}, {Fujimoto}, {Le F{\'e}vre},
  {Faisst}, {Schaerer}, {Cassata}, {Silverman}, {Yan}, {Capak}, {Bardelli},
  {Boquien}, {Carraro}, {Dessauges-Zavadsky}, {Giavalisco}, {Gruppioni},
  {Ibar}, {Khusanova}, {Lemaux}, {Maiolino}, {Narayanan}, {Oesch}, {Pozzi},
  {Rodighiero}, {Talia}, {Toft}, {Vallini}, {Vergani}, \&
  {Zamorani}}]{ginolfi20a}
{Ginolfi}, M., {Jones}, G.~C., {B{\'e}thermin}, M., {et~al.}
  2020{\natexlab{b}}, \aap, 633, A90

\bibitem[{{Gobat} {et~al.}(2011){Gobat}, {Daddi}, {Onodera}, {Finoguenov},
  {Renzini}, {Arimoto}, {Bouwens}, {Brusa}, {Chary}, {Cimatti}, {Dickinson},
  {Kong}, \& {Mignoli}}]{gobat11}
{Gobat}, R., {Daddi}, E., {Onodera}, M., {et~al.} 2011, \aap, 526, A133

\bibitem[{{G{\'o}mez} {et~al.}(2003){G{\'o}mez}, {Nichol}, {Miller}, {Balogh},
  {Goto}, {Zabludoff}, {Romer}, {Bernardi}, {Sheth}, {Hopkins}, {Castander},
  {Connolly}, {Schneider}, {Brinkmann}, {Lamb}, {SubbaRao}, \&
  {York}}]{gomez03}
{G{\'o}mez}, P.~L., {Nichol}, R.~C., {Miller}, C.~J., {et~al.} 2003, \apj, 584,
  210

\bibitem[{{G{\'o}mez-Guijarro} {et~al.}(2019){G{\'o}mez-Guijarro}, {Riechers},
  {Pavesi}, {Magdis}, {Leung}, {Valentino}, {Toft}, {Aravena}, {Chapman},
  {Clements}, {Dannerbauer}, {Oliver}, {P{\'e}rez-Fournon}, \&
  {Valtchanov}}]{gg19}
{G{\'o}mez-Guijarro}, C., {Riechers}, D.~A., {Pavesi}, R., {et~al.} 2019, \apj,
  872, 117

\bibitem[{{Greenslade} {et~al.}(2018){Greenslade}, {Clements}, {Cheng}, {De
  Zotti}, {Scott}, {Valiante}, {Eales}, {Bremer}, {Dannerbauer}, {Birkinshaw},
  {Farrah}, {Harrison}, {Micha{\l}owski}, {Valtchanov}, {Oteo}, {Baes},
  {Cooray}, {Negrello}, {Wang}, {van der Werf}, {Dunne}, \&
  {Dye}}]{greenslade18}
{Greenslade}, J., {Clements}, D.~L., {Cheng}, T., {et~al.} 2018, \mnras, 476,
  3336

\bibitem[{{Grogin} {et~al.}(2011){Grogin}, {Kocevski}, {Faber}, {Ferguson},
  {Koekemoer}, {Riess}, {Acquaviva}, {Alexander}, {Almaini}, {Ashby}, {Barden},
  {Bell}, {Bournaud}, {Brown}, {Caputi}, {Casertano}, {Cassata}, {Castellano},
  {Challis}, {Chary}, {Cheung}, {Cirasuolo}, {Conselice}, {Roshan Cooray},
  {Croton}, {Daddi}, {Dahlen}, {Dav{\'e}}, {de Mello}, {Dekel}, {Dickinson},
  {Dolch}, {Donley}, {Dunlop}, {Dutton}, {Elbaz}, {Fazio}, {Filippenko},
  {Finkelstein}, {Fontana}, {Gardner}, {Garnavich}, {Gawiser}, {Giavalisco},
  {Grazian}, {Guo}, {Hathi}, {H{\"a}ussler}, {Hopkins}, {Huang}, {Huang},
  {Jha}, {Kartaltepe}, {Kirshner}, {Koo}, {Lai}, {Lee}, {Li}, {Lotz}, {Lucas},
  {Madau}, {McCarthy}, {McGrath}, {McIntosh}, {McLure}, {Mobasher},
  {Moustakas}, {Mozena}, {Nandra}, {Newman}, {Niemi}, {Noeske}, {Papovich},
  {Pentericci}, {Pope}, {Primack}, {Rajan}, {Ravindranath}, {Reddy}, {Renzini},
  {Rix}, {Robaina}, {Rodney}, {Rosario}, {Rosati}, {Salimbeni}, {Scarlata},
  {Siana}, {Simard}, {Smidt}, {Somerville}, {Spinrad}, {Straughn}, {Strolger},
  {Telford}, {Teplitz}, {Trump}, {van der Wel}, {Villforth}, {Wechsler},
  {Weiner}, {Wiklind}, {Wild}, {Wilson}, {Wuyts}, {Yan}, \& {Yun}}]{grogin11}
{Grogin}, N.~A., {Kocevski}, D.~D., {Faber}, S.~M., {et~al.} 2011, \apjs, 197,
  35

\bibitem[{{Gruppioni} {et~al.}(2020){Gruppioni}, {B{\'e}thermin}, {Loiacono},
  {Le F{\`e}vre}, {Capak}, {Cassata}, {Faisst}, {Schaerer}, {Silverman}, {Yan},
  {Bardelli}, {Boquien}, {Carraro}, {Cimatti}, {Dessauges-Zavadsky}, {Ginolfi},
  {Fujimoto}, {Hathi}, {Jones}, {Khusanova}, {Koekemoer}, {Lagache}, {Lemaux},
  {Oesch}, {Pozzi}, {Riechers}, {Rodighiero}, {Romano}, {Talia}, {Vallini},
  {Vergani}, {Zamorani}, \& {Zucca}}]{gruppioni20}
{Gruppioni}, C., {B{\'e}thermin}, M., {Loiacono}, F., {et~al.} 2020, \aap, 643,
  A8

\bibitem[{{Guaita} {et~al.}(2020){Guaita}, {Pompei}, {Castellano},
  {Pentericci}, {Cucciati}, {Zamorani}, {Zoldan}, {Fontanot}, {Bauer},
  {Amorin}, {Bolzonella}, {de Lucia}, {Gargiulo}, {Hathi}, {Hibon},
  {Hirschmann}, {Koekemoer}, {McLure}, {Pozzetti}, {Talia}, {Thomas}, \&
  {Xie}}]{loopylucia20}
{Guaita}, L., {Pompei}, E., {Castellano}, M., {et~al.} 2020, \aap, 640, A107

\bibitem[{{Hansen} {et~al.}(2009){Hansen}, {Sheldon}, {Wechsler}, \&
  {Koester}}]{hansen09}
{Hansen}, S.~M., {Sheldon}, E.~S., {Wechsler}, R.~H., \& {Koester}, B.~P. 2009,
  \apj, 699, 1333

\bibitem[{{Hatch} {et~al.}(2014){Hatch}, {Wylezalek}, {Kurk}, {Stern}, {De
  Breuck}, {Jarvis}, {Galametz}, {Gonzalez}, {Hartley}, {Mortlock}, {Seymour},
  \& {Stevens}}]{hatch14}
{Hatch}, N.~A., {Wylezalek}, D., {Kurk}, J.~D., {et~al.} 2014, \mnras, 445, 280

\bibitem[{{Hathi} {et~al.}(2009){Hathi}, {Ferreras}, {Pasquali}, {Malhotra},
  {Rhoads}, {Pirzkal}, {Windhorst}, \& {Xu}}]{nimish09}
{Hathi}, N.~P., {Ferreras}, I., {Pasquali}, A., {et~al.} 2009, \apj, 690, 1866

\bibitem[{{Heckman} {et~al.}(2015){Heckman}, {Alexandroff}, {Borthakur},
  {Overzier}, \& {Leitherer}}]{heckman15}
{Heckman}, T.~M., {Alexandroff}, R.~M., {Borthakur}, S., {Overzier}, R., \&
  {Leitherer}, C. 2015, \apj, 809, 147

\bibitem[{{Hildebrandt} {et~al.}(2006){Hildebrandt}, {Erben}, {Dietrich},
  {Cordes}, {Haberzettl}, {Hetterscheidt}, {Schirmer}, {Schmithuesen},
  {Schneider}, {Simon}, \& {Trachternach}}]{hil06}
{Hildebrandt}, H., {Erben}, T., {Dietrich}, J.~P., {et~al.} 2006, \aap, 452,
  1121

\bibitem[{{Hill} {et~al.}(2021){Hill}, {Chapman}, {Phadke}, {Aravena},
  {Archipley}, {Bethermin}, {Canning}, {Gonzalez}, {Greve}, {Gururajan},
  {Hayward}, {Hezaveh}, {Jarugula}, {Marrone}, {Miller}, {Reuter}, {Rotermund},
  {Scott}, {Spilker}, {Vieira}, {Wang}, \& {Weiss}}]{ryley21}
{Hill}, R., {Chapman}, S., {Phadke}, K.~A., {et~al.} 2021, arXiv e-prints,
  arXiv:2109.04534, MNRAS, submitted

\bibitem[{{Hill} {et~al.}(2020){Hill}, {Chapman}, {Scott}, {Apostolovski},
  {Aravena}, {B{\'e}thermin}, {Bradford}, {Canning}, {De Breuck}, {Dong},
  {Gonzalez}, {Greve}, {Hayward}, {Hezaveh}, {Litke}, {Malkan}, {Marrone},
  {Phadke}, {Reuter}, {Rotermund}, {Spilker}, {Vieira}, \&
  {Wei{\ss}}}]{ryley20}
{Hill}, R., {Chapman}, S., {Scott}, D., {et~al.} 2020, \mnras, 495, 3124

\bibitem[{{Hilton} {et~al.}(2009){Hilton}, {Stanford}, {Stott}, {Collins},
  {Hoyle}, {Davidson}, {Hosmer}, {Kay}, {Liddle}, {Lloyd-Davies}, {Mann},
  {Mehrtens}, {Miller}, {Nichol}, {Romer}, {Sabirli}, {Sahl{\'e}n}, {Viana},
  {West}, {Barbary}, {Dawson}, {Meyers}, {Perlmutter}, {Rubin}, \&
  {Suzuki}}]{hilton09}
{Hilton}, M., {Stanford}, S.~A., {Stott}, J.~P., {et~al.} 2009, \apj, 697, 436

\bibitem[{{Hopkins} {et~al.}(2008){Hopkins}, {Hernquist}, {Cox}, \& {Kere{\v
  s}}}]{hopkins08}
{Hopkins}, P.~F., {Hernquist}, L., {Cox}, T.~J., \& {Kere{\v s}}, D. 2008,
  \apjs, 175, 356

\bibitem[{{Hung} {et~al.}(2020){Hung}, {Lemaux}, {Gal}, {Tomczak}, {Lubin},
  {Cucciati}, {Pelliccia}, {Shen}, {Le F{\`e}vre}, {Wu}, {Kocevski}, {Mei}, \&
  {Squires}}]{denise20}
{Hung}, D., {Lemaux}, B.~C., {Gal}, R.~R., {et~al.} 2020, \mnras, 491, 5524

\bibitem[{{Hung} {et~al.}(2021){Hung}, {Lemaux}, {Gal}, {Tomczak}, {Lubin},
  {Cucciati}, {Pelliccia}, {Shen}, {Le F{\`e}vre}, {Zamorani}, {Wu},
  {Kocevski}, {Fassnacht}, \& {Squires}}]{denise21}
{Hung}, D., {Lemaux}, B.~C., {Gal}, R.~R., {et~al.} 2021, \mnras, 502, 3942

\bibitem[{{Hwang} {et~al.}(2019){Hwang}, {Shin}, \& {Song}}]{hwang19}
{Hwang}, H.~S., {Shin}, J., \& {Song}, H. 2019, \mnras, 489, 339

\bibitem[{{Ilbert} {et~al.}(2006){Ilbert}, {Arnouts}, {McCracken},
  {Bolzonella}, {Bertin}, {Le F{\`e}vre}, {Mellier}, {Zamorani}, {Pell{\`o}},
  {Iovino}, {Tresse}, {Le Brun}, {Bottini}, {Garilli}, {Maccagni}, {Picat},
  {Scaramella}, {Scodeggio}, {Vettolani}, {Zanichelli}, {Adami}, {Bardelli},
  {Cappi}, {Charlot}, {Ciliegi}, {Contini}, {Cucciati}, {Foucaud}, {Franzetti},
  {Gavignaud}, {Guzzo}, {Marano}, {Marinoni}, {Mazure}, {Meneux}, {Merighi},
  {Paltani}, {Pollo}, {Pozzetti}, {Radovich}, {Zucca}, {Bondi}, {Bongiorno},
  {Busarello}, {de La Torre}, {Gregorini}, {Lamareille}, {Mathez}, {Merluzzi},
  {Ripepi}, {Rizzo}, \& {Vergani}}]{dreadolivier06}
{Ilbert}, O., {Arnouts}, S., {McCracken}, H.~J., {et~al.} 2006, \aap, 457, 841

\bibitem[{{Ilbert} {et~al.}(2009){Ilbert}, {Capak}, {Salvato}, {Aussel},
  {McCracken}, {Sanders}, {Scoville}, {Kartaltepe}, {Arnouts}, {Le Floc'h},
  {Mobasher}, {Taniguchi}, {Lamareille}, {Leauthaud}, {Sasaki}, {Thompson},
  {Zamojski}, {Zamorani}, {Bardelli}, {Bolzonella}, {Bongiorno}, {Brusa},
  {Caputi}, {Carollo}, {Contini}, {Cook}, {Coppa}, {Cucciati}, {de la Torre},
  {de Ravel}, {Franzetti}, {Garilli}, {Hasinger}, {Iovino}, {Kampczyk},
  {Kneib}, {Knobel}, {Kovac}, {Le Borgne}, {Le Brun}, {F{\`e}vre}, {Lilly},
  {Looper}, {Maier}, {Mainieri}, {Mellier}, {Mignoli}, {Murayama}, {Pell{\`o}},
  {Peng}, {P{\'e}rez-Montero}, {Renzini}, {Ricciardelli}, {Schiminovich},
  {Scodeggio}, {Shioya}, {Silverman}, {Surace}, {Tanaka}, {Tasca}, {Tresse},
  {Vergani}, \& {Zucca}}]{dreadolivier09}
{Ilbert}, O., {Capak}, P., {Salvato}, M., {et~al.} 2009, \apj, 690, 1236

\bibitem[{{Ilbert} {et~al.}(2013){Ilbert}, {McCracken}, {Le F{\`e}vre},
  {Capak}, {Dunlop}, {Karim}, {Renzini}, {Caputi}, {Boissier}, {Arnouts},
  {Aussel}, {Comparat}, {Guo}, {Hudelot}, {Kartaltepe}, {Kneib}, {Krogager},
  {Le Floc'h}, {Lilly}, {Mellier}, {Milvang-Jensen}, {Moutard}, {Onodera},
  {Richard}, {Salvato}, {Sanders}, {Scoville}, {Silverman}, {Taniguchi},
  {Tasca}, {Thomas}, {Toft}, {Tresse}, {Vergani}, {Wolk}, \&
  {Zirm}}]{dreadolivier13}
{Ilbert}, O., {McCracken}, H.~J., {Le F{\`e}vre}, O., {et~al.} 2013, \aap, 556,
  A55

\bibitem[{{Ito} {et~al.}(2020){Ito}, {Kashikawa}, {Toshikawa}, {Overzier},
  {Kubo}, {Uchiyama}, {Liang}, {Onoue}, {Tanaka}, {Komiyama}, {Lee}, {Lin},
  {Marinello}, {Martin}, \& {Shibuya}}]{ito20}
{Ito}, K., {Kashikawa}, N., {Toshikawa}, J., {et~al.} 2020, \apj, 899, 5

\bibitem[{{Jones} {et~al.}(2020){Jones}, {B{\'e}thermin}, {Fudamoto},
  {Ginolfi}, {Capak}, {Cassata}, {Faisst}, {Le F{\`e}vre}, {Schaerer},
  {Silverman}, {Yan}, {Bardelli}, {Boquien}, {Cimatti}, {Dessauges-Zavadsky},
  {Giavalisco}, {Gruppioni}, {Ibar}, {Khusanova}, {Koekemoer}, {Lemaux},
  {Loiacono}, {Maiolino}, {Oesch}, {Pozzi}, {Riechers}, {Rodighiero}, {Talia},
  {Vallini}, {Vergani}, {Zamorani}, \& {Zucca}}]{jones19}
{Jones}, G.~C., {B{\'e}thermin}, M., {Fudamoto}, Y., {et~al.} 2020, \mnras,
  491, L18

\bibitem[{{Juneau} {et~al.}(2013){Juneau}, {Dickinson}, {Bournaud},
  {Alexander}, {Daddi}, {Mullaney}, {Magnelli}, {Kartaltepe}, {Hwang},
  {Willner}, {Coil}, {Rosario}, {Trump}, {Weiner}, {Willmer}, {Cooper},
  {Elbaz}, {Faber}, {Frayer}, {Kocevski}, {Laird}, {Monkiewicz}, {Nandra},
  {Newman}, {Salim}, \& {Symeonidis}}]{juneau13}
{Juneau}, S., {Dickinson}, M., {Bournaud}, F., {et~al.} 2013, \apj, 764, 176

\bibitem[{{Kitzbichler} \& {White}(2007)}]{kdubz07}
{Kitzbichler}, M.~G. \& {White}, S.~D.~M. 2007, \mnras, 376, 2

\bibitem[{{Kodama} {et~al.}(2007){Kodama}, {Tanaka}, {Kajisawa}, {Kurk},
  {Venemans}, {De Breuck}, {Vernet}, \& {Lidman}}]{kodama07}
{Kodama}, T., {Tanaka}, I., {Kajisawa}, M., {et~al.} 2007, \mnras, 377, 1717

\bibitem[{{Koekemoer} {et~al.}(2011){Koekemoer}, {Faber}, {Ferguson}, {Grogin},
  {Kocevski}, {Koo}, {Lai}, {Lotz}, {Lucas}, {McGrath}, {Ogaz}, {Rajan},
  {Riess}, {Rodney}, {Strolger}, {Casertano}, {Castellano}, {Dahlen},
  {Dickinson}, {Dolch}, {Fontana}, {Giavalisco}, {Grazian}, {Guo}, {Hathi},
  {Huang}, {van der Wel}, {Yan}, {Acquaviva}, {Alexander}, {Almaini}, {Ashby},
  {Barden}, {Bell}, {Bournaud}, {Brown}, {Caputi}, {Cassata}, {Challis},
  {Chary}, {Cheung}, {Cirasuolo}, {Conselice}, {Roshan Cooray}, {Croton},
  {Daddi}, {Dav{\'e}}, {de Mello}, {de Ravel}, {Dekel}, {Donley}, {Dunlop},
  {Dutton}, {Elbaz}, {Fazio}, {Filippenko}, {Finkelstein}, {Frazer}, {Gardner},
  {Garnavich}, {Gawiser}, {Gruetzbauch}, {Hartley}, {H{\"a}ussler},
  {Herrington}, {Hopkins}, {Huang}, {Jha}, {Johnson}, {Kartaltepe},
  {Khostovan}, {Kirshner}, {Lani}, {Lee}, {Li}, {Madau}, {McCarthy},
  {McIntosh}, {McLure}, {McPartland}, {Mobasher}, {Moreira}, {Mortlock},
  {Moustakas}, {Mozena}, {Nandra}, {Newman}, {Nielsen}, {Niemi}, {Noeske},
  {Papovich}, {Pentericci}, {Pope}, {Primack}, {Ravindranath}, {Reddy},
  {Renzini}, {Rix}, {Robaina}, {Rosario}, {Rosati}, {Salimbeni}, {Scarlata},
  {Siana}, {Simard}, {Smidt}, {Snyder}, {Somerville}, {Spinrad}, {Straughn},
  {Telford}, {Teplitz}, {Trump}, {Vargas}, {Villforth}, {Wagner}, {Wandro},
  {Wechsler}, {Weiner}, {Wiklind}, {Wild}, {Wilson}, {Wuyts}, \&
  {Yun}}]{anton11}
{Koekemoer}, A.~M., {Faber}, S.~M., {Ferguson}, H.~C., {et~al.} 2011, \apjs,
  197, 36

\bibitem[{{Kova{\v c}} {et~al.}(2014){Kova{\v c}}, {Lilly}, {Knobel},
  {Bschorr}, {Peng}, {Carollo}, {Contini}, {Kneib}, {Le F{\'e}vre}, {Mainieri},
  {Renzini}, {Scodeggio}, {Zamorani}, {Bardelli}, {Bolzonella}, {Bongiorno},
  {Caputi}, {Cucciati}, {de la Torre}, {de Ravel}, {Franzetti}, {Garilli},
  {Iovino}, {Kampczyk}, {Lamareille}, {Le Borgne}, {Le Brun}, {Maier},
  {Mignoli}, {Oesch}, {Pello}, {Montero}, {Presotto}, {Silverman}, {Tanaka},
  {Tasca}, {Tresse}, {Vergani}, {Zucca}, {Aussel}, {Koekemoer}, {Le Floc'h},
  {Moresco}, \& {Pozzetti}}]{kovac14}
{Kova{\v c}}, K., {Lilly}, S.~J., {Knobel}, C., {et~al.} 2014, \mnras, 438, 717

\bibitem[{{Kubo} {et~al.}(2013){Kubo}, {Uchimoto}, {Yamada}, {Kajisawa},
  {Ichikawa}, {Matsuda}, {Akiyama}, {Hayashino}, {Konishi}, {Nishimura},
  {Omata}, {Suzuki}, {Tanaka}, {Yoshikawa}, {Alexander}, {Fazio}, {Huang}, \&
  {Lehmer}}]{kubo13}
{Kubo}, M., {Uchimoto}, Y.~K., {Yamada}, T., {et~al.} 2013, \apj, 778, 170

\bibitem[{{Kurk} {et~al.}(2013){Kurk}, {Cimatti}, {Daddi}, {Mignoli},
  {Pozzetti}, {Dickinson}, {Bolzonella}, {Zamorani}, {Cassata}, \&
  {Rodighiero}}]{kurk13}
{Kurk}, J., {Cimatti}, A., {Daddi}, E., {et~al.} 2013, \aap, 549, A63

\bibitem[{{Laigle} {et~al.}(2016){Laigle}, {McCracken}, {Ilbert}, {Hsieh},
  {Davidzon}, {Capak}, {Hasinger}, {Silverman}, {Pichon}, {Coupon}, {Aussel},
  {Le Borgne}, {Caputi}, {Cassata}, {Chang}, {Civano}, {Dunlop}, {Fynbo},
  {Kartaltepe}, {Koekemoer}, {Le F{\`e}vre}, {Le Floc'h}, {Leauthaud}, {Lilly},
  {Lin}, {Marchesi}, {Milvang-Jensen}, {Salvato}, {Sanders}, {Scoville},
  {Smolcic}, {Stockmann}, {Taniguchi}, {Tasca}, {Toft}, {Vaccari}, \&
  {Zabl}}]{laigle16}
{Laigle}, C., {McCracken}, H.~J., {Ilbert}, O., {et~al.} 2016, \apjs, 224, 24

\bibitem[{{Le F{\`e}vre} {et~al.}(2020){Le F{\`e}vre}, {B{\'e}thermin},
  {Faisst}, {Jones}, {Capak}, {Cassata}, {Silverman}, {Schaerer}, {Yan},
  {Amorin}, {Bardelli}, {Boquien}, {Cimatti}, {Dessauges-Zavadsky},
  {Giavalisco}, {Hathi}, {Fudamoto}, {Fujimoto}, {Ginolfi}, {Gruppioni},
  {Hemmati}, {Ibar}, {Koekemoer}, {Khusanova}, {Lagache}, {Lemaux}, {Loiacono},
  {Maiolino}, {Mancini}, {Narayanan}, {Morselli}, {M{\'e}ndez-Hern{\`a}ndez},
  {Oesch}, {Pozzi}, {Romano}, {Riechers}, {Scoville}, {Talia}, {Tasca},
  {Thomas}, {Toft}, {Vallini}, {Vergani}, {Walter}, {Zamorani}, \&
  {Zucca}}]{dong20}
{Le F{\`e}vre}, O., {B{\'e}thermin}, M., {Faisst}, A., {et~al.} 2020, \aap,
  643, A1

\bibitem[{{Le F{\`e}vre} {et~al.}(2013){Le F{\`e}vre}, {Cassata}, {Cucciati},
  {Garilli}, {Ilbert}, {Le Brun}, {Maccagni}, {Moreau}, {Scodeggio}, {Tresse},
  {Zamorani}, {Adami}, {Arnouts}, {Bardelli}, {Bolzonella}, {Bondi},
  {Bongiorno}, {Bottini}, {Cappi}, {Charlot}, {Ciliegi}, {Contini}, {de la
  Torre}, {Foucaud}, {Franzetti}, {Gavignaud}, {Guzzo}, {Iovino}, {Lemaux},
  {L{\'o}pez-Sanjuan}, {McCracken}, {Marano}, {Marinoni}, {Mazure}, {Mellier},
  {Merighi}, {Merluzzi}, {Paltani}, {Pell{\`o}}, {Pollo}, {Pozzetti},
  {Scaramella}, {Tasca}, {Vergani}, {Vettolani}, {Zanichelli}, \&
  {Zucca}}]{dong13}
{Le F{\`e}vre}, O., {Cassata}, P., {Cucciati}, O., {et~al.} 2013, \aap, 559,
  A14

\bibitem[{{Le F{\`e}vre} {et~al.}(2003){Le F{\`e}vre}, {Saisse}, {Mancini},
  {Brau-Nogue}, {Caputi}, {Castinel}, {D'Odorico}, {Garilli}, {Kissler-Patig},
  {Lucuix}, {Mancini}, {Pauget}, {Sciarretta}, {Scodeggio}, {Tresse}, \&
  {Vettolani}}]{dong03}
{Le F{\`e}vre}, O., {Saisse}, M., {Mancini}, D., {et~al.} 2003, in \procspie,
  Vol. 4841, Instrument Design and Performance for Optical/Infrared
  Ground-based Telescopes, ed. M.~{Iye} \& A.~F.~M. {Moorwood}, 1670--1681

\bibitem[{{Le F{\`e}vre} {et~al.}(2015){Le F{\`e}vre}, {Tasca}, {Cassata},
  {Garilli}, {Le Brun}, {Maccagni}, {Pentericci}, {Thomas}, {Vanzella},
  {Zamorani}, {Zucca}, {Amorin}, {Bardelli}, {Capak}, {Cassar{\`a}},
  {Castellano}, {Cimatti}, {Cuby}, {Cucciati}, {de la Torre}, {Durkalec},
  {Fontana}, {Giavalisco}, {Grazian}, {Hathi}, {Ilbert}, {Lemaux}, {Moreau},
  {Paltani}, {Ribeiro}, {Salvato}, {Schaerer}, {Scodeggio}, {Sommariva},
  {Talia}, {Taniguchi}, {Tresse}, {Vergani}, {Wang}, {Charlot}, {Contini},
  {Fotopoulou}, {L{\'o}pez-Sanjuan}, {Mellier}, \& {Scoville}}]{dong15}
{Le F{\`e}vre}, O., {Tasca}, L.~A.~M., {Cassata}, P., {et~al.} 2015, \aap, 576,
  A79

\bibitem[{{Le F{\`e}vre} {et~al.}(2005){Le F{\`e}vre}, {Vettolani}, {Garilli},
  {Tresse}, {Bottini}, {Le Brun}, {Maccagni}, {Picat}, {Scaramella},
  {Scodeggio}, {Zanichelli}, {Adami}, {Arnaboldi}, {Arnouts}, {Bardelli},
  {Bolzonella}, {Cappi}, {Charlot}, {Ciliegi}, {Contini}, {Foucaud},
  {Franzetti}, {Gavignaud}, {Guzzo}, {Ilbert}, {Iovino}, {McCracken}, {Marano},
  {Marinoni}, {Mathez}, {Mazure}, {Meneux}, {Merighi}, {Paltani}, {Pell{\`o}},
  {Pollo}, {Pozzetti}, {Radovich}, {Zamorani}, {Zucca}, {Bondi}, {Bongiorno},
  {Busarello}, {Lamareille}, {Mellier}, {Merluzzi}, {Ripepi}, \&
  {Rizzo}}]{dong05}
{Le F{\`e}vre}, O., {Vettolani}, G., {Garilli}, B., {et~al.} 2005, \aap, 439,
  845

\bibitem[{{Lee} {et~al.}(2016){Lee}, {Hennawi}, {White}, {Prochaska},
  {Font-Ribera}, {Schlegel}, {Rich}, {Suzuki}, {Stark}, {Le F{\`e}vre},
  {Nugent}, {Salvato}, \& {Zamorani}}]{lee16}
{Lee}, K.-G., {Hennawi}, J.~F., {White}, M., {et~al.} 2016, \apj, 817, 160

\bibitem[{{Lee} {et~al.}(2017){Lee}, {Tanaka}, {Kawabe}, {Kohno}, {Kodama},
  {Kajisawa}, {Yun}, {Nakanishi}, {Iono}, {Tamura}, {Hatsukade}, {Umehata},
  {Saito}, {Izumi}, {Aretxaga}, {Tadaki}, {Zeballos}, {Ikarashi}, {Wilson},
  {Hughes}, \& {Ivison}}]{lee17}
{Lee}, M.~M., {Tanaka}, I., {Kawabe}, R., {et~al.} 2017, \apj, 842, 55

\bibitem[{{Lehmer} {et~al.}(2005){Lehmer}, {Brandt}, {Alexander}, {Bauer},
  {Schneider}, {Tozzi}, {Bergeron}, {Garmire}, {Giacconi}, {Gilli}, {Hasinger},
  {Hornschemeier}, {Koekemoer}, {Mainieri}, {Miyaji}, {Nonino}, {Rosati},
  {Silverman}, {Szokoly}, \& {Vignali}}]{lehmer05}
{Lehmer}, B.~D., {Brandt}, W.~N., {Alexander}, D.~M., {et~al.} 2005, \apjs,
  161, 21

\bibitem[{{Leja} {et~al.}(2019){Leja}, {Carnall}, {Johnson}, {Conroy}, \&
  {Speagle}}]{leja19}
{Leja}, J., {Carnall}, A.~C., {Johnson}, B.~D., {Conroy}, C., \& {Speagle},
  J.~S. 2019, \apj, 876, 3

\bibitem[{{Lemaux} {et~al.}(2014{\natexlab{a}}){Lemaux}, {Cucciati}, {Tasca},
  {Le F{\`e}vre}, {Zamorani}, {Cassata}, {Garilli}, {Le Brun}, {Maccagni},
  {Pentericci}, {Thomas}, {Vanzella}, {Zucca}, {Amor{\'{\i}}n}, {Bardelli},
  {Capak}, {Cassar{\`a}}, {Castellano}, {Cimatti}, {Cuby}, {de la Torre},
  {Durkalec}, {Fontana}, {Giavalisco}, {Grazian}, {Hathi}, {Ilbert}, {Moreau},
  {Paltani}, {Ribeiro}, {Salvato}, {Schaerer}, {Scodeggio}, {Sommariva},
  {Talia}, {Taniguchi}, {Tresse}, {Vergani}, {Wang}, {Charlot}, {Contini},
  {Fotopoulou}, {Gal}, {Kocevski}, {L{\'o}pez-Sanjuan}, {Lubin}, {Mellier},
  {Sadibekova}, \& {Scoville}}]{lem14a}
{Lemaux}, B.~C., {Cucciati}, O., {Tasca}, L.~A.~M., {et~al.}
  2014{\natexlab{a}}, \aap, 572, A41

\bibitem[{{Lemaux} {et~al.}(2012){Lemaux}, {Gal}, {Lubin}, {Kocevski},
  {Fassnacht}, {McGrath}, {Squires}, {Surace}, \& {Lacy}}]{lem12}
{Lemaux}, B.~C., {Gal}, R.~R., {Lubin}, L.~M., {et~al.} 2012, \apj, 745, 106

\bibitem[{{Lemaux} {et~al.}(2018){Lemaux}, {Le F{\`e}vre}, {Cucciati},
  {Ribeiro}, {Tasca}, {Zamorani}, {Ilbert}, {Thomas}, {Bardelli}, {Cassata},
  {Hathi}, {Pforr}, {Smol{\v c}i{\'c}}, {Delvecchio}, {Novak}, {Berta},
  {McCracken}, {Koekemoer}, {Amor{\'{\i}}n}, {Garilli}, {Maccagni}, {Schaerer},
  \& {Zucca}}]{lem18}
{Lemaux}, B.~C., {Le F{\`e}vre}, O., {Cucciati}, O., {et~al.} 2018, \aap, 615,
  A77

\bibitem[{{Lemaux} {et~al.}(2014{\natexlab{b}}){Lemaux}, {Le Floc'h}, {Le
  F{\`e}vre}, {Ilbert}, {Tresse}, {Lubin}, {Zamorani}, {Gal}, {Ciliegi},
  {Cassata}, {Kocevski}, {McGrath}, {Bardelli}, {Zucca}, \& {Squires}}]{lem14b}
{Lemaux}, B.~C., {Le Floc'h}, E., {Le F{\`e}vre}, O., {et~al.}
  2014{\natexlab{b}}, \aap, 572, A90

\bibitem[{{Lemaux} {et~al.}(2009){Lemaux}, {Lubin}, {Sawicki}, {Martin},
  {Lagattuta}, {Gal}, {Kocevski}, {Fassnacht}, \& {Squires}}]{lem09}
{Lemaux}, B.~C., {Lubin}, L.~M., {Sawicki}, M., {et~al.} 2009, \apj, 700, 20

\bibitem[{{Lemaux} {et~al.}(2019){Lemaux}, {Tomczak}, {Lubin}, {Gal}, {Shen},
  {Pelliccia}, {Wu}, {Hung}, {Mei}, {Le F{\`e}vre}, {Rumbaugh}, {Kocevski}, \&
  {Squires}}]{lem19}
{Lemaux}, B.~C., {Tomczak}, A.~R., {Lubin}, L.~M., {et~al.} 2019, \mnras, 490,
  1231

\bibitem[{{Lemaux} {et~al.}(2017){Lemaux}, {Tomczak}, {Lubin}, {Wu}, {Gal},
  {Rumbaugh}, {Kocevski}, \& {Squires}}]{lem17}
{Lemaux}, B.~C., {Tomczak}, A.~R., {Lubin}, L.~M., {et~al.} 2017, \mnras, 472,
  419

\bibitem[{{Lewis} {et~al.}(2018){Lewis}, {Ivison}, {Best}, {Simpson}, {Weiss},
  {Oteo}, {Zhang}, {Arumugam}, {Bremer}, {Chapman}, {Clements}, {Dannerbauer},
  {Dunne}, {Eales}, {Maddox}, {Oliver}, {Omont}, {Riechers}, {Serjeant},
  {Valiante}, {Wardlow}, {van der Werf}, \& {De Zotti}}]{lewis18}
{Lewis}, A.~J.~R., {Ivison}, R.~J., {Best}, P.~N., {et~al.} 2018, \apj, 862, 96

\bibitem[{{Lilly} {et~al.}(2009){Lilly}, {Le Brun}, {Maier}, {Mainieri},
  {Mignoli}, {Scodeggio}, {Zamorani}, {Carollo}, {Contini}, {Kneib}, {Le
  F{\`e}vre}, {Renzini}, {Bardelli}, {Bolzonella}, {Bongiorno}, {Caputi},
  {Coppa}, {Cucciati}, {de la Torre}, {de Ravel}, {Franzetti}, {Garilli},
  {Iovino}, {Kampczyk}, {Kovac}, {Knobel}, {Lamareille}, {Le Borgne}, {Pello},
  {Peng}, {P{\'e}rez-Montero}, {Ricciardelli}, {Silverman}, {Tanaka}, {Tasca},
  {Tresse}, {Vergani}, {Zucca}, {Ilbert}, {Salvato}, {Oesch}, {Abbas},
  {Bottini}, {Capak}, {Cappi}, {Cassata}, {Cimatti}, {Elvis}, {Fumana},
  {Guzzo}, {Hasinger}, {Koekemoer}, {Leauthaud}, {Maccagni}, {Marinoni},
  {McCracken}, {Memeo}, {Meneux}, {Porciani}, {Pozzetti}, {Sanders},
  {Scaramella}, {Scarlata}, {Scoville}, {Shopbell}, \& {Taniguchi}}]{lilly09}
{Lilly}, S.~J., {Le Brun}, V., {Maier}, C., {et~al.} 2009, \apjs, 184, 218

\bibitem[{{Lilly} {et~al.}(2007){Lilly}, {Le F{\`e}vre}, {Renzini}, {Zamorani},
  {Scodeggio}, {Contini}, {Carollo}, {Hasinger}, {Kneib}, {Iovino}, {Le Brun},
  {Maier}, {Mainieri}, {Mignoli}, {Silverman}, {Tasca}, {Bolzonella},
  {Bongiorno}, {Bottini}, {Capak}, {Caputi}, {Cimatti}, {Cucciati}, {Daddi},
  {Feldmann}, {Franzetti}, {Garilli}, {Guzzo}, {Ilbert}, {Kampczyk}, {Kovac},
  {Lamareille}, {Leauthaud}, {Borgne}, {McCracken}, {Marinoni}, {Pello},
  {Ricciardelli}, {Scarlata}, {Vergani}, {Sanders}, {Schinnerer}, {Scoville},
  {Taniguchi}, {Arnouts}, {Aussel}, {Bardelli}, {Brusa}, {Cappi}, {Ciliegi},
  {Finoguenov}, {Foucaud}, {Franceschini}, {Halliday}, {Impey}, {Knobel},
  {Koekemoer}, {Kurk}, {Maccagni}, {Maddox}, {Marano}, {Marconi}, {Meneux},
  {Mobasher}, {Moreau}, {Peacock}, {Porciani}, {Pozzetti}, {Scaramella},
  {Schiminovich}, {Shopbell}, {Smail}, {Thompson}, {Tresse}, {Vettolani},
  {Zanichelli}, \& {Zucca}}]{lilly07}
{Lilly}, S.~J., {Le F{\`e}vre}, O., {Renzini}, A., {et~al.} 2007, \apjs, 172,
  70

\bibitem[{{Loiacono} {et~al.}(2021){Loiacono}, {Decarli}, {Gruppioni}, {Talia},
  {Cimatti}, {Zamorani}, {Pozzi}, {Yan}, {Lemaux}, {Riechers}, {Le F{\`e}vre},
  {B{\`e}thermin}, {Capak}, {Cassata}, {Faisst}, {Schaerer}, {Silverman},
  {Bardelli}, {Boquien}, {Burkutean}, {Dessauges-Zavadsky}, {Fudamoto},
  {Fujimoto}, {Ginolfi}, {Hathi}, {Jones}, {Khusanova}, {Koekemoer}, {Lagache},
  {Lubin}, {Massardi}, {Oesch}, {Romano}, {Vallini}, {Vergani}, \&
  {Zucca}}]{federica21}
{Loiacono}, F., {Decarli}, R., {Gruppioni}, C., {et~al.} 2021, \aap, 646, A76

\bibitem[{{Long} {et~al.}(2020){Long}, {Cooray}, {Ma}, {Casey}, {Wardlow},
  {Nayyeri}, {Ivison}, {Farrah}, \& {Dannerbauer}}]{long20}
{Long}, A.~S., {Cooray}, A., {Ma}, J., {et~al.} 2020, \apj, 898, 133

\bibitem[{{L{\'o}pez-Sanjuan} {et~al.}(2013){L{\'o}pez-Sanjuan}, {Le
  F{\`e}vre}, {Tasca}, {Epinat}, {Amram}, {Contini}, {Garilli},
  {Kissler-Patig}, {Moultaka}, {Paioro}, {Perret}, {Queyrel}, {Tresse},
  {Vergani}, \& {Divoy}}]{carlos13}
{L{\'o}pez-Sanjuan}, C., {Le F{\`e}vre}, O., {Tasca}, L.~A.~M., {et~al.} 2013,
  \aap, 553, A78

\bibitem[{{Lower} {et~al.}(2020){Lower}, {Narayanan}, {Leja}, {Johnson},
  {Conroy}, \& {Dav{\'e}}}]{lower20}
{Lower}, S., {Narayanan}, D., {Leja}, J., {et~al.} 2020, \apj, 904, 33

\bibitem[{{Lubin} {et~al.}(2009){Lubin}, {Gal}, {Lemaux}, {Kocevski}, \&
  {Squires}}]{lub09}
{Lubin}, L.~M., {Gal}, R.~R., {Lemaux}, B.~C., {Kocevski}, D.~D., \& {Squires},
  G.~K. 2009, \aj, 137, 4867

\bibitem[{{Macklin}(1982)}]{macklin82}
{Macklin}, J.~T. 1982, \mnras, 199, 1119

\bibitem[{{Madau} \& {Dickinson}(2014)}]{lunabarnfriendz14}
{Madau}, P. \& {Dickinson}, M. 2014, \araa, 52, 415

\bibitem[{{Maraston} {et~al.}(2010){Maraston}, {Pforr}, {Renzini}, {Daddi},
  {Dickinson}, {Cimatti}, \& {Tonini}}]{maraston10}
{Maraston}, C., {Pforr}, J., {Renzini}, A., {et~al.} 2010, \mnras, 407, 830

\bibitem[{{Martin} {et~al.}(2005){Martin}, {Fanson}, {Schiminovich},
  {Morrissey}, {Friedman}, {Barlow}, {Conrow}, {Grange}, {Jelinsky},
  {Milliard}, {Siegmund}, {Bianchi}, {Byun}, {Donas}, {Forster}, {Heckman},
  {Lee}, {Madore}, {Malina}, {Neff}, {Rich}, {Small}, {Surber}, {Szalay},
  {Welsh}, \& {Wyder}}]{martin05}
{Martin}, D.~C., {Fanson}, J., {Schiminovich}, D., {et~al.} 2005, \apjl, 619,
  L1

\bibitem[{{Mauduit} {et~al.}(2012){Mauduit}, {Lacy}, {Farrah}, {Surace},
  {Jarvis}, {Oliver}, {Maraston}, {Vaccari}, {Marchetti}, {Zeimann},
  {Gonz{\'a}les-Solares}, {Pforr}, {Petric}, {Henriques}, {Thomas}, {Afonso},
  {Rettura}, {Wilson}, {Falder}, {Geach}, {Huynh}, {Norris}, {Seymour},
  {Richards}, {Stanford}, {Alexander}, {Becker}, {Best}, {Bizzocchi},
  {Bonfield}, {Castro}, {Cava}, {Chapman}, {Christopher}, {Clements}, {Covone},
  {Dubois}, {Dunlop}, {Dyke}, {Edge}, {Ferguson}, {Foucaud}, {Franceschini},
  {Gal}, {Grant}, {Grossi}, {Hatziminaoglou}, {Hickey}, {Hodge}, {Huang},
  {Ivison}, {Kim}, {LeFevre}, {Lehnert}, {Lonsdale}, {Lubin}, {McLure},
  {Messias}, {Mart{\'{\i}}nez-Sansigre}, {Mortier}, {Nielsen}, {Ouchi},
  {Parish}, {Perez-Fournon}, {Pierre}, {Rawlings}, {Readhead}, {Ridgway},
  {Rigopoulou}, {Romer}, {Rosebloom}, {Rottgering}, {Rowan-Robinson}, {Sajina},
  {Simpson}, {Smail}, {Squires}, {Stevens}, {Taylor}, {Trichas}, {Urrutia},
  {van Kampen}, {Verma}, \& {Xu}}]{mauduit12}
{Mauduit}, J.-C., {Lacy}, M., {Farrah}, D., {et~al.} 2012, \pasp, 124, 714

\bibitem[{{Mawatari} {et~al.}(2016){Mawatari}, {Yamada}, {Fazio}, {Huang}, \&
  {Ashby}}]{maw16}
{Mawatari}, K., {Yamada}, T., {Fazio}, G.~G., {Huang}, J.-S., \& {Ashby},
  M.~L.~N. 2016, \pasj, 68, 46

\bibitem[{{McCracken} {et~al.}(2012){McCracken}, {Milvang-Jensen}, {Dunlop},
  {Franx}, {Fynbo}, {Le F{\`e}vre}, {Holt}, {Caputi}, {Goranova}, {Buitrago},
  {Emerson}, {Freudling}, {Hudelot}, {L{\'o}pez-Sanjuan}, {Magnard}, {Mellier},
  {M{\o}ller}, {Nilsson}, {Sutherland}, {Tasca}, \& {Zabl}}]{mccracken12}
{McCracken}, H.~J., {Milvang-Jensen}, B., {Dunlop}, J., {et~al.} 2012, \aap,
  544, A156

\bibitem[{{McLean} {et~al.}(2012){McLean}, {Steidel}, {Epps}, {Konidaris},
  {Matthews}, {Adkins}, {Aliado}, {Brims}, {Canfield}, {Cromer}, {Fucik},
  {Kulas}, {Mace}, {Magnone}, {Rodriguez}, {Rudie}, {Trainor}, {Wang}, {Weber},
  \& {Weiss}}]{mclean12}
{McLean}, I.~S., {Steidel}, C.~C., {Epps}, H.~W., {et~al.} 2012, in \procspie,
  Vol. 8446, Ground-based and Airborne Instrumentation for Astronomy IV, 84460J

\bibitem[{{Miller} {et~al.}(2018){Miller}, {Chapman}, {Aravena}, {Ashby},
  {Hayward}, {Vieira}, {Wei{\ss}}, {Babul}, {B{\'e}thermin}, {Bradford},
  {Brodwin}, {Carlstrom}, {Chen}, {Cunningham}, {De Breuck}, {Gonzalez},
  {Greve}, {Harnett}, {Hezaveh}, {Lacaille}, {Litke}, {Ma}, {Malkan},
  {Marrone}, {Morningstar}, {Murphy}, {Narayanan}, {Pass}, {Perry}, {Phadke},
  {Rennehan}, {Rotermund}, {Simpson}, {Spilker}, {Sreevani}, {Stark},
  {Strandet}, \& {Strom}}]{miller18}
{Miller}, T.~B., {Chapman}, S.~C., {Aravena}, M., {et~al.} 2018, \nat, 556, 469

\bibitem[{{Miller} {et~al.}(2015){Miller}, {Hayward}, {Chapman}, \&
  {Behroozi}}]{miller15}
{Miller}, T.~B., {Hayward}, C.~C., {Chapman}, S.~C., \& {Behroozi}, P.~S. 2015,
  \mnras, 452, 878

\bibitem[{{Miyazaki} {et~al.}(2012){Miyazaki}, {Komiyama}, {Nakaya}, {Kamata},
  {Doi}, {Hamana}, {Karoji}, {Furusawa}, {Kawanomoto}, {Morokuma}, {Ishizuka},
  {Nariai}, {Tanaka}, {Uraguchi}, {Utsumi}, {Obuchi}, {Okura}, {Oguri},
  {Takata}, {Tomono}, {Kurakami}, {Namikawa}, {Usuda}, {Yamanoi}, {Terai},
  {Uekiyo}, {Yamada}, {Koike}, {Aihara}, {Fujimori}, {Mineo}, {Miyatake},
  {Yasuda}, {Nishizawa}, {Saito}, {Tanaka}, {Uchida}, {Katayama}, {Wang},
  {Chen}, {Lupton}, {Loomis}, {Bickerton}, {Price}, {Gunn}, {Suzuki},
  {Miyazaki}, {Muramatsu}, {Yamamoto}, {Endo}, {Ezaki}, {Itoh}, {Miwa},
  {Yokota}, {Matsuda}, {Ebinuma}, \& {Takeshi}}]{miyazaki12}
{Miyazaki}, S., {Komiyama}, Y., {Nakaya}, H., {et~al.} 2012, in \procspie, Vol.
  8446, Ground-based and Airborne Instrumentation for Astronomy IV, 84460Z

\bibitem[{{Miyazaki} {et~al.}(2002){Miyazaki}, {Komiyama}, {Sekiguchi},
  {Okamura}, {Doi}, {Furusawa}, {Hamabe}, {Imi}, {Kimura}, {Nakata}, {Okada},
  {Ouchi}, {Shimasaku}, {Yagi}, \& {Yasuda}}]{miyazaki02}
{Miyazaki}, S., {Komiyama}, Y., {Sekiguchi}, M., {et~al.} 2002, \pasj, 54, 833

\bibitem[{{Moretti} {et~al.}(2018){Moretti}, {Poggianti}, {Gullieuszik},
  {Mapelli}, {Jaff{\'e}}, {Fritz}, {Biviano}, {Fasano}, {Bettoni}, {Vulcani},
  \& {D'Onofrio}}]{moretti18}
{Moretti}, A., {Poggianti}, B.~M., {Gullieuszik}, M., {et~al.} 2018, \mnras,
  475, 4055

\bibitem[{{Morris} {et~al.}(2015){Morris}, {Kocevski}, {Trump}, {Weiner},
  {Hathi}, {Barro}, {Dahlen}, {Faber}, {Finkelstein}, \& {Fontana}}]{morris15}
{Morris}, A.~M., {Kocevski}, D.~D., {Trump}, J.~R., {et~al.} 2015, \aj, 149,
  178

\bibitem[{{Mostek} {et~al.}(2012){Mostek}, {Coil}, {Moustakas}, {Salim}, \&
  {Weiner}}]{mostek12}
{Mostek}, N., {Coil}, A.~L., {Moustakas}, J., {Salim}, S., \& {Weiner}, B.~J.
  2012, \apj, 746, 124

\bibitem[{{Moutard} {et~al.}(2016){Moutard}, {Arnouts}, {Ilbert}, {Coupon},
  {Davidzon}, {Guzzo}, {Hudelot}, {McCracken}, {Van Werbaeke}, {Morrison}, {Le
  F{\`e}vre}, {Comte}, {Bolzonella}, {Fritz}, {Garilli}, \&
  {Scodeggio}}]{thibaud16}
{Moutard}, T., {Arnouts}, S., {Ilbert}, O., {et~al.} 2016, \aap, 590, A103

\bibitem[{{Moy} {et~al.}(2003){Moy}, {Barmby}, {Rigopoulou}, {Huang},
  {Willner}, \& {Fazio}}]{moy03}
{Moy}, E., {Barmby}, P., {Rigopoulou}, D., {et~al.} 2003, \aap, 403, 493

\bibitem[{{Muldrew} {et~al.}(2018){Muldrew}, {Hatch}, \& {Cooke}}]{muldrew18}
{Muldrew}, S.~I., {Hatch}, N.~A., \& {Cooke}, E.~A. 2018, \mnras, 473, 2335

\bibitem[{{Muzzin} {et~al.}(2013){Muzzin}, {Marchesini}, {Stefanon}, {Franx},
  {McCracken}, {Milvang-Jensen}, {Dunlop}, {Fynbo}, {Brammer}, {Labb{\'e}}, \&
  {van Dokkum}}]{muz13}
{Muzzin}, A., {Marchesini}, D., {Stefanon}, M., {et~al.} 2013, \apj, 777, 18

\bibitem[{{Muzzin} {et~al.}(2012){Muzzin}, {Wilson}, {Yee}, {Gilbank},
  {Hoekstra}, {Demarco}, {Balogh}, {van Dokkum}, {Franx}, {Ellingson}, {Hicks},
  {Nantais}, {Noble}, {Lacy}, {Lidman}, {Rettura}, {Surace}, \&
  {Webb}}]{muzz12}
{Muzzin}, A., {Wilson}, G., {Yee}, H.~K.~C., {et~al.} 2012, \apj, 746, 188

\bibitem[{{Nantais} {et~al.}(2017){Nantais}, {Muzzin}, {van der Burg},
  {Wilson}, {Lidman}, {Foltz}, {DeGroot}, {Noble}, {Cooper}, \&
  {Demarco}}]{nantais17}
{Nantais}, J.~B., {Muzzin}, A., {van der Burg}, R.~F.~J., {et~al.} 2017,
  \mnras, 465, L104

\bibitem[{{Nantais} {et~al.}(2016){Nantais}, {van der Burg}, {Lidman},
  {Demarco}, {Noble}, {Wilson}, {Muzzin}, {Foltz}, {DeGroot}, \&
  {Cooper}}]{nantais16}
{Nantais}, J.~B., {van der Burg}, R.~F.~J., {Lidman}, C., {et~al.} 2016, \aap,
  592, A161

\bibitem[{{Nelson} {et~al.}(2019){Nelson}, {Springel}, {Pillepich},
  {Rodriguez-Gomez}, {Torrey}, {Genel}, {Vogelsberger}, {Pakmor}, {Marinacci},
  {Weinberger}, {Kelley}, {Lovell}, {Diemer}, \& {Hernquist}}]{nelson19}
{Nelson}, D., {Springel}, V., {Pillepich}, A., {et~al.} 2019, Computational
  Astrophysics and Cosmology, 6, 2

\bibitem[{{Newman} {et~al.}(2014){Newman}, {Ellis}, {Andreon}, {Treu},
  {Raichoor}, \& {Trinchieri}}]{newman14}
{Newman}, A.~B., {Ellis}, R.~S., {Andreon}, S., {et~al.} 2014, \apj, 788, 51

\bibitem[{{Newman} {et~al.}(2020){Newman}, {Rudie}, {Blanc}, {Kelson},
  {Rhoades}, {Hare}, {P{\'e}rez}, {Benson}, {Dressler}, {Gonzalez},
  {Kollmeier}, {Konidaris}, {Mulchaey}, {Rauch}, {Le F{\`e}vre}, {Lemaux},
  {Cucciati}, \& {Lilly}}]{drew20}
{Newman}, A.~B., {Rudie}, G.~C., {Blanc}, G.~A., {et~al.} 2020, \apj, 891, 147

\bibitem[{{Newman} {et~al.}(2013){Newman}, {Cooper}, {Davis}, {Faber}, {Coil},
  {Guhathakurta}, {Koo}, {Phillips}, {Conroy}, {Dutton}, {Finkbeiner}, {Gerke},
  {Rosario}, {Weiner}, {Willmer}, {Yan}, {Harker}, {Kassin}, {Konidaris},
  {Lai}, {Madgwick}, {Noeske}, {Wirth}, {Connolly}, {Kaiser}, {Kirby},
  {Lemaux}, {Lin}, {Lotz}, {Luppino}, {Marinoni}, {Matthews}, {Metevier}, \&
  {Schiavon}}]{new13}
{Newman}, J.~A., {Cooper}, M.~C., {Davis}, M., {et~al.} 2013, \apjs, 208, 5

\bibitem[{{Noirot} {et~al.}(2018){Noirot}, {Stern}, {Mei}, {Wylezalek},
  {Cooke}, {De Breuck}, {Galametz}, {Hatch}, {Vernet}, {Brodwin}, {Eisenhardt},
  {Gonzalez}, {Jarvis}, {Rettura}, {Seymour}, \& {Stanford}}]{noirot18}
{Noirot}, G., {Stern}, D., {Mei}, S., {et~al.} 2018, \apj, 859, 38

\bibitem[{{Noll} {et~al.}(2009){Noll}, {Burgarella}, {Giovannoli}, {Buat},
  {Marcillac}, \& {Mu{\~n}oz-Mateos}}]{noll09}
{Noll}, S., {Burgarella}, D., {Giovannoli}, E., {et~al.} 2009, \aap, 507, 1793

\bibitem[{{Oke} \& {Gunn}(1983)}]{okengunn83}
{Oke}, J.~B. \& {Gunn}, J.~E. 1983, \apj, 266, 713

\bibitem[{{Old} {et~al.}(2020){Old}, {Balogh}, {van der Burg}, {Biviano},
  {Yee}, {Pintos-Castro}, {Webb}, {Muzzin}, {Rudnick}, {Vulcani}, {Poggianti},
  {Cooper}, {Zaritsky}, {Cerulo}, {Wilson}, {Chan}, {Lidman}, {McGee},
  {Demarco}, {Forrest}, {De Lucia}, {Gilbank}, {Kukstas}, {McCarthy},
  {Jablonka}, {Nantais}, {Noble}, {Reeves}, \& {Shipley}}]{old20}
{Old}, L.~J., {Balogh}, M.~L., {van der Burg}, R. F.~J., {et~al.} 2020, \mnras,
  493, 5987

\bibitem[{{Oteo} {et~al.}(2018){Oteo}, {Ivison}, {Dunne}, {Manilla-Robles},
  {Maddox}, {Lewis}, {de Zotti}, {Bremer}, {Clements}, {Cooray}, {Dannerbauer},
  {Eales}, {Greenslade}, {Omont}, {Perez─Fourn{\'o}n}, {Riechers}, {Scott},
  {van der Werf}, {Weiss}, \& {Zhang}}]{oteo18}
{Oteo}, I., {Ivison}, R.~J., {Dunne}, L., {et~al.} 2018, \apj, 856, 72

\bibitem[{{Overzier}(2016)}]{overzier16}
{Overzier}, R.~A. 2016, \aapr, 24, 14

\bibitem[{{Owers} {et~al.}(2019){Owers}, {Hudson}, {Oman}, {Bland -Hawthorn},
  {Brough}, {Bryant}, {Cortese}, {Couch}, {Croom}, {van de Sande}, {Federrath},
  {Groves}, {Hopkins}, {Lawrence}, {Lorente}, {McDermid}, {Medling},
  {Richards}, {Scott}, {Taranu}, {Welker}, \& {Yi}}]{owers19}
{Owers}, M.~S., {Hudson}, M.~J., {Oman}, K.~A., {et~al.} 2019, \apj, 873, 52

\bibitem[{{Paccagnella} {et~al.}(2019){Paccagnella}, {Vulcani}, {Poggianti},
  {Moretti}, {Fritz}, {Gullieuszik}, \& {Fasano}}]{paccagnella19}
{Paccagnella}, A., {Vulcani}, B., {Poggianti}, B.~M., {et~al.} 2019, \mnras,
  482, 881

\bibitem[{{Peng} {et~al.}(2010){Peng}, {Lilly}, {Kova{\v c}}, {Bolzonella},
  {Pozzetti}, {Renzini}, {Zamorani}, {Ilbert}, {Knobel}, {Iovino}, {Maier},
  {Cucciati}, {Tasca}, {Carollo}, {Silverman}, {Kampczyk}, {de Ravel},
  {Sanders}, {Scoville}, {Contini}, {Mainieri}, {Scodeggio}, {Kneib}, {Le
  F{\`e}vre}, {Bardelli}, {Bongiorno}, {Caputi}, {Coppa}, {de la Torre},
  {Franzetti}, {Garilli}, {Lamareille}, {Le Borgne}, {Le Brun}, {Mignoli},
  {Perez Montero}, {Pello}, {Ricciardelli}, {Tanaka}, {Tresse}, {Vergani},
  {Welikala}, {Zucca}, {Oesch}, {Abbas}, {Barnes}, {Bordoloi}, {Bottini},
  {Cappi}, {Cassata}, {Cimatti}, {Fumana}, {Hasinger}, {Koekemoer},
  {Leauthaud}, {Maccagni}, {Marinoni}, {McCracken}, {Memeo}, {Meneux}, {Nair},
  {Porciani}, {Presotto}, \& {Scaramella}}]{peng10}
{Peng}, Y.-j., {Lilly}, S.~J., {Kova{\v c}}, K., {et~al.} 2010, \apj, 721, 193

\bibitem[{{Pr\'{e}vot} {et~al.}(1984){Pr\'{e}vot}, {Lequeux}, {Prevot},
  {Maurice}, \& {Rocca-Volmerange}}]{prevot84}
{Pr\'{e}vot}, M.~L., {Lequeux}, J., {Prevot}, L., {Maurice}, E., \&
  {Rocca-Volmerange}, B. 1984, \aap, 132, 389

\bibitem[{{Puget} {et~al.}(2004){Puget}, {Stadler}, {Doyon}, {Gigan},
  {Thibault}, {Luppino}, {Barrick}, {Benedict}, {Forveille}, {Rambold},
  {Thomas}, {Vermeulen}, {Ward}, {Beuzit}, {Feautrier}, {Magnard}, {Mella},
  {Preis}, {Vallee}, {Wang}, {Lin}, {Hall}, \& {Hodapp}}]{puget04}
{Puget}, P., {Stadler}, E., {Doyon}, R., {et~al.} 2004, in \procspie, Vol.
  5492, Ground-based Instrumentation for Astronomy, ed. A.~F.~M. {Moorwood} \&
  M.~{Iye}, 978--987

\bibitem[{{Quadri} {et~al.}(2012){Quadri}, {Williams}, {Franx}, \&
  {Hildebrandt}}]{quadri12}
{Quadri}, R.~F., {Williams}, R.~J., {Franx}, M., \& {Hildebrandt}, H. 2012,
  \apj, 744, 88

\bibitem[{{Raichoor} {et~al.}(2011){Raichoor}, {Mei}, {Nakata}, {Stanford},
  {Holden}, {Rettura}, {Huertas-Company}, {Postman}, {Rosati}, {Blakeslee},
  {Demarco}, {Eisenhardt}, {Illingworth}, {Jee}, {Kodama}, {Tanaka}, \&
  {White}}]{rai11}
{Raichoor}, A., {Mei}, S., {Nakata}, F., {et~al.} 2011, \apj, 732, 12

\bibitem[{{Rettura} {et~al.}(2010){Rettura}, {Rosati}, {Nonino}, {Fosbury},
  {Gobat}, {Menci}, {Strazzullo}, {Mei}, {Demarco}, \& {Ford}}]{ret10}
{Rettura}, A., {Rosati}, P., {Nonino}, M., {et~al.} 2010, \apj, 709, 512

\bibitem[{{Rodighiero} {et~al.}(2014){Rodighiero}, {Renzini}, {Daddi},
  {Baronchelli}, {Berta}, {Cresci}, {Franceschini}, {Gruppioni}, {Lutz},
  {Mancini}, {Santini}, {Zamorani}, {Silverman}, {Kashino}, {Andreani},
  {Cimatti}, {S{\'a}nchez}, {Le Floch}, {Magnelli}, {Popesso}, \&
  {Pozzi}}]{rodighiero14}
{Rodighiero}, G., {Renzini}, A., {Daddi}, E., {et~al.} 2014, \mnras, 443, 19

\bibitem[{{Romano} {et~al.}(2020){Romano}, {Cassata}, {Morselli}, {Lemaux},
  {B{\'e}thermin}, {Capak}, {Faisst}, {Le F{\`e}vre}, {Schaerer}, {Silverman},
  {Yan}, {Bardelli}, {Boquien}, {Cimatti}, {Dessauges-Zavadsky}, {Enia},
  {Fudamoto}, {Fujimoto}, {Ginolfi}, {Gruppioni}, {Hathi}, {Ibar}, {Jones},
  {Koekemoer}, {Loiacono}, {Mancini}, {Riechers}, {Rodighiero},
  {Rodr{\'\i}guez-Mu{\~n}oz}, {Talia}, {Vallini}, {Vergani}, {Zamorani}, \&
  {Zucca}}]{romano20}
{Romano}, M., {Cassata}, P., {Morselli}, L., {et~al.} 2020, \mnras, 496, 875

\bibitem[{{Ryan} {et~al.}(2014){Ryan}, {Gonzalez}, {Lemaux}, {Brada{\v{c}}},
  {Casertano}, {Allen}, {Cain}, {Gladders}, {Hall}, {Hildebradt}, {Hinz},
  {Huang}, {Lubin}, {Schrabback}, {Stiavelli}, {Treu}, {von der Linden}, \&
  {Zaritsky}}]{wrustle14}
{Ryan}, R.~E., J., {Gonzalez}, A.~H., {Lemaux}, B.~C., {et~al.} 2014, \apjl,
  786, L4

\bibitem[{{Santos} {et~al.}(2014){Santos}, {Altieri}, {Tanaka}, {Valtchanov},
  {Saintonge}, {Dickinson}, {Foucaud}, {Kodama}, {Rawle}, \&
  {Tadaki}}]{santos14}
{Santos}, J.~S., {Altieri}, B., {Tanaka}, M., {et~al.} 2014, \mnras, 438, 2565

\bibitem[{{Santos} {et~al.}(2015){Santos}, {Altieri}, {Valtchanov}, {Nastasi},
  {B{\"o}hringer}, {Cresci}, {Elbaz}, {Fassbender}, {Rosati}, {Tozzi}, \&
  {Verdugo}}]{santos15}
{Santos}, J.~S., {Altieri}, B., {Valtchanov}, I., {et~al.} 2015, \mnras, 447,
  L65

\bibitem[{{Schaerer} {et~al.}(2013){Schaerer}, {de Barros}, \&
  {Sklias}}]{schaerer13}
{Schaerer}, D., {de Barros}, S., \& {Sklias}, P. 2013, \aap, 549, A4

\bibitem[{{Schaerer} {et~al.}(2020){Schaerer}, {Ginolfi}, {B{\'e}thermin},
  {Fudamoto}, {Oesch}, {Le F{\`e}vre}, {Faisst}, {Capak}, {Cassata},
  {Silverman}, {Yan}, {Jones}, {Amorin}, {Bardelli}, {Boquien}, {Cimatti},
  {Dessauges-Zavadsky}, {Giavalisco}, {Hathi}, {Fujimoto}, {Ibar}, {Koekemoer},
  {Lagache}, {Lemaux}, {Loiacono}, {Maiolino}, {Narayanan}, {Morselli},
  {M{\'e}ndez-Hern{\`a}ndez}, {Pozzi}, {Riechers}, {Talia}, {Toft}, {Vallini},
  {Vergani}, {Zamorani}, \& {Zucca}}]{schaerer20}
{Schaerer}, D., {Ginolfi}, M., {B{\'e}thermin}, M., {et~al.} 2020, \aap, 643,
  A3

\bibitem[{{Schinnerer} {et~al.}(2016){Schinnerer}, {Groves}, {Sargent},
  {Karim}, {Oesch}, {Magnelli}, {LeFevre}, {Tasca}, {Civano}, {Cassata}, \&
  {Smol{\v c}i{\'c}}}]{schinnerer16}
{Schinnerer}, E., {Groves}, B., {Sargent}, M.~T., {et~al.} 2016, \apj, 833, 112

\bibitem[{{Scoville} {et~al.}(2013){Scoville}, {Arnouts}, {Aussel}, {Benson},
  {Bongiorno}, {Bundy}, {Calvo}, {Capak}, {Carollo}, {Civano}, {Dunlop},
  {Elvis}, {Faisst}, {Finoguenov}, {Fu}, {Giavalisco}, {Guo}, {Ilbert},
  {Iovino}, {Kajisawa}, {Kartaltepe}, {Leauthaud}, {Le F{\`e}vre}, {LeFloch},
  {Lilly}, {Liu}, {Manohar}, {Massey}, {Masters}, {McCracken}, {Mobasher},
  {Peng}, {Renzini}, {Rhodes}, {Salvato}, {Sanders}, {Sarvestani}, {Scarlata},
  {Schinnerer}, {Sheth}, {Shopbell}, {Smol{\v c}i{\'c}}, {Taniguchi}, {Taylor},
  {White}, \& {Yan}}]{scoville13}
{Scoville}, N., {Arnouts}, S., {Aussel}, H., {et~al.} 2013, \apjs, 206, 3

\bibitem[{{Scoville} {et~al.}(2007){Scoville}, {Aussel}, {Brusa}, {Capak},
  {Carollo}, {Elvis}, {Giavalisco}, {Guzzo}, {Hasinger}, {Impey}, {Kneib},
  {LeFevre}, {Lilly}, {Mobasher}, {Renzini}, {Rich}, {Sanders}, {Schinnerer},
  {Schminovich}, {Shopbell}, {Taniguchi}, \& {Tyson}}]{scoville07a}
{Scoville}, N., {Aussel}, H., {Brusa}, M., {et~al.} 2007, \apjs, 172, 1

\bibitem[{{Scoville} {et~al.}(2017){Scoville}, {Lee}, {Vanden Bout},
  {Diaz-Santos}, {Sanders}, {Darvish}, {Bongiorno}, {Casey}, {Murchikova},
  {Koda}, {Capak}, {Vlahakis}, {Ilbert}, {Sheth}, {Morokuma-Matsui}, {Ivison},
  {Aussel}, {Laigle}, {McCracken}, {Armus}, {Pope}, {Toft}, \&
  {Masters}}]{smokynicky17}
{Scoville}, N., {Lee}, N., {Vanden Bout}, P., {et~al.} 2017, \apj, 837, 150

\bibitem[{{Scoville} {et~al.}(2016){Scoville}, {Sheth}, {Aussel}, {Vanden
  Bout}, {Capak}, {Bongiorno}, {Casey}, {Murchikova}, {Koda},
  {{\'A}lvarez-M{\'a}rquez}, {Lee}, {Laigle}, {McCracken}, {Ilbert}, {Pope},
  {Sanders}, {Chu}, {Toft}, {Ivison}, \& {Manohar}}]{smokynicky16}
{Scoville}, N., {Sheth}, K., {Aussel}, H., {et~al.} 2016, \apj, 820, 83

\bibitem[{{Shen} {et~al.}(2021){Shen}, {Lemaux}, {Lubin}, {Cucciati}, {Le
  F{\`e}vre}, {Liu}, {Fang}, {Pelliccia}, {Tomczak}, {McKean}, {Miller},
  {Fassnacht}, {Gal}, {Hung}, {Hathi}, {Bardelli}, {Vergani}, \&
  {Zucca}}]{shen21}
{Shen}, L., {Lemaux}, B.~C., {Lubin}, L.~M., {et~al.} 2021, \apj, 912, 60

\bibitem[{{Shi} {et~al.}(2019){Shi}, {Lee}, {Dey}, {Huang}, {Malavasi}, {Hung},
  {Inami}, {Ashby}, {Duncan}, {Xue}, {Reddy}, {Hong}, {Jannuzi}, {Cooper},
  {Gonzalez}, {R{\"o}ttgering}, {Best}, \& {Tasse}}]{shi19}
{Shi}, K., {Lee}, K.-S., {Dey}, A., {et~al.} 2019, \apj, 871, 83

\bibitem[{{Shi} {et~al.}(2020){Shi}, {Toshikawa}, {Cai}, {Lee}, \&
  {Fang}}]{shi20}
{Shi}, K., {Toshikawa}, J., {Cai}, Z., {Lee}, K.-S., \& {Fang}, T. 2020, \apj,
  899, 79

\bibitem[{{Shi} {et~al.}(2021){Shi}, {Toshikawa}, {Lee}, {Wang}, {Cai}, \&
  {Fang}}]{shi21}
{Shi}, K., {Toshikawa}, J., {Lee}, K.-S., {et~al.} 2021, \apj, 911, 46

\bibitem[{{Shimakawa} {et~al.}(2018{\natexlab{a}}){Shimakawa}, {Kodama},
  {Hayashi}, {Prochaska}, {Tanaka}, {Cai}, {Suzuki}, {Tadaki}, \&
  {Koyama}}]{rhythm18}
{Shimakawa}, R., {Kodama}, T., {Hayashi}, M., {et~al.} 2018{\natexlab{a}},
  \mnras, 473, 1977

\bibitem[{{Shimakawa} {et~al.}(2017){Shimakawa}, {Kodama}, {Hayashi}, {Tanaka},
  {Matsuda}, {Kashikawa}, {Shibuya}, {Tadaki}, {Koyama}, {Suzuki}, \&
  {Yamamoto}}]{rhythm17}
{Shimakawa}, R., {Kodama}, T., {Hayashi}, M., {et~al.} 2017, \mnras, 468, L21

\bibitem[{{Shimakawa} {et~al.}(2018{\natexlab{b}}){Shimakawa}, {Koyama},
  {R{\"o}ttgering}, {Kodama}, {Hayashi}, {Hatch}, {Dannerbauer}, {Tanaka},
  {Tadaki}, {Suzuki}, {Fukagawa}, {Cai}, \& {Kurk}}]{rhythm18b}
{Shimakawa}, R., {Koyama}, Y., {R{\"o}ttgering}, H. J.~A., {et~al.}
  2018{\natexlab{b}}, \mnras, 481, 5630

\bibitem[{{Shivaei} {et~al.}(2016){Shivaei}, {Kriek}, {Reddy}, {Shapley},
  {Barro}, {Conroy}, {Coil}, {Freeman}, {Mobasher}, {Siana}, {Sanders},
  {Price}, {Azadi}, {Pasha}, \& {Inami}}]{shivaei16}
{Shivaei}, I., {Kriek}, M., {Reddy}, N.~A., {et~al.} 2016, \apjl, 820, L23

\bibitem[{{Silverman} {et~al.}(2018){Silverman}, {Rujopakarn}, {Daddi},
  {Renzini}, {Rodighiero}, {Liu}, {Puglisi}, {Sargent}, {Mancini},
  {Kartaltepe}, {Kashino}, {Koekemoer}, {Arimoto}, {B{\'e}thermin}, {Jin},
  {Magdis}, {Nagao}, {Onodera}, {Sand ers}, \& {Valentino}}]{silverman18}
{Silverman}, J.~D., {Rujopakarn}, W., {Daddi}, E., {et~al.} 2018, \apj, 867, 92

\bibitem[{{Smol{\v c}i{\'c}} {et~al.}(2017{\natexlab{a}}){Smol{\v c}i{\'c}},
  {Miettinen}, {Tomi{\v c}i{\'c}}, {Zamorani}, {Finoguenov}, {Lemaux},
  {Aravena}, {Capak}, {Chiang}, {Civano}, {Delvecchio}, {Ilbert}, {Jurlin},
  {Karim}, {Laigle}, {Le F{\`e}vre}, {Marchesi}, {McCracken}, {Riechers},
  {Salvato}, {Schinnerer}, {Tasca}, \& {Toft}}]{vernesa17}
{Smol{\v c}i{\'c}}, V., {Miettinen}, O., {Tomi{\v c}i{\'c}}, N., {et~al.}
  2017{\natexlab{a}}, \aap, 597, A4

\bibitem[{{Smol{\v c}i{\'c}} {et~al.}(2017{\natexlab{b}}){Smol{\v c}i{\'c}},
  {Novak}, {Bondi}, {Ciliegi}, {Mooley}, {Schinnerer}, {Zamorani}, {Navarrete},
  {Bourke}, {Karim}, {Vardoulaki}, {Leslie}, {Delhaize}, {Carilli}, {Myers},
  {Baran}, {Delvecchio}, {Miettinen}, {Banfield}, {Balokovi{\'c}}, {Bertoldi},
  {Capak}, {Frail}, {Hallinan}, {Hao}, {Herrera Ruiz}, {Horesh}, {Ilbert},
  {Intema}, {Jeli{\'c}}, {Kl{\"o}ckner}, {Krpan}, {Kulkarni}, {McCracken},
  {Laigle}, {Middleberg}, {Murphy}, {Sargent}, {Scoville}, \&
  {Sheth}}]{vernesa17b}
{Smol{\v c}i{\'c}}, V., {Novak}, M., {Bondi}, M., {et~al.} 2017{\natexlab{b}},
  \aap, 602, A1

\bibitem[{{Socolovsky} {et~al.}(2018){Socolovsky}, {Almaini}, {Hatch}, {Wild},
  {Maltby}, {Hartley}, \& {Simpson}}]{mildmanneredmiguel18}
{Socolovsky}, M., {Almaini}, O., {Hatch}, N.~A., {et~al.} 2018, \mnras, 476,
  1242

\bibitem[{{Springel} {et~al.}(2005){Springel}, {White}, {Jenkins}, {Frenk},
  {Yoshida}, {Gao}, {Navarro}, {Thacker}, {Croton}, {Helly}, {Peacock}, {Cole},
  {Thomas}, {Couchman}, {Evrard}, {Colberg}, \& {Pearce}}]{springel05}
{Springel}, V., {White}, S. D.~M., {Jenkins}, A., {et~al.} 2005, \nat, 435, 629

\bibitem[{{Steinhauser} {et~al.}(2016){Steinhauser}, {Schindler}, \&
  {Springel}}]{steinhauser16}
{Steinhauser}, D., {Schindler}, S., \& {Springel}, V. 2016, \aap, 591, A51

\bibitem[{{Strait} {et~al.}(2021){Strait}, {Brada{\v{c}}}, {Coe}, {Lemaux},
  {Carnall}, {Bradley}, {Pelliccia}, {Sharon}, {Zitrin}, {Acebron}, {Neufeld},
  {Andrade-Santos}, {Avila}, {Frye}, {Mahler}, {Nonino}, {Ogaz}, {Oguri},
  {Ouchi}, {Paterno-Mahler}, {Stark}, {Mainali}, {Oesch}, {Trenti}, {Carrasco},
  {Dawson}, {Jones}, {Umetsu}, \& {Vulcani}}]{victoria21}
{Strait}, V., {Brada{\v{c}}}, M., {Coe}, D., {et~al.} 2021, \apj, 910, 135

\bibitem[{{Straughn} {et~al.}(2009){Straughn}, {Pirzkal}, {Meurer}, {Cohen},
  {Windhorst}, {Malhotra}, {Rhoads}, {Gardner}, {Hathi}, \&
  {Jansen}}]{straughn09}
{Straughn}, A.~N., {Pirzkal}, N., {Meurer}, G.~R., {et~al.} 2009, \aj, 138,
  1022

\bibitem[{{Strazzullo} {et~al.}(2018){Strazzullo}, {Coogan}, {Daddi},
  {Sargent}, {Gobat}, {Valentino}, {Bethermin}, {Pannella}, {Dickinson},
  {Renzini}, {Arimoto}, {Cimatti}, {Dannerbauer}, {Finoguenov}, {Liu}, \&
  {Onodera}}]{strazzullo18}
{Strazzullo}, V., {Coogan}, R.~T., {Daddi}, E., {et~al.} 2018, \apj, 862, 64

\bibitem[{{Strazzullo} {et~al.}(2013){Strazzullo}, {Gobat}, {Daddi}, {Onodera},
  {Carollo}, {Dickinson}, {Renzini}, {Arimoto}, {Cimatti}, {Finoguenov}, \&
  {Chary}}]{strazz13}
{Strazzullo}, V., {Gobat}, R., {Daddi}, E., {et~al.} 2013, \apj, 772, 118

\bibitem[{{Sugai} {et~al.}(2012){Sugai}, {Karoji}, {Takato}, {Tamura},
  {Shimono}, {Ohyama}, {Ueda}, {Ling}, {Vital de Arruda}, {Barkhouser},
  {Bennett}, {Bickerton}, {Braun}, {Bruno}, {Carr}, {Batista de Carvalho
  Oliveira}, {Chang}, {Chen}, {Dekany}, {Pereira Dominici}, {Ellis}, {Fisher},
  {Gunn}, {Heckman}, {Ho}, {Hu}, {Jaquet}, {Karr}, {Kimura}, {Le F{\`e}vre},
  {Le Mignant}, {Loomis}, {Lupton}, {Madec}, {Marrara}, {Martin}, {Murayama},
  {Cesar de Oliveira}, {Mendes de Oliveira}, {Souza de Oliveira}, {Orndorff},
  {de Paiva Vila{\c{c}}a}, {Macanhan}, {Prieto}, {Bispo dos Santos},
  {Seiffert}, {Smee}, {Smith}, {Sodr{\'e}}, {Spergel}, {Surace}, {Vives},
  {Wang}, \& {Yan}}]{sugai12}
{Sugai}, H., {Karoji}, H., {Takato}, N., {et~al.} 2012, in Society of
  Photo-Optical Instrumentation Engineers (SPIE) Conference Series, Vol. 8446,
  Ground-based and Airborne Instrumentation for Astronomy IV, 84460Y

\bibitem[{{Tacconi} {et~al.}(2018){Tacconi}, {Genzel}, {Saintonge}, {Combes},
  {Garc{\'\i}a-Burillo}, {Neri}, {Bolatto}, {Contini}, {F{\"o}rster Schreiber},
  {Lilly}, {Lutz}, {Wuyts}, {Accurso}, {Boissier}, {Boone}, {Bouch{\'e}},
  {Bournaud}, {Burkert}, {Carollo}, {Cooper}, {Cox}, {Feruglio}, {Freundlich},
  {Herrera-Camus}, {Juneau}, {Lippa}, {Naab}, {Renzini}, {Salome}, {Sternberg},
  {Tadaki}, {{\"U}bler}, {Walter}, {Weiner}, \& {Weiss}}]{tacconi18}
{Tacconi}, L.~J., {Genzel}, R., {Saintonge}, A., {et~al.} 2018, \apj, 853, 179

\bibitem[{{Tacconi} {et~al.}(2020){Tacconi}, {Genzel}, \&
  {Sternberg}}]{tacconi20}
{Tacconi}, L.~J., {Genzel}, R., \& {Sternberg}, A. 2020, \araa, 58, 157

\bibitem[{{Tadaki} {et~al.}(2019){Tadaki}, {Kodama}, {Hayashi}, {Shimakawa},
  {Koyama}, {Lee}, {Tanaka}, {Hatsukade}, {Iono}, {Kohno}, {Matsuda}, {Suzuki},
  {Tamura}, {Toshikawa}, \& {Umehata}}]{tadaki19}
{Tadaki}, K.-i., {Kodama}, T., {Hayashi}, M., {et~al.} 2019, \pasj, 71, 40

\bibitem[{{Talia} {et~al.}(2015){Talia}, {Cimatti}, {Pozzetti}, {Rodighiero},
  {Gruppioni}, {Pozzi}, {Daddi}, {Maraston}, {Mignoli}, \& {Kurk}}]{marghe15}
{Talia}, M., {Cimatti}, A., {Pozzetti}, L., {et~al.} 2015, \aap, 582, A80

\bibitem[{{Tasca} {et~al.}(2015){Tasca}, {Le F{\`e}vre}, {Hathi}, {Schaerer},
  {Ilbert}, {Zamorani}, {Lemaux}, {Cassata}, {Garilli}, {Le Brun}, {Maccagni},
  {Pentericci}, {Thomas}, {Vanzella}, {Zucca}, {Amorin}, {Bardelli},
  {Cassar{\`a}}, {Castellano}, {Cimatti}, {Cucciati}, {Durkalec}, {Fontana},
  {Giavalisco}, {Grazian}, {Paltani}, {Ribeiro}, {Scodeggio}, {Sommariva},
  {Talia}, {Tresse}, {Vergani}, {Capak}, {Charlot}, {Contini}, {de la Torre},
  {Dunlop}, {Fotopoulou}, {Koekemoer}, {L{\'o}pez-Sanjuan}, {Mellier}, {Pforr},
  {Salvato}, {Scoville}, {Taniguchi}, \& {Wang}}]{tizzytasca15}
{Tasca}, L.~A.~M., {Le F{\`e}vre}, O., {Hathi}, N.~P., {et~al.} 2015, \aap,
  581, A54

\bibitem[{{Tasca} {et~al.}(2014){Tasca}, {Le F{\`e}vre}, {L{\'o}pez-Sanjuan},
  {Wang}, {Cassata}, {Garilli}, {Ilbert}, {Le Brun}, {Lemaux}, {Maccagni},
  {Tresse}, {Bardelli}, {Contini}, {Charlot}, {Cucciati}, {Fontana},
  {Giavalisco}, {Kneib}, {Salvato}, {Taniguchi}, {Vergani}, {Zamorani}, \&
  {Zucca}}]{tizzytasca14}
{Tasca}, L.~A.~M., {Le F{\`e}vre}, O., {L{\'o}pez-Sanjuan}, C., {et~al.} 2014,
  \aap, 565, A10

\bibitem[{{Tasca} {et~al.}(2017){Tasca}, {Le F{\`e}vre}, {Ribeiro}, {Thomas},
  {Moreau}, {Cassata}, {Garilli}, {Le Brun}, {Lemaux}, {Maccagni},
  {Pentericci}, {Schaerer}, {Vanzella}, {Zamorani}, {Zucca}, {Amorin},
  {Bardelli}, {Cassar{\`a}}, {Castellano}, {Cimatti}, {Cucciati}, {Durkalec},
  {Fontana}, {Giavalisco}, {Grazian}, {Hathi}, {Ilbert}, {Paltani}, {Pforr},
  {Scodeggio}, {Sommariva}, {Talia}, {Tresse}, {Vergani}, {Capak}, {Charlot},
  {Contini}, {de la Torre}, {Dunlop}, {Fotopoulou}, {Guaita}, {Koekemoer},
  {L{\'o}pez-Sanjuan}, {Mellier}, {Salvato}, {Scoville}, {Taniguchi}, \&
  {Wang}}]{tizzytasca17}
{Tasca}, L.~A.~M., {Le F{\`e}vre}, O., {Ribeiro}, B., {et~al.} 2017, \aap, 600,
  A110

\bibitem[{{Tomczak} {et~al.}(2017){Tomczak}, {Lemaux}, {Lubin}, {Gal}, {Wu},
  {Holden}, {Kocevski}, {Mei}, {Pelliccia}, {Rumbaugh}, \& {Shen}}]{AA17}
{Tomczak}, A.~R., {Lemaux}, B.~C., {Lubin}, L.~M., {et~al.} 2017, \mnras, 472,
  3512

\bibitem[{{Tomczak} {et~al.}(2019){Tomczak}, {Lemaux}, {Lubin}, {Pelliccia},
  {Shen}, {Gal}, {Hung}, {Kocevski}, {Le F{\`e}vre}, \& {Mei}}]{AA19}
{Tomczak}, A.~R., {Lemaux}, B.~C., {Lubin}, L.~M., {et~al.} 2019, \mnras, 484,
  4695

\bibitem[{{Tomczak} {et~al.}(2016){Tomczak}, {Quadri}, {Tran}, {Labb{\'e}},
  {Straatman}, {Papovich}, {Glazebrook}, {Allen}, {Brammer}, {Cowley},
  {Dickinson}, {Elbaz}, {Inami}, {Kacprzak}, {Morrison}, {Nanayakkara},
  {Persson}, {Rees}, {Salmon}, {Schreiber}, {Spitler}, \& {Whitaker}}]{AA16}
{Tomczak}, A.~R., {Quadri}, R.~F., {Tran}, K.-V.~H., {et~al.} 2016, \apj, 817,
  118

\bibitem[{{Tomczak} {et~al.}(2014){Tomczak}, {Quadri}, {Tran}, {Labb{\'e}},
  {Straatman}, {Papovich}, {Glazebrook}, {Allen}, {Brammer}, {Kacprzak},
  {Kawinwanichakij}, {Kelson}, {McCarthy}, {Mehrtens}, {Monson}, {Persson},
  {Spitler}, {Tilvi}, \& {van Dokkum}}]{AA14}
{Tomczak}, A.~R., {Quadri}, R.~F., {Tran}, K.-V.~H., {et~al.} 2014, \apj, 783,
  85

\bibitem[{{Tonnesen} \& {Cen}(2014)}]{tonneson14}
{Tonnesen}, S. \& {Cen}, R. 2014, \apj, 788, 133

\bibitem[{{Toshikawa} {et~al.}(2016){Toshikawa}, {Kashikawa}, {Overzier},
  {Malkan}, {Furusawa}, {Ishikawa}, {Onoue}, {Ota}, {Tanaka}, {Niino}, \&
  {Uchiyama}}]{tersetoshikawa16}
{Toshikawa}, J., {Kashikawa}, N., {Overzier}, R., {et~al.} 2016, \apj, 826, 114

\bibitem[{{Toshikawa} {et~al.}(2014){Toshikawa}, {Kashikawa}, {Overzier},
  {Shibuya}, {Ishikawa}, {Ota}, {Shimasaku}, {Tanaka}, {Hayashi}, {Niino}, \&
  {Onoue}}]{tersetoshikawa14}
{Toshikawa}, J., {Kashikawa}, N., {Overzier}, R., {et~al.} 2014, \apj, 792, 15

\bibitem[{{Toshikawa} {et~al.}(2020){Toshikawa}, {Malkan}, {Kashikawa},
  {Overzier}, {Uchiyama}, {Ota}, {Ishikawa}, \& {Ito}}]{tersetoshikawa20}
{Toshikawa}, J., {Malkan}, M.~A., {Kashikawa}, N., {et~al.} 2020, \apj, 888, 89

\bibitem[{{Toshikawa} {et~al.}(2018){Toshikawa}, {Uchiyama}, {Kashikawa},
  {Ouchi}, {Overzier}, {Ono}, {Harikane}, {Ishikawa}, {Kodama}, {Matsuda},
  {Lin}, {Onoue}, {Tanaka}, {Nagao}, {Akiyama}, {Komiyama}, {Goto}, \&
  {Lee}}]{tersetoshikawa18}
{Toshikawa}, J., {Uchiyama}, H., {Kashikawa}, N., {et~al.} 2018, \pasj, 70, S12

\bibitem[{{Tran} {et~al.}(2003){Tran}, {Franx}, {Illingworth}, {Kelson}, \&
  {van Dokkum}}]{tran03}
{Tran}, K.-V.~H., {Franx}, M., {Illingworth}, G., {Kelson}, D.~D., \& {van
  Dokkum}, P. 2003, \apj, 599, 865

\bibitem[{{Tran} {et~al.}(2010){Tran}, {Papovich}, {Saintonge}, {Brodwin},
  {Dunlop}, {Farrah}, {Finkelstein}, {Finkelstein}, {Lotz}, {McLure},
  {Momcheva}, \& {Willmer}}]{tran10}
{Tran}, K.-V.~H., {Papovich}, C., {Saintonge}, A., {et~al.} 2010, \apjl, 719,
  L126

\bibitem[{{Trump} {et~al.}(2013){Trump}, {Konidaris}, {Barro}, {Koo},
  {Kocevski}, {Juneau}, {Weiner}, {Faber}, {McLean}, \& {Yan}}]{trump13}
{Trump}, J.~R., {Konidaris}, N.~P., {Barro}, G., {et~al.} 2013, \apj, 763, L6

\bibitem[{{van der Burg} {et~al.}(2020){van der Burg}, {Rudnick}, {Balogh},
  {Muzzin}, {Lidman}, {Old}, {Shipley}, {Gilbank}, {McGee}, {Biviano},
  {Cerulo}, {Chan}, {Cooper}, {De Lucia}, {Demarco}, {Forrest}, {Gwyn},
  {Jablonka}, {Kukstas}, {Marchesini}, {Nantais}, {Noble}, {Pintos-Castro},
  {Poggianti}, {Reeves}, {Stefanon}, {Vulcani}, {Webb}, {Wilson}, {Yee}, \&
  {Zaritsky}}]{vanderberg20}
{van der Burg}, R. F.~J., {Rudnick}, G., {Balogh}, M.~L., {et~al.} 2020, \aap,
  638, A112

\bibitem[{{Vanzella} {et~al.}(2008){Vanzella}, {Cristiani}, {Dickinson},
  {Giavalisco}, {Kuntschner}, {Haase}, {Nonino}, {Rosati}, {Cesarsky}, \&
  {Ferguson}}]{eros08}
{Vanzella}, E., {Cristiani}, S., {Dickinson}, M., {et~al.} 2008, \aap, 478, 83

\bibitem[{{Vanzella} {et~al.}(2009){Vanzella}, {Giavalisco}, {Dickinson},
  {Cristiani}, {Nonino}, {Kuntschner}, {Popesso}, {Rosati}, {Renzini}, \&
  {Stern}}]{eros09}
{Vanzella}, E., {Giavalisco}, M., {Dickinson}, M., {et~al.} 2009, \apj, 695,
  1163

\bibitem[{{von der Linden} {et~al.}(2010){von der Linden}, {Wild}, {Kauffmann},
  {White}, \& {Weinmann}}]{vonderlinden10}
{von der Linden}, A., {Wild}, V., {Kauffmann}, G., {White}, S.~D.~M., \&
  {Weinmann}, S. 2010, \mnras, 404, 1231

\bibitem[{{Wang} {et~al.}(2016){Wang}, {Elbaz}, {Daddi}, {Finoguenov}, {Liu},
  {Schreiber}, {Mart{\'\i}n}, {Strazzullo}, {Valentino}, {van der Burg},
  {Zanella}, {Ciesla}, {Gobat}, {Le Brun}, {Pannella}, {Sargent}, {Shu}, {Tan},
  {Cappelluti}, \& {Li}}]{taowang16}
{Wang}, T., {Elbaz}, D., {Daddi}, E., {et~al.} 2016, \apj, 828, 56

\bibitem[{{Wang} {et~al.}(2018){Wang}, {Elbaz}, {Daddi}, {Liu}, {Kodama},
  {Tanaka}, {Schreiber}, {Zanella}, {Valentino}, \& {Sargent}}]{taowang18}
{Wang}, T., {Elbaz}, D., {Daddi}, E., {et~al.} 2018, \apj, 867, L29

\bibitem[{{Wetzel}(2011)}]{wetzel11}
{Wetzel}, A.~R. 2011, \mnras, 412, 49

\bibitem[{{Wetzel} {et~al.}(2013){Wetzel}, {Tinker}, {Conroy}, \& {van den
  Bosch}}]{wetzel13}
{Wetzel}, A.~R., {Tinker}, J.~L., {Conroy}, C., \& {van den Bosch}, F.~C. 2013,
  \mnras, 432, 336

\bibitem[{{Wu} {et~al.}(2014){Wu}, {Gal}, {Lemaux}, {Kocevski}, {Lubin},
  {Rumbaugh}, \& {Squires}}]{pwu14}
{Wu}, P.-F., {Gal}, R.~R., {Lemaux}, B.~C., {et~al.} 2014, \apj, 792, 16

\bibitem[{{Wuyts} {et~al.}(2011){Wuyts}, {F{\"o}rster Schreiber}, {Lutz},
  {Nordon}, {Berta}, {Altieri}, {Andreani}, {Aussel}, {Bongiovanni}, {Cepa},
  {Cimatti}, {Daddi}, {Elbaz}, {Genzel}, {Koekemoer}, {Magnelli}, {Maiolino},
  {McGrath}, {P{\'e}rez Garc{\'\i}a}, {Poglitsch}, {Popesso}, {Pozzi},
  {Sanchez-Portal}, {Sturm}, {Tacconi}, \& {Valtchanov}}]{wuyts11}
{Wuyts}, S., {F{\"o}rster Schreiber}, N.~M., {Lutz}, D., {et~al.} 2011, \apj,
  738, 106

\bibitem[{{Zavala} {et~al.}(2019){Zavala}, {Casey}, {Scoville}, {Champagne},
  {Chiang}, {Dannerbauer}, {Drew}, {Fu}, {Spilker}, {Spitler}, {Tran},
  {Treister}, \& {Toft}}]{zavala19}
{Zavala}, J.~A., {Casey}, C.~M., {Scoville}, N., {et~al.} 2019, \apj, 887, 183

\bibitem[{{Zhou} {et~al.}(2020){Zhou}, {Elbaz}, {Franco}, {Magnelli},
  {Schreiber}, {Wang}, {Ciesla}, {Daddi}, {Dickinson}, {Nagar}, {Magdis},
  {Alexander}, {B{\'e}thermin}, {Demarco}, {Mullaney}, {Bournaud}, {Ferguson},
  {Finkelstein}, {Giavalisco}, {Inami}, {Iono}, {Juneau}, {Lagache}, {Messias},
  {Motohara}, {Okumura}, {Pannella}, {Papovich}, {Pope}, {Rujopakarn}, {Shi},
  {Shu}, \& {Silverman}}]{wenjia20}
{Zhou}, L., {Elbaz}, D., {Franco}, M., {et~al.} 2020, \aap, 642, A155

\end{thebibliography}

\appendix

\section{Modification to the VMC method in the VUDS+ mapping}
\label{appendix:A}

In \S\ref{voronoi} we described the base VMC method used for the mapping, which is identical to that adopted by earlier works that implemented VMC-based density mapping. In this Appendix, we discuss 
the slight modification that we made to this method for the purposes of this study. For each Monte Carlo realization of each redshift slice in the VMC map, we first selected the spectroscopic
subsample by drawing from a uniform distribution ranging from 0 to 1 and retaining those $z_{spec}$ measurements where the number drawn was below a certain likelihood threshold set
for that slice and iteration for all galaxies flags=X2/X9 and X3/X4. As a reminder, those galaxies with flag=X1 never had their spectral information used. The likelihood threshold
of each flag was determined by the reliability of a $z_{spec}$ measurement with a given flag, which was estimated from $\sim$200 objects that were independently observed multiple
times as part of the VUDS survey. From these $\sim$200 objects, we estimated, for each flag, from the number of observations that yielded statistically identical
redshift measures both a likelihood of a redshift of a given flag to be reproducible in independent observations with independent measurements as well as its associated uncertainty.
Due to small numbers, objects assigned a flag=X9 were treated in combination with those assigned a flag=X2 and objects assigned a flag=X3 were grouped with flag=X4 objects.
The likelihood values and their associated
uncertainties were $0.70^{+0.04}_{-0.06}$ and $0.993^{+0.004}_{-0.006}$ for flags=X2/X9 and flags=X3/X4, respectively. Though we do not use them in this study,
this same exercise was performed for flag=X1 objects and returned a likelihood value of $0.41^{+0.06}_{-0.08}$, a relatively high value given this is the lowest confidence
flag assigned to objects that have some detectable signal. These thresholds depart slightly from the values listed in \citet{dong15} due to the differing methodology used to
assess the reliability of the flags. We note that the final VUDS+ spectroscopic sample contains roughly equal numbers of flag=X2/X9 and flag=X3/X4 objects (47\% to 53\%, respectively).

To determine the threshold for flags=X2/X9 and X3/X4 objects for a given iteration and slice, a Gaussian was sampled with a mean and dispersion corresponding to the likelihood
values and their associated uncertainties above. For each flag combination, this sampling would set the threshold for a given iteration/slice. For example, a sampling value
of +$1\sigma$ for the flag=X2/X9 objects would set the threshold for those objects to 0.70 + 0.04 = 0.74 in that iteration/slice. For each object, if the 
determined threshold for the appropriate flag value was not exceeded, $z_{spec}$ would be retained for that realization, otherwise the $z_{phot}$ information was instead used.
For each object where the $z_{phot}$ information was adopted, or was the only information available, the original $z_{phot}$ for that object was perturbed by sampling from an
asymmetric Gaussian distribution with $\sigma$ values that correspond to the lower and upper effective $1\sigma$ $z_{phot}$ uncertainties. This treatment generated a unique
set of redshifts for our entire sample for each iteration of each redshift slice from which to select the objects to Voronoi tessellate over. Results of preliminary tests
on the accuracy and precision of various incarnations of this method are discussed in \citet{AA17} and \citet{lem18}. Additionally, the effectiveness of a version of this
method to recover groups and clusters in the intermediate redshift ORELSE \citep{lub09} survey is
discussed in detail in \cite{denise20} and will be discussed in the context of the VUDS and C3VO surveys in a future work. 

\end{document}